\documentclass[twocolumn,twocolappendix]{aastex7}

\usepackage{amsmath}
\usepackage{booktabs,tabularx}
\usepackage{booktabs}   
\usepackage{multirow} 
\shorttitle{AT2025ulz and S250818k: Investigating Early Time Observations}
\shortauthors{X. J. Hall, M. Busmann et al.}
\submitjournal{ApJL}

\usepackage{siunitx}
\usepackage{booktabs}
\usepackage{longtable}
\usepackage{caption}

\definecolor{blazeorange}{rgb}{1.0, 0.4, 0.0}
\definecolor{seagreen}{rgb}{0.18, 0.55, 0.34}
\definecolor{darkgreen}{rgb}{0.08, 0.45, 0.2}
\definecolor{rufous}{rgb}{0.66, 0.11, 0.03}
\definecolor{royalfuchsia}{rgb}{0.79, 0.17, 0.57}
\definecolor{scarlet}{rgb}{1.0, 0.13, 0.0}
\definecolor{royalpurple}{rgb}{0.47, 0.32, 0.66}

\newcommand{\mcwilliams}{
    McWilliams Center for Cosmology and Astrophysics,
    Department of Physics,
    Carnegie Mellon University,
    5000 Forbes Avenue, Pittsburgh, PA 15213, USA
}

\newcommand{\lmu}{
    University Observatory, 
    Faculty of Physics, 
    Ludwig-Maximilians-Universität München, 
    Scheinerstr. 1, 81679 Munich, Germany
}

\begin{document}

\title{AT2025ulz and S250818k: Investigating early time observations of a subsolar mass gravitational-wave binary neutron star merger candidate
}

\correspondingauthor{Xander J. Hall, Malte Busmann}

\author[0000-0002-9364-5419]{Xander J. Hall}
\affiliation{\mcwilliams}
\email[show]{xhall@cmu.edu}

\author[0009-0001-0574-2332]{Malte Busmann}\thanks{Recipient of a Wübben Stiftung Wissenschaft Student Grant}
    \affiliation{\lmu}
    \email[show]{m.busmann@physik.lmu.de}

\author[0009-0001-5350-7468]{Hauke Koehn}
\email{hauke.koehn@uni-potsdam.de}
\affiliation{Institut für Physik und Astronomie, Universität Potsdam, Haus 28, Karl-Liebknecht-Str. 24/25, 14476, Potsdam, Germany}

\author[0009-0000-4830-1484]{Keerthi Kunnumkai}
\affiliation{\mcwilliams}
\email{kkunnumk@andrew.cmu.edu}

\author[0000-0002-6011-0530]{Antonella Palmese}
\affiliation{\mcwilliams}
\email{palmese@cmu.edu}

\author[0000-0002-9700-0036]{Brendan O'Connor}
    \altaffiliation{McWilliams Fellow}
    \affiliation{\mcwilliams}
    \email{boconno2@andrew.cmu.edu}  

\author[0009-0006-7990-0547]{James Freeburn}
	\affiliation{University of North Carolina at Chapel Hill, 120 E. Cameron Ave., Chapel Hill, NC 27514, USA}
	\email{jamesfreeburn54@gmail.com}

\author[0000-0001-7201-1938]{Lei Hu}
\affiliation{\mcwilliams}
\email{leihu@andrew.cmu.edu}

\author[0000-0003-3270-7644]{Daniel Gruen}
	\affiliation{\lmu}
	\affiliation{Excellence Cluster ORIGINS, Boltzmannstr. 2, 85748 Garching, Germany}
	\email{daniel.gruen@lmu.de}

\author[0000-0003-2374-307X]{Tim Dietrich}
\email{tim.dietrich@aei.mpg.de}
\affiliation{Institut für Physik und Astronomie, Universität Potsdam, Haus 28, Karl-Liebknecht-Str. 24/25, 14476, Potsdam, Germany}
\affiliation{Max Planck Institute for Gravitational Physics (Albert Einstein Institute), Am Mühlenberg 1, Potsdam 14476, Germany}

\author[0000-0002-8255-5127]{Mattia Bulla}
\email{mattia.bulla@unife.it}
\affiliation{Department of Physics and Earth Science, University of Ferrara, via Saragat 1, I-44122 Ferrara, Italy}
\affiliation{INFN, Sezione di Ferrara, via Saragat 1, I-44122 Ferrara, Italy}
\affiliation{INAF, Osservatorio Astronomico d’Abruzzo, via Mentore Maggini snc, 64100 Teramo, Italy}

\author[0000-0002-8262-2924]{Michael W. Coughlin}
\affiliation{School of Physics and Astronomy, University of Minnesota, Minneapolis, MN 55414, USA}
\email{cough052@umn.edu}

\author[0000-0002-7686-3334]{Sarah Antier}\email{antier@ijclab.in2p3.fr}
\affiliation{IJCLab, Univ Paris-Saclay, CNRS/IN2P3, Orsay, France}

\author[0000-0003-3224-2146]{Marion Pillas}\email{marion.pillas@iap.fr}
\affiliation{Institut d’Astrophysique de Paris, Sorbonne Université and CNRS, UMR 7095, 98 bis bd Arago, 75014 Paris, France}

\author[0000-0003-0511-0228]{Paul A. Price}
\email{price@astro.princeton.edu}
\affiliation{Department of Astrophysical Sciences, Princeton University, Princeton, NJ 08544, USA}

\author[0000-0002-2184-6430]{Tomás Ahumada}\email{tahumada@astro.caltech.edu}
\affiliation{Cahill Center for Astrophysics, California Institute of Technology, MC 249-17, 1216 E California Boulevard, Pasadena, CA, 91125, USA}

\author[0000-0003-3433-2698]{Ariel Amsellem}
\affiliation{\mcwilliams}
\email{aamselle@andrew.cmu.edu}

\author[0000-0002-8977-1498]{Igor Andreoni}
	\affiliation{University of North Carolina at Chapel Hill, 120 E. Cameron Ave., Chapel Hill, NC 27514, USA}
	\email{igor.andreoni@unc.edu}

\author[0009-0002-6662-4900]{Jule Augustin}
    \affiliation{\lmu}
    \email{jule.augustin@physik.lmu.de}

\author[0000-0002-1270-7666]{Tom\'as Cabrera}
\affiliation{\mcwilliams}
\email{tcabrera@andrew.cmu.edu}

\author[0009-0008-0132-0993]{Rasika Deshpande}
\affiliation{\lmu}
\email{deshpanderasika7@gmail.com}

\author[0009-0006-7670-9843]{Jennifer Fabà-Moreno}
\affiliation{\lmu}
\email{J.Faba@campus.lmu.de}

\author[0009-0008-2754-1946]{Julius Gassert}
\affiliation{\lmu}
\affiliation{\mcwilliams}
\email{julius.gassert@campus.lmu.de}

\author[0000-0003-0035-651X]{Sergey Karpov}\email{karpov@fzu.cz}
\affiliation{FZU - Institute of Physics of the Czech Academy of Sciences, Na Slovance 1999/2, CZ-182 21, Praha, Czech Republic}

\author[0000-0002-5619-4938]{Mansi Kasliwal}
\email{mansi@astro.caltech.edu}
\affiliation{Division of Physics, Mathematics and Astronomy, California Institute of Technology, Pasadena, CA 91125, USA}

\author[0000-0003-2362-0459]{Ignacio Maga\~na~Hernandez}
\altaffiliation{McWilliams Fellow}
\affiliation{\mcwilliams}
\email{imhernan@andrew.cmu.edu}

\author[0000-0003-2271-1527]{Rachel Mandelbaum}
\email{rmandelb@andrew.cmu.edu}
\affiliation{\mcwilliams}

\author[0000-0001-7129-1325]{Felipe Fontinele Nunes}
\affiliation{School of Physics and Astronomy, University of Minnesota, Minneapolis, MN 55414, USA}
\affiliation{NSF Institute on Accelerated AI Algorithms for Data-Driven Discovery (A3D3), MIT, Cambridge, MA 02139 and University of Minnesota, Minneapolis, MN 55455, USA}
\email{fonti007@umn.edu}

\author[0000-0001-7041-3239]{Peter T. H. Pang}
\email{thopang@nikhef.nl}
\affiliation{Nikhef, Science Park 105, 1098 XG Amsterdam, The Netherlands}
\affiliation{Institute for Gravitational and Subatomic Physics (GRASP), Utrecht University, Princetonplein 1, 3584 CC Utrecht, The Netherlands}

\author[0000-0002-1154-8317]{Julian Sommer}
\affiliation{\lmu}
\email{julian.sommer@campus.lmu.de}

\author[0000-0003-2434-0387]{Robert Stein}
\email{rdstein@umd.edu}
\affiliation{Department of Astronomy, University of Maryland, College Park, MD 20742, USA} \affiliation{Joint Space-Science Institute, University of Maryland, College Park, MD 20742, USA} \affiliation{Astrophysics Science Division, NASA Goddard Space Flight Center, MC 661, Greenbelt, MD 20771, USA}

\author[0009-0009-0232-9081]{Constantin Tabor}
    \affiliation{\lmu}
    \email{constantin.tabor@campus.lmu.de}

\author[0009-0008-0928-7884]{Pablo Vega}
    \affiliation{\lmu}
    \email{P.Vega@campus.lmu.de}

\author[0009-0006-2797-3808]{Thibeau Wouters}
\email{t.r.i.wouters@uu.nl}
\affiliation{Institute for Gravitational and Subatomic Physics (GRASP), Utrecht University, Princetonplein 1, 3584 CC Utrecht, The Netherlands} \affiliation{Nikhef, Science Park 105, 1098 XG Amsterdam, The Netherlands}


\author[0000-0002-0772-6280]{Xiaoxiong Zuo}
    \affiliation{\lmu}
    \email{zuoxx@bao.ac.cn}


\begin{abstract}

Over the past LIGO/Virgo/KAGRA (LVK) observing runs, it has become increasingly clear that identifying the next electromagnetic counterparts to gravitational-wave (GW) neutron star mergers will likely be more challenging compared to the case of GW170817. The rarity of these GW events, and their electromagnetic counterparts, motivates rapid searches of any candidate binary neutron star (BNS) merger detected by the LVK. We present our extensive photometric and spectroscopic campaign of the candidate counterpart AT2025ulz to the low-significance GW event S250818k, which had a ${\sim} 29\%$ probability of being a BNS merger. We demonstrate that during the first five days, the luminosity and color evolution of AT2025ulz are consistent with both kilonova and shock cooling models, although a Bayesian model comparison shows preference for the shock cooling model, underscoring the ambiguity inherent to early data obtained over only a few days. Continued monitoring beyond this window reveals a rise and color evolution incompatible with kilonova models and instead consistent with a supernova. This event emphasizes the difficulty in identifying the electromagnetic counterparts to BNS mergers and the significant allotment of observing time necessary to robustly differentiate kilonovae from impostors. 

\end{abstract}



\section{Introduction} 
\label{sec:intro}

It has been more than a decade since the first detection of gravitational-waves (GWs) in 2015 \citep{ligo_scientific_collaboration_and_virgo_collaboration_observation_2016}. Since then, many groups of scientists spread across the globe have coordinated exhaustive campaigns to capture the elusive electromagnetic counterparts of compact-object mergers. This effort culminated in the landmark discovery of GW170817 \citep{abbott_bns_2017}, along with its short gamma ray burst \citep[GRB;][]{Goldstein2017,Savchenko2017, hallinan_radio_2017} and kilonova \citep[KN;][]{Coulter2017,Troja2017,Evans2017,Arcavi2017,SoaresSantos2017,Drout2017,Kasliwal2017,Margutti2017}. GW170817 transformed our understanding of kilonova physics and established multimessenger astronomy as a key field of modern astrophysics. Since then, astronomers have been unable to confidently associate any other gravitational-wave event with an electromagnetic source. Several candidates have been suggested for both binary black hole \citep[BBH;][]{Graham2020, Graham2023,Cabrera2024} and binary neutron star \citep[BNS;][]{Moroianu} mergers, but their association to the GW events remain inconclusive \citep{ashton_current_2021,palmese_ligovirgo_2021,2024ApJ...977..122B}.

Even in the absence of confirmed counterparts, gravitational-wave detections alone have revolutionized our understanding of compact-object populations \citep{abbott_population_2023, the_ligo_scientific_collaboration_gwtc-40_2025}. Large samples of mergers provide valuable insights into compact binary formation channels (e.g. \citealt{Andrews2019a,callister2024observedgravitationalwavepopulations}) and power statistical “dark siren” tests of cosmological physics \citep{schutz1986determining,Chen2018, palmese_standard_2023, GWCosmology}. The detection of gravitational-wave Neutron Star–Black Hole (NSBH) mergers, and especially those with black holes in the low-mass gap between $3-5~M_\odot$ has opened new possibilities for tidal disruption signatures and short gamma-ray bursts \citep{abbott_observation_2021, zhu_long-duration_2022,martineau_black_2024,kunnumkai_GW230529,kunnumkai_O5}.

At the same time, each attempted search for electromagnetic (EM) counterparts sharpens our ability to identify promising candidates and to rule out unrelated transients such as fast blue optical transients \citep[FBOT;][]{HoFBOTs,Perley2019}, rapidly evolving supernovae \citep{Agudo2023}, active galactic nuclei flares \citep{ulrich_variability_1997, kelly_are_2009}, and other so-called kilonova impostors \citep{barna_iib_2025}. This practice is invaluable as facilities like the Vera C. Rubin Observatory begin to deliver wide-field, rapid Target of Opportunity (ToO) capabilities \citep{Andreoni2022gw,Stevenson2025}. Rubin has already demonstrated its ability to follow up a well-localized BBH merger \citep{macbride_ligovirgokagra_2025}, and in principle it can detect kilonovae as distant as 400–500 Mpc in regular survey operations \citep{Scolnic2018,Chase2022,Andreoni2022kn,andrade_effect_2025}. Yet, this ability remains largely untested, and an outstanding challenge is deciding which candidates merit scarce time on the world’s most powerful observatories and which can be discarded as unrelated impostors. 

On August 18th, 2025, the LIGO-Virgo-KAGRA (LVK) collaboration reported S250818k \citep{chaudhary_low-latency_2024, ligo_scientific_collaboration_ligovirgokagra_2025}. Based on the GW signal, the LVK collaboration estimated a $71\%$ probability that the alert was due to terrestrial noise and a $29\%$ chance of it being astrophysical in origin and consistent with a BNS merger \citep{ligo_scientific_collaboration_ligovirgokagra_2025}. The report included a chirp mass ($< 0.87~ \text{M}_\odot$) implying a sub-solar-mass compact object \citep{ligo_scientific_collaboration_ligovirgokagra_2025}. 

The detection of a sub-solar mass merger candidate has generated unanswered questions about stellar death pathways and the formation of exotic remnants, for example through the formation of low-mass neutron star mergers in the accretion disk of a core collapse supernova \citep{Metzger2024,ChenMetzger2025}. The observation of an event candidate such as S250818k highlights how GW data can reveal hidden corners of the compact-object zoo that may not be accessible electromagnetically.

The Zwicky Transient Facility \citep[ZTF;][]{Bellm2019} rapidly covered the localization provided by this event and discovered the transient AT2025ulz (ZTF25abjmnps; \citealt{stein_ligovirgokagra_2025,Kasliwal2025sn}). This transient was then rapidly followed up with the Fraunhofer Telescope at Wendelstein Observatory \citep[FTW;][]{busmann_ligovirgokagra_2025}. At first, AT2025ulz appeared to be an unusually bright kilonova, with rapid color evolution and fast fading \citep{hall_ligovirgokagra_2025}. However, within just a few days the light curve plateaued, re-brightened, and developed a pronounced hydrogen P Cygni feature, revealing a supernova-like nature (\citealt{2025GCN.41507....1F,2025GCN.41532....1B,2025GCN.41544....1B,2025GCN.41540....1G,2025GCN.41532....1B,Gillanders25ulz,Kasliwal2025sn,Hall2025desi,Yang2025ulz,Franz2025}). S250818k and AT2025ulz provide an opportune test case for evaluating how future wide-field searches might capture kilonova-like candidates at cosmological distances.

In \S~\ref{sec:obs}, we describe the rapid follow-up campaign undertaken by our team using multiple telescopes.
In \S~\ref{sec:results}, we compare the early data to both kilonova and shock cooling supernova models, highlighting the initial ambiguity and the importance of early-time, high-cadence observations. Finally in \S~\ref{sec:discuss} and \ref{sec:conclusions}, we consider what this event reveals about the feasibility of optical kilonova follow-up at distances of ${\sim}400$ Mpc, and the lessons it offers for maximizing the potential of Rubin and other next-generation surveys in the search for the next multi-messenger gold mine.

Throughout the manuscript, we adopt the host redshift $z = 0.08484$ as reported in \cite{Hall2025desi}, we report all magnitudes in the AB system, and assume a Planck 2018 \citep{aghanim_planck_2020} flat $\Lambda$CDM cosmology with the following parameters: $H_{0} = 67.66~\mathrm{km\,s^{-1}\,Mpc^{-1}}$,  $\Omega_{m} = 0.30966$, $T_{\mathrm{CMB}} = 2.7255~\mathrm{K}$, $N_{\mathrm{eff}} = 3.046$, $m_{\nu} = [0.0,\, 0.0,\, 0.06]~\mathrm{eV}$, and $\Omega_{b} = 0.04897$., $T_{\mathrm{CMB}} = 2.7255~\mathrm{K}$, $N_{\mathrm{eff}} = 3.046$, $m_{\nu} = [0.0,\, 0.0,\, 0.06]~\mathrm{eV}$, and $\Omega_{b} = 0.04897$.

\begin{figure*}
    \centering
    \includegraphics[width=0.98\linewidth]{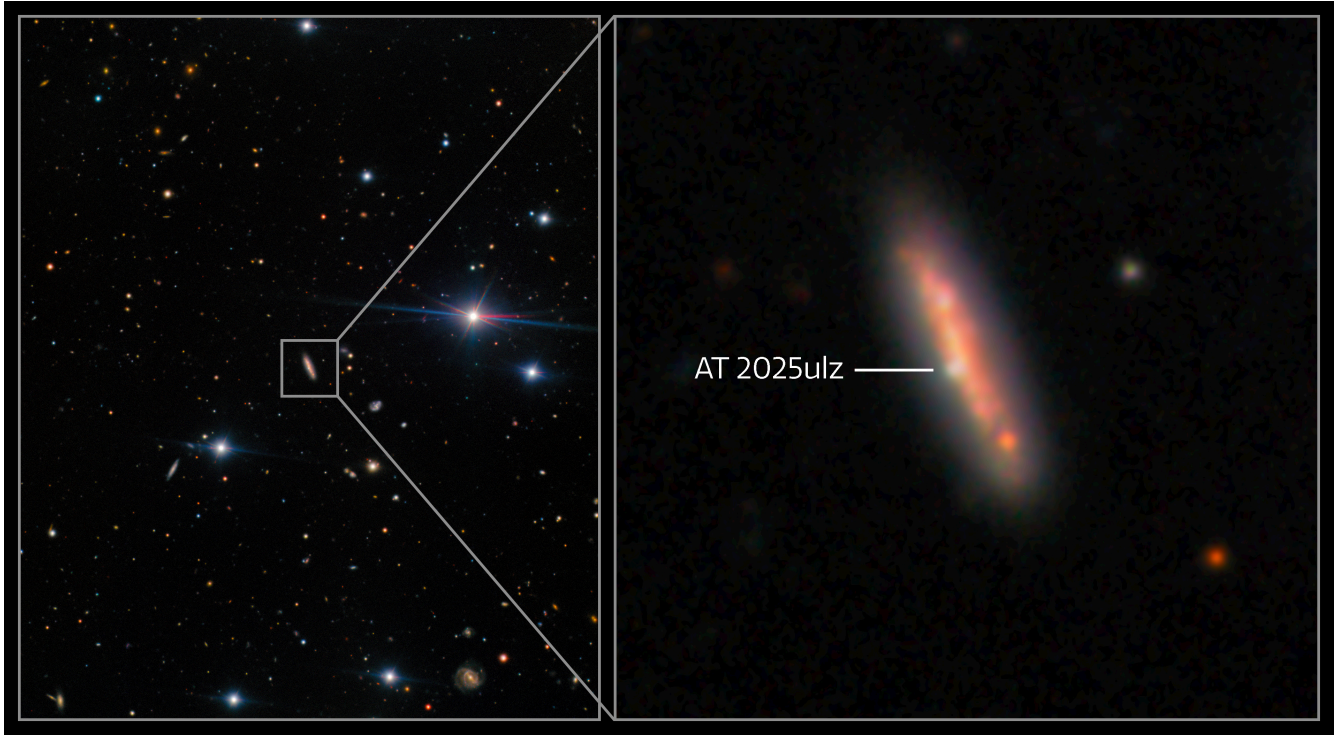}
    \caption{A representative color image of the field of AT2025ulz using the $griY$ images %
    obtained on 2025-08-20 with GMOS mounted on Gemini-North. The right panel zooms in on the host galaxy of AT2025ulz. Clumpy star forming regions can be seen across the galaxy. The transient is visible, even without image subtraction, close to the center of the host (though clearly offset from the nucleus). The image is oriented such that North is up and East is to the left. Image credit: NOIRLab CEE Team.}
    \label{fig:finder}
\end{figure*}





\section{Observations}
\label{sec:obs}


\begin{figure*}
    \centering
    \includegraphics[width=0.98\linewidth]{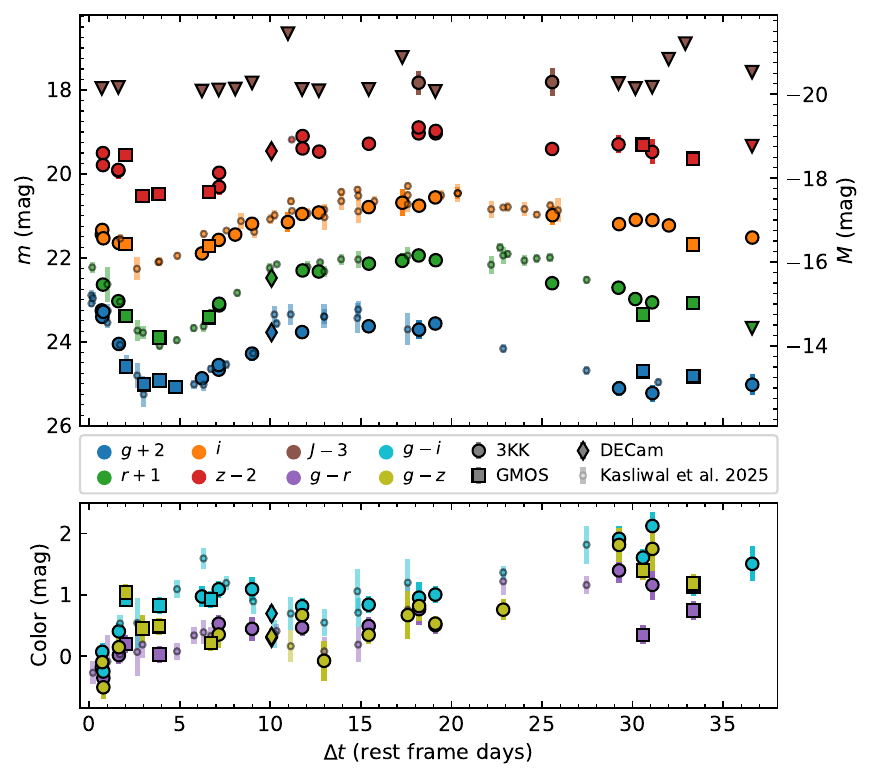}
    \caption{\textbf{Top:} Optical and near-infrared lightcurve ($grizJ$) of AT2025ulz. We supplemented our photometry (3KK and GMOS) with the data presented by \citet{Kasliwal2025sn}. The apparent magnitude $m$ is shown on the left axis and the absolute magnitude $M$ is shown on the right.  \textbf{Bottom:} Color of AT2025ulz as a function of time in $g-r$, $g-i$, and $g-z$. All photometry has been corrected for Galactic extinction \citep{Schlafly2011}. }
    \label{fig:lightcurve}
\end{figure*}

\subsection{Fraunhofer Telescope at Wendelstein Observatory}

We observed with the Three Channel Imager (3KK; \citealt{lang2016wendelstein}) instrument mounted on the 2.1\,m FTW \citep{2014SPIE.9145E..2DH} in the $g$, $r$, $i$, $z$, and $J$ bands betwen 0.7-36.6 days after the GW trigger, see Table \ref{tab:photometry}. The optical CCD and near-infrared (NIR) CMOS data were reduced using a custom pipeline \citep{2002A&A...381.1095G, 2025arXiv250314588B}. For the astrometric calibration of the images, we used the Gaia EDR3 catalog \citep{Gaia2021, 2021A&A...649A...2L, gaiaEDR3}. Tools from the AstrOmatic software suite \citep{1996A&AS..117..393B, 2006ASPC..351..112B, 2002ASPC..281..228B} were used for the coaddition of each epoch's individual exposures.

\subsection{Gemini-North}

\subsubsection{Imaging}


We imaged the location of AT2025ulz using GMOS mounted on the 8.1\,m Gemini-North telescope over numerous epochs in $g$, $r$, $i$, $z$ and $Y$-bands (GN-2025A-FT-114; PI: O'Connor). The images were reduced with the \texttt{DRAGONS} pipeline \citep{2019ASPC..523..321L}. Astometric calibration was determined using Gaia \citep[e.g.,][]{Gaia2021}. The log of observations is reported in Table \ref{tab:photometry}. 

\subsection{Spectroscopy}

We observed the location of AT2025ulz using GMOS mounted on the 8.1\,m Gemini-North telescope on at 2025-08-22 06:51:11 using the B480 grating with $4\times600\,\text{s}$ observations centered at 660nm and 670nm (GN-2025A-FT-216; PI: Palmese). The spectrum was reduced with the \texttt{DRAGONS} pipeline \citep{2019ASPC..523..321L}. The spectrum is presented in Figure \ref{fig:modeled_spectra}.

\subsection{Dark Energy Camera}

We observed with the Dark Energy Camera (DECam) instrument mounted on the Blanco 4\,m Telescope at the Cerro Tololo Inter-American Observatory (CTIO) using the $g$, $r$, and $z$ bands (PI: Palmese; Table \ref{tab:photometry}). The images were astrometric calibrated against Gaia DR3 \citep{Gaia2021, 2021A&A...649A...2L, gaiaEDR3}. We use \texttt{SFFT} \citep{hu_image_2022} for image subtraction against archival DECam images and photometrically calibrate using the Pan-STARRS1 catalog \citep[PS1;][]{2010SPIE.7733E..0EK}. See \citet{Cabrera2024,Hu2025} for further details on the DECam analysis pipeline.

\subsection{Subaru}
\label{hsc}

We retrieved 16 archival exposures of the field of AT2025ulz obtained with the 8.2\,m Subaru Telescope Hyper Suprime-Cam (HSC; \citealt{2018PASJ...70S...1M}) from SMOKA \citep{2002ASPC..281..298B}, and processed them with the LSST Science Pipeline \citep{10.71929/rubin/2570545} version 28 (an updated version of the HSC pipeline; \citealt{2018PASJ...70S...5B}). The exposures 
are all in $g$ band, and were taken between 2016 and 2023. We used bias, dark, and flat frames from the HSC Strategic Survey Program (SSP; \citealt{2022PASJ...74..247A}), and used the Pan-STARRS1 catalog \citep{2020ApJS..251....6M} and Gaia DR2 catalog \citep{2018A&A...616A...1G} for photometric and astrometric calibration, respectively. Due to the small number of exposures and availability of a single band, we disabled the sky correction and multicolor photometric calibration. The result is a coadded image with a depth and image quality approximately matching that of our Gemini observations, suitable for use as a reference image for image subtraction.

\subsection{Hobby-Eberly Telescope}

We observed AT2025ulz with the 11\,m Hobby-Eberly Telescope (HET; \citealt{1998SPIE.3352...34R, 2021AJ....162..298H}) at McDonald Observatory through program M25-3-005 (PI: Gruen) at 2025-08-21 02:59h UTC and 2025-08-23 02:54h UTC, near the minimum of the light curve.
Scheduling used the HET queue system \citep{2007PASP..119..556S}.
We observed the transient with the red (650-1050 nm) low-resolution integral-field spectrograph (LRS2-R; \citealt{Chonis2014,Chonis2016}).
The raw LRS2 data were processed with \texttt{Panacea}\footnote{\url{https://github.com/grzeimann/Panacea}}, which performs detrending, fiber tracing, fiber wavelength evaluation, fiber extraction, fiber-to-fiber normalization, and flux calibration.
The absolute flux calibration comes from standard star measurements and estimates of the mirror illumination as well as the exposure throughput from guider images. Astrometry was determined using the positions of stars detected in the parallel VIRUS observations.
We extracted the flux-calibrated data cube with the \texttt{LRS2Multi}\footnote{\url{https://github.com/grzeimann/LRS2Multi}} package.  

\subsection{Template Image Selection}

For our $grz$-band templates, we use Legacy Survey DR10 images \citep{2019AJ....157..168D}. However, we note that we also identified a deeper $g$-band template taken with the Subaru Telescope (\S \ref{hsc}). We also do subtractions with this template and note that the change in our reported g-band photometry is less than $1\sigma$ in all epochs. In order to ensure photometric consistency with other works \citep{Kasliwal2025sn}, we utilized the LS $g$-band template. For our $i$-band template, we use PS1 imaging \citep{Chambers2016, waters_pan-starrs_2020}. For our $J$-band template, we use observations from the United Kingdom Infrared Telescope \citep[UKIRT;][]{dye_ukirt_2018}. It should be noted that all of our imaging is deeper than the imaging provided by UKIRT leaving the template here limiting the depth of observations. For the $Y$ band imaging taken by GMOS, the use of $y_{P1}$ templates was attempted; however, a high quality difference image was not possible due to the vast difference in band passes. 

\subsection{Image Subtraction and Photometry}

For all instruments, we use the Saccadic Fast Fourier Transform (\texttt{SFFT}; \citealt{hu_image_2022}) algorithm for image subtraction. 
The image subtraction was complicated by the several noticeable star forming regions within the host galaxy and the limited depth of the available templates. After tweaking the \texttt{SFFT} software, we obtain high quality difference images (see Appendix \ref{app:imagesub}, Figure \ref{fig:FC_2025ulz_diff}). All photometry was then performed using forced aperture photometry on the location of the transient. We used the PS1 catalog \citep{2010SPIE.7733E..0EK} for the optical photometric calibration and the 2MASS catalog \citep{Skrutskie2006} for the $J$ band calibrations. This analysis follows the same procedures outlined in \citet{Kasliwal2025sn} for a fully consistent set of AT2025ulz photometry from 0.1 to 40+ days from the GW trigger. All photometry was then corrected for Milky Way extinction using dust maps from \citet{schlafly_measuring_2011} and the extinction curve from \citet{fitzpatrick_correcting_1999}. 
The full lightcurve of AT2025ulz is displayed in Figure \ref{fig:lightcurve} and tabulated in Appendix \ref{app:phot} and Table \ref{tab:photometry}. This lightcurve (Figure \ref{fig:lightcurve}) is complemented with additional data presented in \citet{Kasliwal2025sn}.



\section{Analysis and Results}
\label{sec:results}


\begin{figure*}
    \centering
    \includegraphics[width=0.98\linewidth]{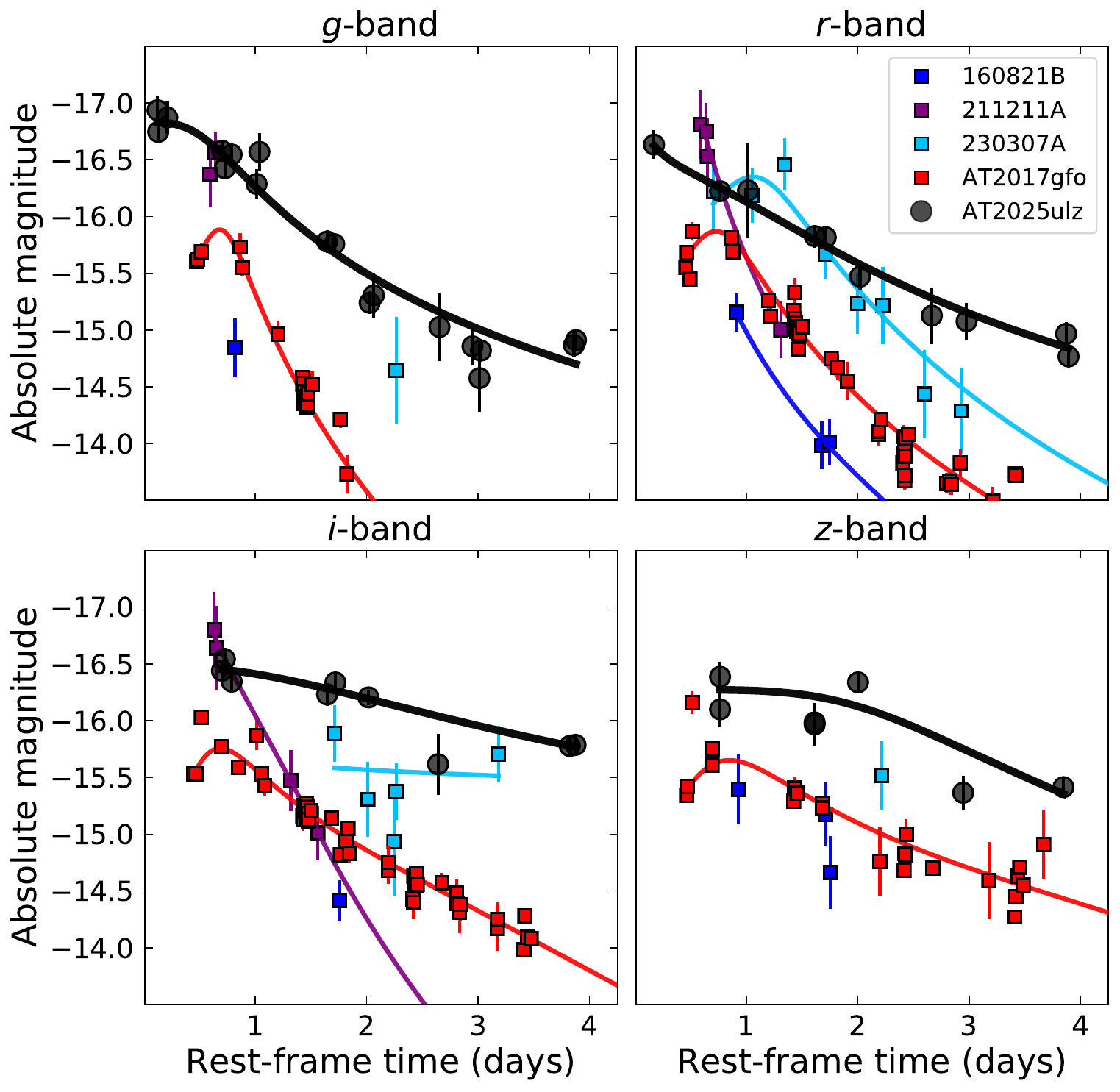}
    \caption{Comparison of the rest-frame lightcurve of AT2025ulz to the lightcurves of GRB-KN (GW170817/AT2017gfo and GRBs 160821B, 211211A, and 230307A) in the $griz$ bands. The afterglow subtracted lightcurves of the GRB-KN were taken from \citet{Rastinejad2025kn}, but originally presented in \citet{Troja2019b,Rastinejad2022,Troja2022,Levan2023,Yang2024}. The lightcurves have been fit with a spline to guide the eye. }
    \label{fig:KNcomparison}
\end{figure*}

\begin{figure}
    \centering
    \includegraphics[width= \linewidth]{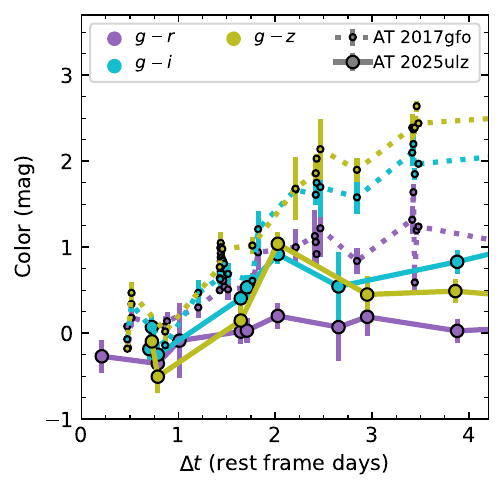}
    \caption{Color comparison of AT 2025ulz (solid lines) and AT2017gfo (dashed lines) over the first few rest frame days. We note that while the color evolution of AT2025ulz is significant it does not redden to the same degree as AT2017gfo.}
    \label{fig:color_comparison}
\end{figure}

\subsection{Phenomenological Comparison to GRB Kilonovae}


The early lightcurve of AT2025ulz (Figure \ref{fig:lightcurve}) shows rapid fading and significant reddening over the first 5 days before plateauing and subsequently rising. The latter behavior clearly proves this source is a supernova, but the first few days of data appeared typical of the expectations for kilonova emission, marking AT2025ulz as an important example of a kilonova impostor. Here we present a phenomenological comparison (Figure \ref{fig:KNcomparison}) between the first few days of AT2025ulz to GW170817/AT2017gfo and other kilonova candidates identified following gamma-ray bursts (GRBs), hereafter GRB-KN. 

We compiled the lightcurve of AT2017gfo from the literature \citep[e.g.,][]{Arcavi2017,Tanvir2017,SoaresSantos2017,Coulter2017,Cowperthwaite2017,Valenti2017,Pian2017,Troja2017,Smartt2017,Drout2017}. We selected GRB-KN from the sample compiled by \citet{Rastinejad2025kn}. This sample includes both short GRBs \citep{Berger2013kilonova,Tanvir2013,Troja150101B,Troja2019b,Fong2021kn,OConnor2021kn} and the two recent long duration GRBs displaying kilonova emission \citep{Rastinejad2022,Troja2022,Levan2023,Yang2024,Gillanders2023}. We focused our comparison on the events with the highest quality lightcurves that included detections in multiple bands. In the end, we include only AT2017gfo and GRBs 160821B, 211211A, and 230307A \citep{Troja2019b,Rastinejad2022,Troja2022,Levan2023,Yang2024} in our comparison shown in Figure \ref{fig:KNcomparison}. We utilize the afterglow subtracted lightcurves derived by \citet{Rastinejad2025kn}. 

It is not expected that every kilonova will look like AT2017gfo and the more luminous peak absolute magnitude of AT2025ulz did not automatically exclude it as a kilonova candidate given the broad range of possibilities that are shown in theoretical kilonova lightcurves. In fact, there is a great diversity in observed GRB-KN lightcurves, peak magnitudes, and colors \citep[Figure \ref{fig:KNcomparison}; e.g.,][]{Rossi2020,OConnor2021kn,Rastinejad2021,Troja2023,Rastinejad2025kn}. Moreover, the peak absolute magnitude ($M \sim -17$) of AT2025ulz is similar to that found in GRBs 211211A and 230307A, which demonstrates observationally that more luminous kilonovae exist in nature. This means that brighter candidates cannot automatically be excluded without further examination. The main conclusion is that there is no uniform, cookie-cutter approach to identify or excluding a kilonova solely based on the early photometry during a wide-field search. The addition of optical or near-infrared spectroscopy is required. 



\subsection{Bayesian Light Curve Analysis}
\label{nmmamodel}

\subsubsection{Model Setup}
\label{subsec:lcanalysis_setup}
\begin{figure*}
    \includegraphics[width=\textwidth]{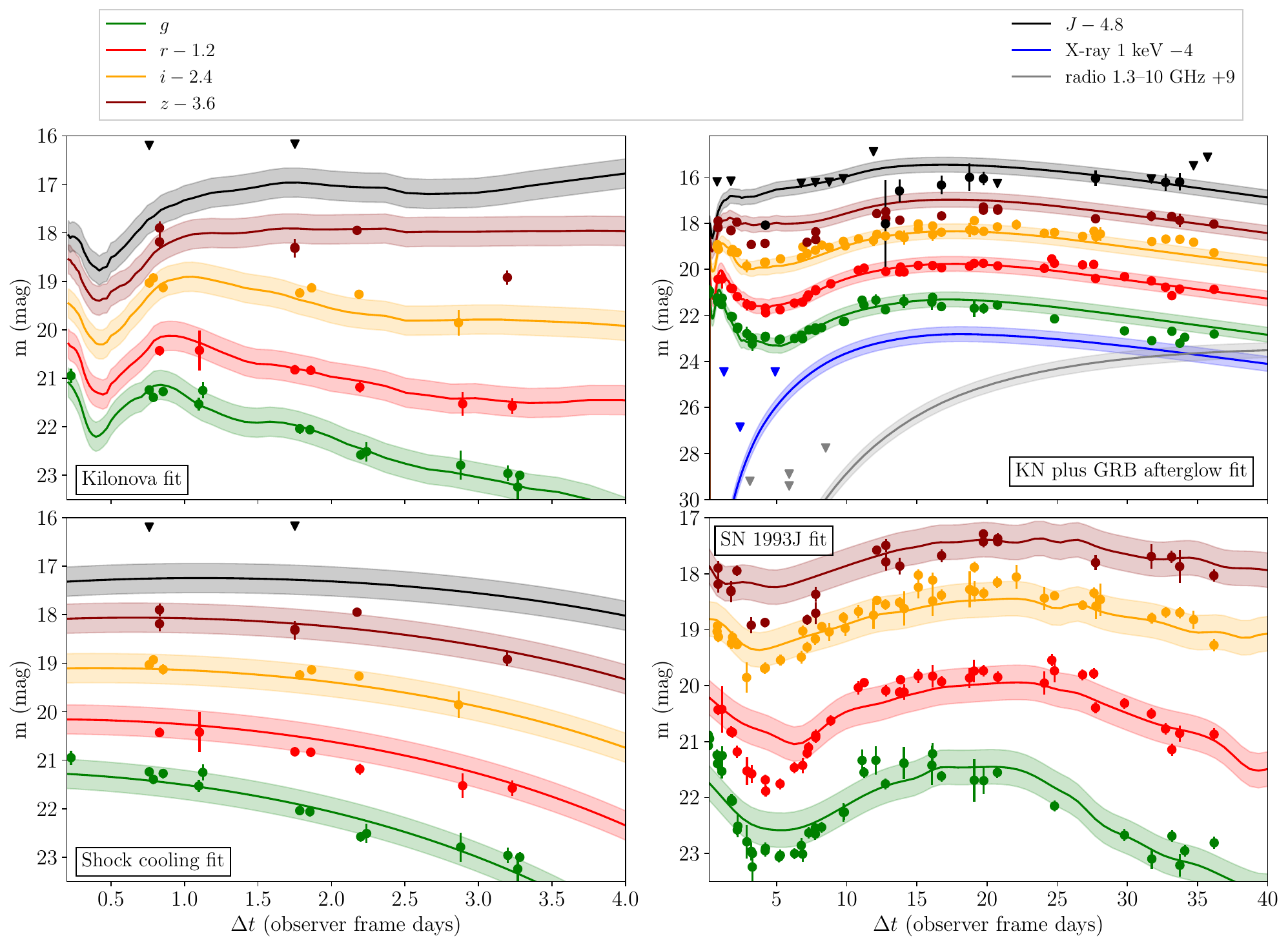}
     \caption{\emph{Top Left:} Best-fit light curves of the kilonova analysis after 4 days of observation. 
    \emph{Top Right:} Best-fit light curves of the joint kilonova plus afterglow analysis after 40 days of observation.
    \emph{Bottom Left:} Best-fit light curves of the shock cooling analysis after 4 days of observation time.
    \emph{Bottom Right:} Best-fit light curves of the SN 1993J template for the same time range, but with a restricted set of passbands.
    The colored bands in indicate the systematic uncertainty inferred during the analysis.
    Upper limits are marked as triangles. $J$-band detections that are below the $3\sigma$ detection threshold are used for these fits.}
    \label{fig:lightcurves_nmma}
\end{figure*}

\begin{figure}
    \centering
    \includegraphics[width= \linewidth]{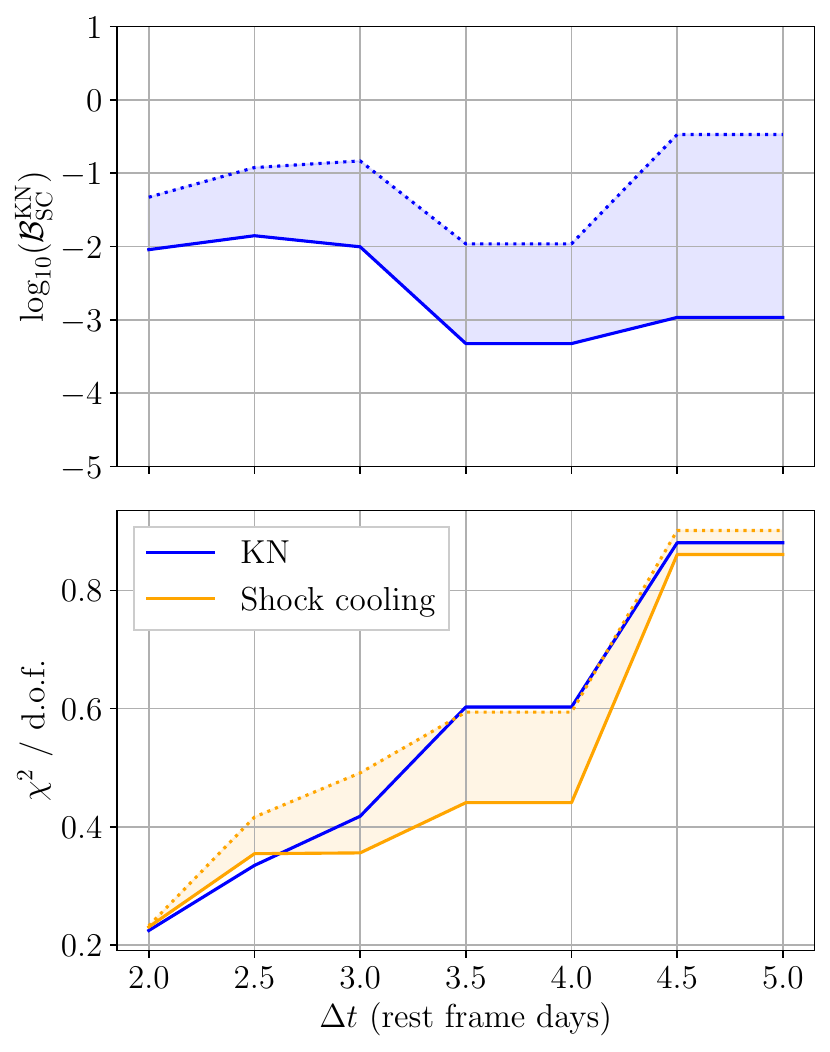}
    \caption{Statistical performance of the kilonova (KN) versus shock cooling model as a function of the number of days $\delta t$ included in the fit, as measured from the GW trigger time. 
    The upper panel shows the evolution of the Bayes factor as more data points are added with time. 
    Similarly, the lower panel displays the evolution for the reduced $\chi^2$ residual.
    The dotted lines indicate the same statistical quantities, when instead a reduced prior range for the shock cooling inference is used.}
    \label{fig:statistical_comparison}
\end{figure}

\begin{table*}[t]
        \centering
    \tabcolsep=0.3cm
    \caption{Parameters and priors used for the Bayesian analyses with the Shock Cooling and kilonova model. 
    The two last columns show the 95\% credible interval for the particular parameter after 2 days and 4 days observation time respectively. Uniform priors are marked with $\mathcal{U}$, the $\text{Sin}$ prior stands for $\cos(\iota)\sim\mathcal{U}(0,1)$. }
    \begin{tabular}{c c c c c c}

\toprule 
\toprule 
Model & Parameter & Label & Prior & 2 days & 4 days \\ 
\midrule 
 \multirow{12}{2.5cm}{Kilonova \citep{Koehn:2025zzb}} & inclination & $\iota$ [rad] & $\text{Sin}(0,\pi/2)$ & $0.59_{-0.43}^{+0.40}$ & $0.42_{-0.33}^{+0.31}$ \\ 
  & dynamical ejecta mass & $\log_{10}(m_{\mathrm{ej,dyn}})$ [$M_\odot$] & $\mathcal{U}(-3, -1.3)$ & $-2.41_{-0.56}^{+0.84}$ & $-2.81_{-0.18}^{+0.34}$ \\ 
  & dynamical ejecta velocity & $\bar{v}_{\mathrm{ej,dyn}}$ [$c$] & $\mathcal{U}(0.12, 0.28)$ & $0.18_{-0.03}^{+0.08}$ & $0.25_{-0.03}^{+0.03}$ \\ 
  & dynamical ejecta electron fraction & $\bar{Y}_{e,\mathrm{dyn}}$ & $\mathcal{U}(0.15, 0.35)$ & $0.30_{-0.07}^{+0.04}$ & $0.33_{-0.04}^{+0.02}$ \\ 
  & wind ejecta mass & $\log_{10}(m_{\mathrm{ej,wind}})$ [$M_\odot$] & $\mathcal{U}(-2, -0.9)$ & $-1.01_{-0.34}^{+0.12}$ & $-0.97_{-0.12}^{+0.07}$ \\ 
  & wind ejecta velocity & $\bar{v}_{\mathrm{ej,wind}}$ [$c$] & $\mathcal{U}(0.05, 0.15)$ & $0.06_{-0.01}^{+0.02}$ & $0.05_{-0.00}^{+0.01}$ \\ 
  & wind ejecta electron fraction & $Y_{e,\mathrm{wind}}$ & $\mathcal{U}(0.2, 0.4)$ & $0.34_{-0.06}^{+0.02}$ & $0.33_{-0.04}^{+0.01}$ \\ 
  & luminosity distance & $d_L$ [Mpc] & $\mathcal{U}(369, 399)$ & $387_{-16}^{+12}$ & $386_{-16}^{+12}$ \\ 
  & color excess & $E(B-V)$ [mag] & $\mathcal{U}(0, 0.6)$ & $0.07_{-0.06}^{+0.12}$ & $0.10_{-0.09}^{+0.10}$ \\ 
  & systematic uncertainty & $\sigma_{\rm sys}$ [mag] & $\mathcal{U}(0.3, 2.0)$ & $0.33_{-0.03}^{+0.12}$ & $0.33_{-0.02}^{+0.09}$ \\ 
 & evidence & $\ln(Z)$ & - & -14.79 & -23.90 \\ 
 & fit residual & $\chi^2$/d.o.f. & - & 0.22 & 0.60 \\ 
\midrule 
 \multirow{9}{2.5cm}{Shock Cooling \citep{Piro2021}} & timeshift & $t_0$ [days] & $\mathcal{U}(-3, 0.1)$ &$-2.04_{-0.90}^{+1.37}$ & $-2.61_{-0.37}^{+1.01}$ \\ 
  & ejecta energy & $\log_{10}(E_e)$ [erg] & $\mathcal{U}(48.4, 51.5)$ &$50.41_{-0.82}^{+0.87}$ & $50.55_{-1.00}^{+0.69}$ \\ 
  & ejecta mass & $\log_{10}(M_{\rm e})$ [$M_\odot$] & $\mathcal{U}(-1.5, -0.1)$ &$-1.17_{-0.31}^{+0.45}$ & $-0.90_{-0.53}^{+0.35}$ \\ 
  & ejecta radius & $\log_{10}(R_{\rm e})$ [cm] & $\mathcal{U}(11.3, 16)$ &$13.79_{-1.48}^{+1.85}$ & $13.43_{-1.04}^{+2.36}$ \\ 
  & luminosity distance & $d_L$ [Mpc] & $\mathcal{U}(369, 399)$ &$384_{-14}^{+14}$ & $384_{-14}^{+14}$ \\ 
  & color excess & $E(B-V)$ [mag] & $\mathcal{U}(0, 0.6)$ &$0.26_{-0.24}^{+0.25}$ & $0.23_{-0.22}^{+0.31}$ \\ 
  & systematic uncertainty & $\sigma_{\rm sys}$ [mag] & $\mathcal{U}(0.3, 2.0)$ &$0.32_{-0.02}^{+0.09}$ & $0.32_{-0.02}^{+0.07}$ \\ 
 & evidence & $\ln(Z)$ & - & -10.09 & -16.24 \\ 
 & fit residual & $\chi^2$/d.o.f. & - & 0.23 & 0.44 \\ 
 \bottomrule
 
\end{tabular}
\label{tab:KN_vs_SC}
\end{table*}

To investigate the cause of the early optical emission of AT2025ulz,  we use state of the art modeling through the Nuclear Multimessenger Astronomy framework \citep[NMMA;][]{pang_updated_2023}. 
NMMA compares the light curve data to predictions of a physical model and samples a posterior distribution on the model's parameters.
To analyze the first 5 days of observation, we use two different models.
First, we fit the data with a KN model~\citep{Koehn:2025zzb} that is based on 3D Monte-Carlo radiation transport simulations from the \textsc{possis} code~\citep{Bulla2019, Bulla2023}.
The second model is an analytical shock cooling model as described in \citet{Piro2021}. We modeled the data compiled in Table \ref{tab:photometry} and supplemented with additional photometry presented by \citet{Kasliwal2025sn}, as shown in Figure \ref{fig:lightcurve}.

Due to the limited time range of the KN model, we only use data after 0.2~days from the GW trigger. 
This limitation to the KN model arises because of high Monte Carlo noise and uncertainties in the thermalization and heating rates for the very early part of the \textsc{possis} light curves~\citep{Koehn:2025zzb, Shenhar:2024rfm, Sarin:2024tja}.
In order to ensure a robust statistical comparison, we adopt the same restriction to the data set used for the shock cooling analysis, but confirmed that our best-fit light curves obey the ZTF upper limits reported before the GW trigger \citep[see][]{Kasliwal2025sn}.

Additionally, we fit the long-term data set with two models that are capable of reproducing the rebrightening observed after $\sim5$~days.
One model assumes joint emission from a KN and a GRB afterglow~\citep{Ryan2019,Ryan2024}.
The other model is the rescaled light curve of the SN 1993J~\citep{SN1993J_circular}.
The latter analysis only fits the $g$, $r$, $i$, and $z$ band data, due to the limited frequency range of the light curve template.
The KN plus GRB afterglow analysis on the other hand even includes data points from the X-ray and radio limits reported in~\citet{2025GCN.41453....1H,2025GCN.41528....1B, 2025GCN.41460....1L,Ricci-1,Ricci-2,2025GCN.41577....1B}. We note that additional, deeper X-ray and radio limits are presented by \citet{OConnor2025ulz}.

In NMMA, the data is represented as a series of multiband magnitude measurements $m(t_j)$ at different times $t_j$.
These measurements are compared to the predicted magnitude $m^{\star}(t_j, \vec{\!\theta}\,)$ of the model, where $\vec{\theta}$ are the model's parameters.
This is done via the likelihood function
\begin{align}
\begin{split}
    \ln \mathcal{L}(\vec{\theta}|d) = - \sum_{t_j} \biggl(&\frac{1}{2} \frac{(m(t_j) - m^{\star}(t_j, \vec{\!\theta}\,))^2}{\sigma^2(t_j) + \sigma_{\text{sys}}^2} \\
    &+ \ln(2\pi (\sigma^2(t_j) + \sigma_{\text{sys}}^2)) \biggr)\ ,
    \label{eq:likelihood}
\end{split}
\end{align}
where $\sigma(t_j)$ is the data uncertainty and $\sigma_{\text{sys}}$ is a nuisance parameter that accounts for systematic uncertainties in the modeling.
Note that $\sigma_{\text{sys}}$ is sampled as a free parameter, so that the sampler will find the lowest $\sigma_{\text{sys}}$ that still covers the data points~\citep{Jhawar:2024ezm}.
The likelihood function is sampled with a parameter prior $\pi(\vec{\theta})$ using the Nested Sampling algorithm as implemented in \textsc{pymultinest}~\citep{Buchner:2014nha}.
This algorithm also allows us to determine the Bayesian evidence 
\begin{align}
    Z = \int d\vec{\theta}\ \mathcal{L}(\vec{\theta}|d) \pi(\vec{\theta})\ ,
\end{align}
which can be used to compare the ability of our models to describe the data through the Bayes factor
\begin{align}
    \mathcal{B}^{\text{KN}}_{\text{SC}} = \frac{Z_{\text{KN}}}{Z_{\text{shock}}}\ .
\end{align}
The parameters and priors used for the KN and shock cooling inference are listed in Table \ref{tab:KN_vs_SC}. 
We point out that in the KN model, the velocity and electron fraction for the dynamical ejecta, as well as the velocity for the wind ejecta, are mass-averaged quantities based on the density profiles implemented in the \textsc{possis} code~\citep{Bulla2023}, thus their labels are marked with a bar.
The prior ranges for the ejecta properties are motivated by BNS numerical relativity simulations and the luminosity distance allows for a Hubble constant prior range spanning from the \citet{Planck2020} to the \citet{Riess_2022} best fit value. 
The reddening $E(B-V)$ accounts for intrinsic dust attenuation from the host galaxy (which in NMMA is assumed to follow a South Magellanic Cloud dust law; \citealt{Pei1992}), and it is found to be similar to the one estimated in the host analysis of \citet{Hall2025desi}, whose posterior peaks close to $E(B-V)\sim 0.1$ mag.
The prior for the systematic uncertainty $\sigma_\textrm{sys}$ is set to be uniform between 0.3 and 2~mag. 
The lower bound on the systematic uncertainty comes from the intrinsic prediction error of the machine learning surrogate that is used to interpolate the light curves from the physical \textsc{possis} base model.
In that manner, $\sigma_{\text{sys}}$ ensures that the posterior contains the physical solution of the model to the data within the uncertainties.
We confirmed explicitly that the surrogate prediction error near the best fit light curves is within 0.3~mag.

\subsubsection{Model Fitting Results}

The results of our early data analysis are summarized in
Table \ref{tab:KN_vs_SC}, where we list the posterior confidence intervals after the first 2 and 4 days of observations, respectively.
The best fit light curves after 4 days of observations are shown in the left panels of Figure~\ref{fig:lightcurves_nmma}.
Plots of the full posterior distributions are provided in Appendix~\ref{app:fits}.
Overall, we find both models a reasonable description of the data obtained over the first 4~days, as indicated by the relatively low reduced $\chi^2$-values.
We note that the $\chi^2$-values include the width introduced by $\sigma_{\text{sys}}$ in addition to the photometric uncertainties (Table \ref{tab:photometry}). 
Beyond 5~days, the onset of the plateau and rebrightening phase in the light curve quickly diminishes the quality of the fit, as both the KN and shock cooling model are incapable to produce such behavior without the addition of another component (e.g., a GRB afterglow or a supernova,  respectively).

While Figure~\ref{fig:lightcurves_nmma} clearly shows that both models are capable of describing the data to a certain extent, the KN posterior in Figure~\ref{fig:corner_KN} displays some features that point toward potential issues when interpreting the AT2025ulz light curve as a KN.
Most notably, the dynamical ejecta masses are relatively low with $\log_{10}(m_{\text{ej, dyn}}/M_\odot) \lesssim -2.5$, while at the same time the wind ejecta mass is estimated to be rather high at $\log_{10}(m_{\text{ej, wind}}/M_\odot) \gtrsim -1.1$, agreeing with the results reported by~\citet{Kasliwal2025sn}. 
In contrast, most numerical-relativity simulations of standard binary neutron stars typically show smaller ejecta masses on the order of a few $10^{-2}M_\odot$~\citep{Hotokezaka:2012ze,Dietrich:2016fpt,Nedora2021, Kruger:2020gig, Foucart:2024kci, Rosswog:2024vfe, Hayashi:2024jwt, Neuweiler:2025klw}. 
Moreover, the analyses result in high electron fractions for the dynamical ejecta, whereas for canonical neutron star mergers one would expect that the dynamical ejecta from the tidal interactions and initial shock interface at merger are weakly to moderately protonized with $Y_{e, \text{dyn}} \lesssim 0.3$~\citep{Kullmann:2021gvo, Nedora2021}.
It should be noted that our KN model samples the average ejecta electron fraction, so even at $\bar{Y}_{e, \text{wind}} \approx 0.3$ a certain amount of dynamical ejecta would have a lower electron fraction.
Nevertheless, high values for $\bar{Y}_{e,\text{dyn}}$ could be obtained in scenarios with high neutrino emission or high shock heating~\citep{Sekiguchi:2015dma, Vincent:2019kor, Radice:2016dwd}. 
Additionally, the posterior indicates a low average velocity for the wind component at $\bar{v}_{\text{ej, wind}} \leq 0.07\, c$. 
This behavior was also observed for KN analyses of AT2017gfo~\citep{Koehn:2025zzb, Breschi:2021tbm, Breschi:2024qlc, Anand:2023jbz} and thus may point to a systematic modeling issue when trying to describe KNe with a long-lasting optical component.
The high masses and electron fractions reflect the need to power a bright light curve peaking around absolute magnitude $-17$ (Figure \ref{fig:KNcomparison}), and while formally our KN model can fit the data to a reasonable degree, the physical implications for a potential BNS progenitor appear hard to reconcile with standard theoretical assumptions. 
On the other hand it is worth noticing that the GW alert, if astrophysical in origin, is unlikely to be a standard BNS merger as it would have contained at least one sub-solar mass component.
In such systems tidal effects are larger due to the lower compactness, which typically results in higher chances of tidal disruption and generally more massive ejecta \citep{fujibayashi_postmerger_2020}, offering a potential explanation for the large ejecta masses inferred from our analyses.
We discuss this further in \S~\ref{sec:discuss}.

In Figure~\ref{fig:statistical_comparison}, we show the time evolution of the Bayes factor and the reduced $\chi^2$-statistic (including $\sigma_{\text{sys}}$), when more data points are added successively.
Across all time ranges, the Bayesian evidence favors the shock cooling model, though the preference increases, as $\mathcal{B}^{\text{KN}}_{\text{SC}}$ decreases from $10^{-2.0}$ initially to $10^{-3.3}$ at the fourth observation day.
Concurrently, the $\chi^2$ statistic of the KN and shock cooling fits grow over the same time span and while initially both models have a similar $\chi^2$~fit value, after four days the residual for the KN is slightly higher.
Both models provide similarly good fits to the data, indicating that the early data of AT2025ulz is consistent both with a KN and a shock breakout.
However, the $z$-band observation at 3.4~days indicates an early decline in the near-infrared, which is difficult to accommodate with the KN model.
The shock cooling is consistent with the behavior in the $z$ band, but tends to slightly overestimate the $r$ band magnitudes.
This shows that additional $z$ band and other NIR observations could have enhanced prospects of excluding a KN as transient source.

We also determine the Bayes factor and $\chi^2$-values for a shock cooling analysis with a smaller prior.
Bayes factors are susceptible to prior choices and tend to prefer models with a smaller number of free parameters.
To assess the impact of this tendency, we show the results from a physically more restrictive shock cooling prior, using $t_0\sim \mathcal{U}(-2, 0.1)$, $\log_{10}(M_e/M_\odot)\sim\mathcal{U}(-1.5, -0.5)$, $\log_{10}(R_e[{\rm cm}])\sim\mathcal{U}(11.3, 14.3)$, and $\log_{10}(E_e[{\rm erg}])\sim\mathcal{U}(48.4, 50.4)$, as dotted lines in Figure~\ref{fig:lightcurves_nmma}.
With this ``small prior'', the Bayes factor increases, making the KN model more plausible, although the preference for the shock cooling model remains.
Likewise, the $\chi^2$ residual is higher than for the larger prior, and exceeds the residual from the KN fit.
This mainly serves as secondary check to provide context to the relatively strong Bayes factor of $10^{-3.3}$ in favor of the smaller shock cooling model.
Overall, the Bayes factors for our analyses of the early time data indicate a preference of the shock cooling over the kilonova model.
We point out however, that the quality of the fit in terms of $\chi^2$ residuals is very similar and that these metrics do not allow an early exclusion of the kilonova hypothesis for this transient, since the different dimensionalities and priors make a direct comparison with Bayes factors difficult.

Figure~\ref{fig:lightcurves_nmma} also shows the best-fit light curves from our analyses of the long-term data with the KN plus GRB afterglow model and the SN 1993J template.
We discuss specifics of their posteriors in the Appendix~\ref{app:fits}, but briefly note that while the GRB afterglow is able to decently reproduce the optical observations, it is not consistent with the NIR data or the presence of spectral features. 
Additionally, more sensitive radio and X-ray constraints at later times were reported by \citet{OConnor2025ulz}. 
These constraints were not used in the present analysis and our best-fit lightcurves violate these additional constraints radio and X-ray upper limits \citep{OConnor2025ulz}, which robustly rules out the GRB afterglow hypothesis.

\subsection{Spectral Analysis}

To study the AT2025ulz spectra, we use the results from the kilonova light curve fitting described in \ref{subsec:lcanalysis_setup}. 
We randomly select 20 samples from the posterior based on 4 days of observation time and for each sample, generate the spectrum using the same KN surrogate used above for the Bayesian light curve fitting~\citep{Koehn:2025zzb}.
This surrogate is based on a grid of 17899 \textsc{possis} computations and trained directly on the flux density arrays.
Therefore our surrogate is capable of predicting both photometric light curves and spectra.

\begin{figure*}
    \centering
    \includegraphics[width=1\linewidth]{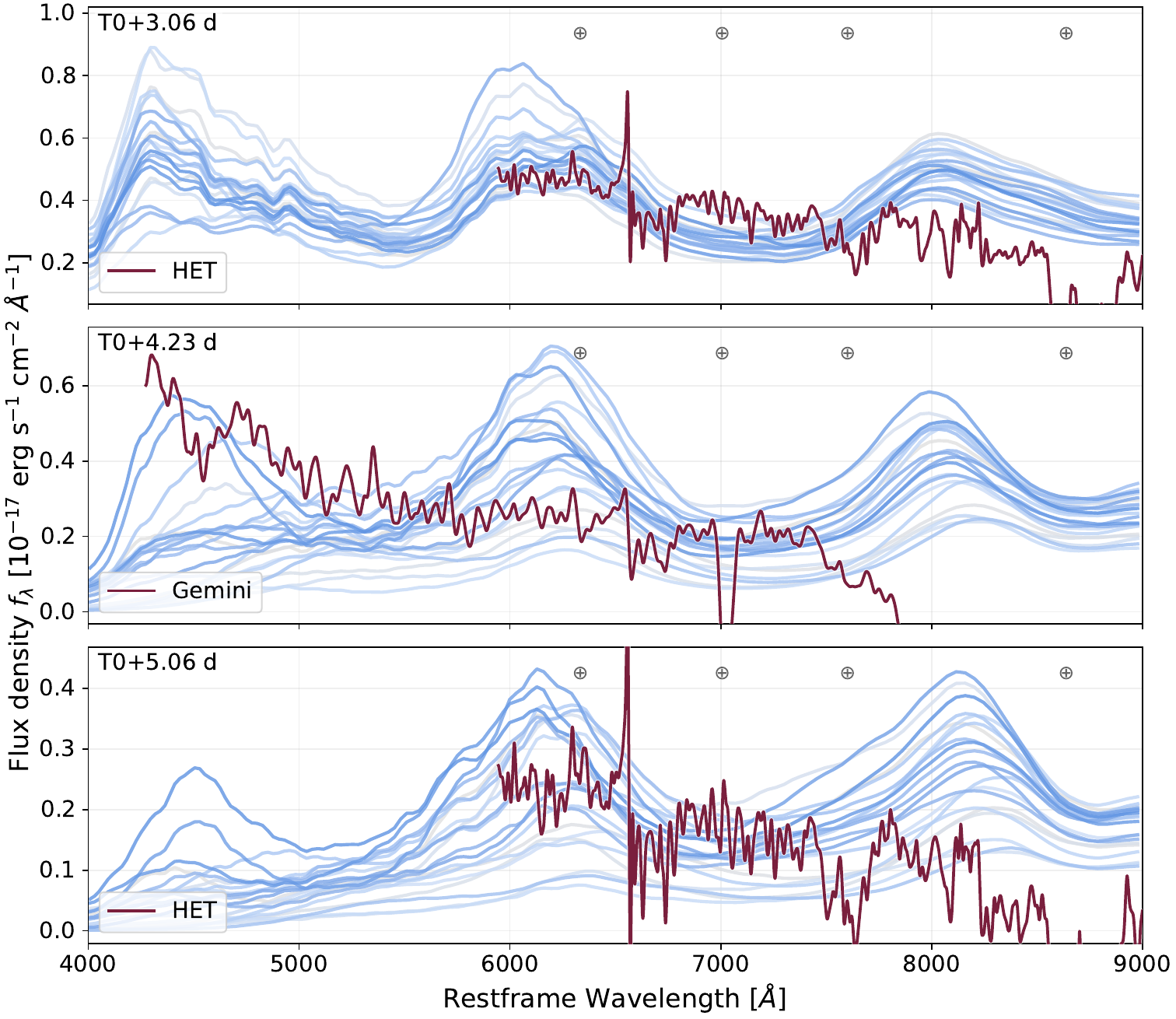}
    \caption{Model spectra of various KN models using the input parameters provided by NMMA (see Table \ref{tab:KN_vs_SC}). HET and Gemini spectra taken for this work are subtracted using a DESI spectrum (see method in \citealt{Hall2025desi}; \citealt{desi_collaboration_desi_2025}.) Features appearing from Tellurics are marked with $\oplus$. The subtracted HET and Gemini spectra are normalized to units of relative flux.}
    \label{fig:modeled_spectra}
\end{figure*}

We then plotted the spectra corresponding to these models at relevant epochs of early time GMOS and HET spectra as shown in Figure \ref{fig:modeled_spectra}. Most of the pixel to pixel noise in the spectra arises from the numerical noise associate with the finite number of Monte Carlo photons. Furthermore, we plot modeled spectra consistent with the observation time of the spectra presented in \citet{Kasliwal2025sn} in Appendix \ref{app:spec} and Figure \ref{fig:modeled_spectra_kasliwal}. These early spectroscopic models, while they may not be representative of a subsolar mass BNS kilonova, demonstrate in comparison to both HET and Gemini spectra that perhaps the observed continuum at these times was in fact far too blue to be a classical KN. Such a conclusion is supported by the color comparison between AT2025ulz and AT2017gfo in Figure \ref{fig:color_comparison}.

\section{Discussion}
\label{sec:discuss}

\subsection{Modeling and Phenomenological Comparison}

Modeling of AT2025ulz’s optical and NIR photometry indicates that both a kilonova and an early shock cooling component can describe the first four days of observations (see Figure \ref{fig:lightcurves_nmma} and \S~\ref{nmmamodel}). Although Figure~\ref{fig:statistical_comparison} shows a slight statistical preference for shock cooling, the KN fit remains plausible. 
Consequently, discriminating events like AT2025ulz requires rapid, continued multi-band follow-up. Our analysis shows that AT2025ulz could not be rejected as a kilonova candidate for at least 5 days after discovery, and that continued monitoring was critical to demonstrate it was a supernova (Figue \ref{fig:lightcurve}). In particular, spectroscopic observations were critical to proving this \citep{Gillanders25ulz,Franz2025,Yang2025ulz, Kasliwal2025sn} and are a clear limitation of kilonova searches at 400 Mpc (or beyond) due to the requirement of significant time needed on the limited number of large aperture facilities (e.g., Gemini). 

A comparison of long-term SN~1993J–like modeling with an afterglow plus KN model shows that the latter model matches the data better during the decline and subsequent rise, while the SN~1993J–like modeling outperforms the KN + afterglow model with the color evolution of the second peak (see Figure \ref{fig:lightcurves_nmma}). Note that such inconsistencies in the SN~1993J–like modeling are expected due to the diversity of core-collapse supernovae and that our early time shock cooling modeling is able to better describe the observations.   
Although continued deep X-ray and radio observations ultimately exclude the GRB afterglow scenario \citep{OConnor2025ulz}, early X-ray limits from \textit{Swift}/XRT \citep{2025GCN.41453....1H,2025GCN.41528....1B} and EP/FXT \citep{2025GCN.41460....1L} were not sufficient to rule it out. Additionally, this scenario (KN + afterglow) is incompatible with the spectroscopic features which are typical of Type IIb supernovae \citep{Gillanders25ulz,Franz2025,Yang2025ulz,Kasliwal2025sn}. 
In any case, the implication for future searches is clear: apparent plateauing or mild rebrightening should not be taken as \textit{a priori} inconsistent with a standard KN; premature rejection of KN scenarios is unwarranted without spectroscopic evidence to the contrary. Thus, the multimessenger community should be prepared for binary neutron star (BNS) mergers that deviate noticeably (e.g., in absolute magnitude) from the AT2017gfo “template”.

Even among short-GRB–associated KNe, color evolution and luminosity exhibit substantial diversity \citep{Rastinejad2021,Rastinejad2025kn}. The initially bright ZTF measurement of AT2025ulz is therefore not disqualifying. As shown in Figure~\ref{fig:KNcomparison}, a KN with $M \sim -17$ is not unprecedented and, if AT2025ulz were a KN, it would not be the most luminous observed.

\subsection{Near-infrared Diagnostics}

NIR data were likely the most powerful diagnostic. Shock cooling and KN light curves can appear similar in the optical, but redder wavelengths should be more strongly affected by heavy-element opacities in a KN. With deeper, wide-area NIR templates, a decisive test would have been possible, for example, a steep continued decline in the $J$ band when KN models predict a very slow decline would have offered a sharp constraint. With improved NIR templates to reach the true depths of our observations, the FTW $J$-band and Gemini $Y$-band observations could have provided a firmer verdict earlier. Even now, only a handful of NIR datapoints exist \citep{Kasliwal2025sn, Yang2025ulz} and their photometry is performed by galaxy modeling rather than direct template subtraction.

AT2025ulz, in particular, highlights a gap in multimessenger follow-up at $0.9$–$2.4\,\mathrm{\mu m}$ when the background host is bright. If the BNS-merger channel is common in luminous, star-forming galaxies\footnote{Though only a minor fraction of short GRBs lies on top of their hosts at offsets as small as AT2025ulz \citep[e.g.,][]{FongBerger2013,Fong2022,OConnor2022}.}, robust NIR subtraction across the sky becomes essential—either via improved two-dimensional galaxy modeling or a modern, truly all-sky NIR effort. The last comprehensive all-sky NIR survey (2MASS; \citealt{skrutskie_two_2006}) dates to 2001 and reaches only ${\sim}17$~mag; deeper UKIRT imaging exists \citep{dye_ukirt_2018}, but its coverage is not uniformly complete nor deep for this purpose. NIR photometry carries significant diagnostic power for GW follow-up and KN identification, yet remains under-resourced \citep[see, e.g.,][]{Chase2022}. In the Northern hemisphere, the Wide-field Infrared Transient Explorer \citep[WINTER;][]{lourie_wide-field_2020} regularly pursues MMA follow-up but is limited to $\sim\!19$~mag at best \citep{frostig_winter_2025}. In the Southern Hemisphere, the PRime-focus Infrared Microlensing Experiment \citep[PRIME;][]{Sumi2025} provides increased sensitivity and a larger field-of-view, but is still limited in its sensitivity to KN beyond 200 Mpc \citep[see][]{Chase2022}. In the future, the \textit{Nancy Grace Roman Space Telescope} \citep{roman} and Cryoscope \citep{kasliwal_cryoscope_2025} will offer the wide-field capability and necessary sensitivity to push KN discovery to cosmological distances \citep{Scolnic2018,Chase2022,AndreoniRoman}.
We emphasize that deep all-sky NIR templates remain woefully insufficient for MMA in LIGO-Virgo-KAGRA's fifth observing run (O5) and beyond, and their acquisition should be prioritized.

\subsection{Sub-solar mass mergers}

It is important to note that if astrophysical in origin, S250818k is likely to have harbored a sub-solar mass compact object. As such, the EM counterpart of this object may be extremely different from our typical KN expectations. Little work has been dedicated to predicting the EM signal from these kinds of sources, but the numerical relativity simulations of \citet{Markin_2023} show that a subsolar mass black hole merging with a $1.4~M_\odot$ neutron star (which may be consistent with the event in question) would produce ejecta and remnant properties extremely different from what one would predict extrapolating fitting relations from higher mass BNS systems. In that case, however, the expected kilonova emission would be orders of magnitude fainter than AT 2025ulz. On the other hand, for a less massive neutron star in a BNS system, it would be reasonable to expect a larger tidal disruption and a brighter kilonova \citep{dietrich_interpreting_2021}.

We also explore the possibility that S250818k could have originated from a Superkilonova (for a more extensive discussion, see \citealt{Kasliwal2025sn}) first coined by \citet{siegel_collapsars_2019, siegel_super-kilonovae_2022} as a method for collapsars to engage in r-processing. The collapse of a massive (${\sim}130~ \text{M}_\odot$), rapidly rotating, helium star results in the formation of a hyperaccreting black hole. In this extreme environment, \citet{PiroPfahl2007, Metzger2024, lerner_fragmentation_2025} demonstrate the potential for the formation of sub-solar mass neutron stars. This is possible at the outer edges of such an accretion disk due to the immense gravitational instability. Such instability combined with such high angular velocity causes the disk to fragment and form gravitationally bound protoclumps \citep{Metzger2024}. These protoclumps radiate temperature rapidly through Neutrino cooling \citep{chen_neutrino-cooled_2007} and the dissociation of alpha particles \citep{Metzger2024}. Furthermore, because of the rapid orbital period these clumps begin to merge until they have sufficiently cooled and exceeded the Chandrasekhar mass, at which point they collapse into sub-solar mass neutron stars. It is important to note that the Chandrasekhar mass is subsolar because of the low electron fraction in the disk. 

In this black hole accretion disk, BNS systems form from these collapsing clumps and then merge on relatively rapid timescales \citep[$10$ s;][]{Metzger2024}. Such a system could produce multiple gravitational-wave events as multiple neutron star systems merge and as neutron stars merge with the central stellar mass black hole. The resulting lightcurve would then result from a combination of a KN and a SN. One notable issue given our current classical modeling with such an explanation is that we find an ejecta velocity of $0.17-0.24 c$ (see Table \ref{tab:KN_vs_SC}). Yet, the hydrogen line velocity was reported to be $0.04 - 0.06 c$ \citep{banerjee_engrave_2025, kasliwal_ligovirgokagra_2025}. If this hydrogen line is purported to have been achieved by a KN within the accretion disk of a collapsar, such velocity discrepancies would need to be resolved.

A superkilonova interpretation is a tantalizing explanation for AT2025ulz. Firm evidence would be clear r-process signatures (potentially detectable in late-time NIR photometric and spectroscopic observations), an associated long- or short-GRB \citep{de_barra_ligovirgokagra_2025} (potentially detectable in late-time radio or X-ray observations), other spatially and temporally coincident BNS, NSBH or even BBH mergers causing multiple gravitational-wave events (predicted in some but not all superkilonova scenarios). Therefore, such events should be included in our \textit{a priori} tests when searching for candidate EM counterparts in O4 and beyond. Consequently, we underscore the need for more fleshed-out modeling to better assess the possibility of such unique events in the future.

\section{Conclusions}
\label{sec:conclusions}

We present an analysis of the first few days of AT2025ulz's lightcurve and color evolution using optical and near-infrared imaging obtained with FTW, DECam, and Gemini. We compare AT2025ulz to known kilonovae and demonstrate that the first few days of data had a good agreement to their lightcurve evolution. Our theoretical modeling shows that both kilonova and shock cooling models are capable of matching the first few days of data. Moreover, we compare our Gemini and HET spectra with modeled spectra from the high probability region of our kilonova posterior, finding that our observations are significantly bluer than the models, consistent with the photometry. Continued monitoring was essential to rule out a bona fide kilonova; with only a few days of data, AT2025ulz would have remained a strong classical KN candidate electromagnetic counterpart to a BNS event, and rejecting it solely based on its peak absolute magnitude would be misleading relative to known GRB-KNe (e.g., GRBs 211211A and 230307A) and theoretical expectations. On the other hand, little is known about what EM signal is to be expected from binaries containing a sub-solar mass compact object, which could arise from exotic binaries containing a primordial black hole or low-mass neutron stars formed in the accretion disk of a core-collapse supernova.

Future GW searches to $>400$ Mpc, will require deep spectroscopy and well-sampled light curves, as sparse early photometry cannot reliably exclude kilonova impostors. This event highlights the need for new strategies and large-aperture facilities for deep imaging and spectroscopy (particularly in the NIR), along with improved predictions for the formation and EM emission of sub-solar mass compact object binaries, in the next phase of multimessenger astronomy at cosmological distances.

\begin{acknowledgments}

X. J. H., B. O., and A. P. acknowledge the NOIRLab Communications, Education \& Engagement (CEE) Team for their creation of the representative color image of AT2025ulz. In particular, we thank Mahdi Zamani, Jen Miller, Lars Christensen, Matias Rodriguez, Jesse Ball, and Josie Fenske for their fantastic assistance and support. X. J. H., B. O., and A. P. further thank the staff of the Gemini Observatory for their aid in rapidly obtaining imaging and spectroscopic observations of AT2025ulz. In particular, we thank Brian Lemaux, Jennifer Andrews, Eunchong Kim, Hyewon Suh, Atsuko Nitta, Jen Miller, Aleksandar Cikota, and Teo Mocnik for their assistance in preparing and scheduling these time critical observations.

This work is supported by NSF Grant No. 2308193. B. O. is supported by the McWilliams Postdoctoral Fellowship in the McWilliams Center for Cosmology and Astrophysics at Carnegie Mellon University. M. B. is supported by a Student Grant from the Wübben Stiftung Wissenschaft. 
M. W. C. acknowledges support from the National Science Foundation with grant numbers PHY-2117997, PHY-2308862 and PHY-2409481.
H. K. and T. D. acknowledge funding from the EU Horizon under ERC Starting Grant, no.\ SMArt-101076369. Views and opinions expressed are those of the authors only and do not necessarily reflect those of the European Union or the European Research Council. Neither the European Union nor the granting authority can be held responsible for them.
M. B. acknowledges the Department of Physics and Earth Science of the University of Ferrara for the financial support through the FIRD 2024 grant.
T. W. is supported by the research program of the Netherlands Organization for Scientific Research (NWO) under grant number OCENW.XL21.XL21.038.

This paper contains data obtained at the Wendelstein Observatory of the Ludwig-Maximilians University Munich. We thank Christoph Ries, Michael Schmidt and Silona Wilke for performing the observations. Funded by the Deutsche Forschungsgemeinschaft (DFG, German Research Foundation) under Germany's Excellence Strategy – EXC-2094 – 390783311.

This work used resources on the Vera Cluster at the Pittsburgh Supercomputing Center (PSC). Vera is a dedicated cluster for the McWilliams Center for Cosmology and Astrophysics at Carnegie Mellon University. We thank the PSC staff for their support of the Vera Cluster.

Based on observations obtained at the international Gemini Observatory, a program of NSF's OIR Lab, which is managed by the Association of Universities for Research in Astronomy (AURA) under a cooperative agreement with the National Science Foundation on behalf of the Gemini Observatory partnership: the National Science Foundation (United States), National Research Council (Canada), Agencia Nacional de Investigaci\'{o}n y Desarrollo (Chile), Ministerio de Ciencia, Tecnolog\'{i}a e Innovaci\'{o}n (Argentina), Minist\'{e}rio da Ci\^{e}ncia, Tecnologia, Inova\c{c}\~{o}es e Comunica\c{c}\~{o}es (Brazil), and Korea Astronomy and Space Science Institute (Republic of Korea). The data were acquired through the Gemini Observatory Archive at NSF NOIRLab and processed using DRAGONS (Data Reduction for Astronomy from Gemini Observatory North and South). The authors wish to recognize and acknowledge the very significant cultural role and reverence that the summit of Maunakea has always had within the indigenous Hawaiian community. 

This research has made use of data and/or software provided by the High Energy Astrophysics Science Archive Research Center (HEASARC), which is a service of the Astrophysics Science Division at NASA/GSFC. 

Based in part on observations obtained with the Hobby-Eberly Telescope (HET), which is a joint project of the University of Texas at Austin, the Pennsylvania State University, Ludwig-Maximillians-Universitaet Muenchen, and Georg-August Universitaet Goettingen. The HET is named in honor of its principal benefactors, William P. Hobby and Robert E. Eberly. We acknowledge the Texas Advanced Computing Center (TACC) at The University of Texas at Austin for providing high performance computing, visualization, and storage resources that have contributed to the results reported within this paper. The Low Resolution Spectrograph 2 (LRS2) was developed and funded by the University of Texas at Austin McDonald Observatory and Department of Astronomy, and by Pennsylvania State University. We thank Sergey Rostopchin, Amy Westfall, Cassie Crowe, and Justen Pautzke from the HET staff for obtaining these observations. We thank the Leibniz-Institut fur Astrophysik Potsdam (AIP) and the Institut fur Astrophysik Goettingen (IAG) for their contributions to the construction of the integral field units. We would like to acknowledge that the HET is built on Indigenous land. Moreover, we would like to acknowledge and pay our respects to the Carrizo \& Comecrudo, Coahuiltecan, Caddo, Tonkawa, Comanche, Lipan Apache, Alabama-Coushatta, Kickapoo, Tigua Pueblo, and all the American Indian and Indigenous Peoples and communities who have been or have become a part of these lands and territories in Texas, here on Turtle Island.

This research is based on data obtained from the Astro Data Archive at NSF NOIRLab. NOIRLab is managed by the Association of Universities for Research in Astronomy (AURA) under a cooperative agreement with the U.S. National Science Foundation. This project used data obtained with the Dark Energy Camera (DECam), which was constructed by the Dark Energy Survey (DES) collaboration. Funding for the DES Projects has been provided by the U.S. Department of Energy, the U.S. National Science Foundation, the Ministry of Science and Education of Spain, the Science and Technology Facilities Council of the United Kingdom, the Higher Education Funding Council for England, the National Center for Supercomputing Applications at the University of Illinois at Urbana-Champaign, the Kavli Institute of Cosmological Physics at the University of Chicago, Center for Cosmology and Astro-Particle Physics at the Ohio State University, the Mitchell Institute for Fundamental Physics and Astronomy at Texas A\&M University, Financiadora de Estudos e Projetos, Fundacao Carlos Chagas Filho de Amparo, Financiadora de Estudos e Projetos, Fundacao Carlos Chagas Filho de Amparo a Pesquisa do Estado do Rio de Janeiro, Conselho Nacional de Desenvolvimento Cientifico e Tecnologico and the Ministerio da Ciencia, Tecnologia e Inovacao, the Deutsche Forschungsgemeinschaft and the Collaborating Institutions in the Dark Energy Survey. 
The Collaborating Institutions are Argonne National Laboratory, the University of California at Santa Cruz, the University of Cambridge, Centro de Investigaciones Energeticas, Medioambientales y Tecnologicas-Madrid, the University of Chicago, University College London, the DES-Brazil Consortium, the University of Edinburgh, the Eidgenossische Technische Hochschule (ETH) Zurich, Fermi National Accelerator Laboratory, the University of Illinois at Urbana-Champaign, the Institut de Ciencies de l’Espai (IEEC/CSIC), the Institut de Fisica d’Altes Energies, Lawrence Berkeley National Laboratory, the Ludwig Maximilians Universitat Munchen and the associated Excellence Cluster Universe, the University of Michigan, NSF NOIRLab, the University of Nottingham, the Ohio State University, the University of Pennsylvania, the University of Portsmouth, SLAC National Accelerator Laboratory, Stanford University, the University of Sussex, and Texas A\&M University.

Based in part on data collected at Subaru Telescope and obtained from the SMOKA, which is operated by the Astronomy Data Center, National Astronomical Observatory of Japan. We are honored and grateful for the opportunity of observing the Universe from Maunakea, which has cultural, historical, and natural significance in Hawaii.

This research used data obtained with the Dark Energy Spectroscopic Instrument (DESI). DESI construction and operations is managed by the Lawrence Berkeley National Laboratory. This material is based upon work supported by the U.S. Department of Energy, Office of Science, Office of High-Energy Physics, under Contract No. DE–AC02–05CH11231, and by the National Energy Research Scientific Computing Center, a DOE Office of Science User Facility under the same contract. Additional support for DESI was provided by the U.S. National Science Foundation (NSF), Division of Astronomical Sciences under Contract No. AST-0950945 to the NSF’s National Optical-Infrared Astronomy Research Laboratory; the Science and Technology Facilities Council of the United Kingdom; the Gordon and Betty Moore Foundation; the Heising-Simons Foundation; the French Alternative Energies and Atomic Energy Commission (CEA); the National Council of Humanities, Science and Technology of Mexico (CONAHCYT); the Ministry of Science and Innovation of Spain (MICINN), and by the DESI Member Institutions: \url{www.desi.lbl.gov/collaborating-institutions}. The DESI collaboration is honored to be permitted to conduct scientific research on I’oligam Du’ag (Kitt Peak), a mountain with particular significance to the Tohono O’odham Nation. Any opinions, findings, and conclusions or recommendations expressed in this material are those of the author(s) and do not necessarily reflect the views of the U.S. National Science Foundation, the U.S. Department of Energy, or any of the listed funding agencies.

This paper makes use of LSST Science Pipelines software developed by the Vera C. Rubin Observatory. We thank the Rubin Observatory for making their code available as free software at \url{https://pipelines.lsst.io}.

\end{acknowledgments}





%
\facilities{Fraunhofer Telescope at Wendelstein Observatory, Gemini-North Telescope, Dark Energy Camera at Blanco, Subaru Telescope}

\software{
  \texttt{Astropy} \citep{2018AJ....156..123A,2022ApJ...935..167A},
  \texttt{NMMA} \citep{pang_updated_2023}, \texttt{afterglowpy} \citep{Ryan2019,Ryan2024}, 
  \texttt{SFFT} \citep{hu_image_2022},
  \texttt{LSST Science Pipeline} \citep{10.71929/rubin/2570545},
  \texttt{AstrOmatic} \citep{Bertin1996,Bertin2006,Bertin2010,2002ASPC..281..228B},
  \texttt{DRAGONS} \citep{Labrie2019,Labrie2023},
  \texttt{GNU Parallel} \citep{2024zndo..14550073T}
}


\appendix

\section{Image Subtraction}\label{app:imagesub}

\begin{figure*}
    \centering
    \includegraphics[width=0.98\linewidth]{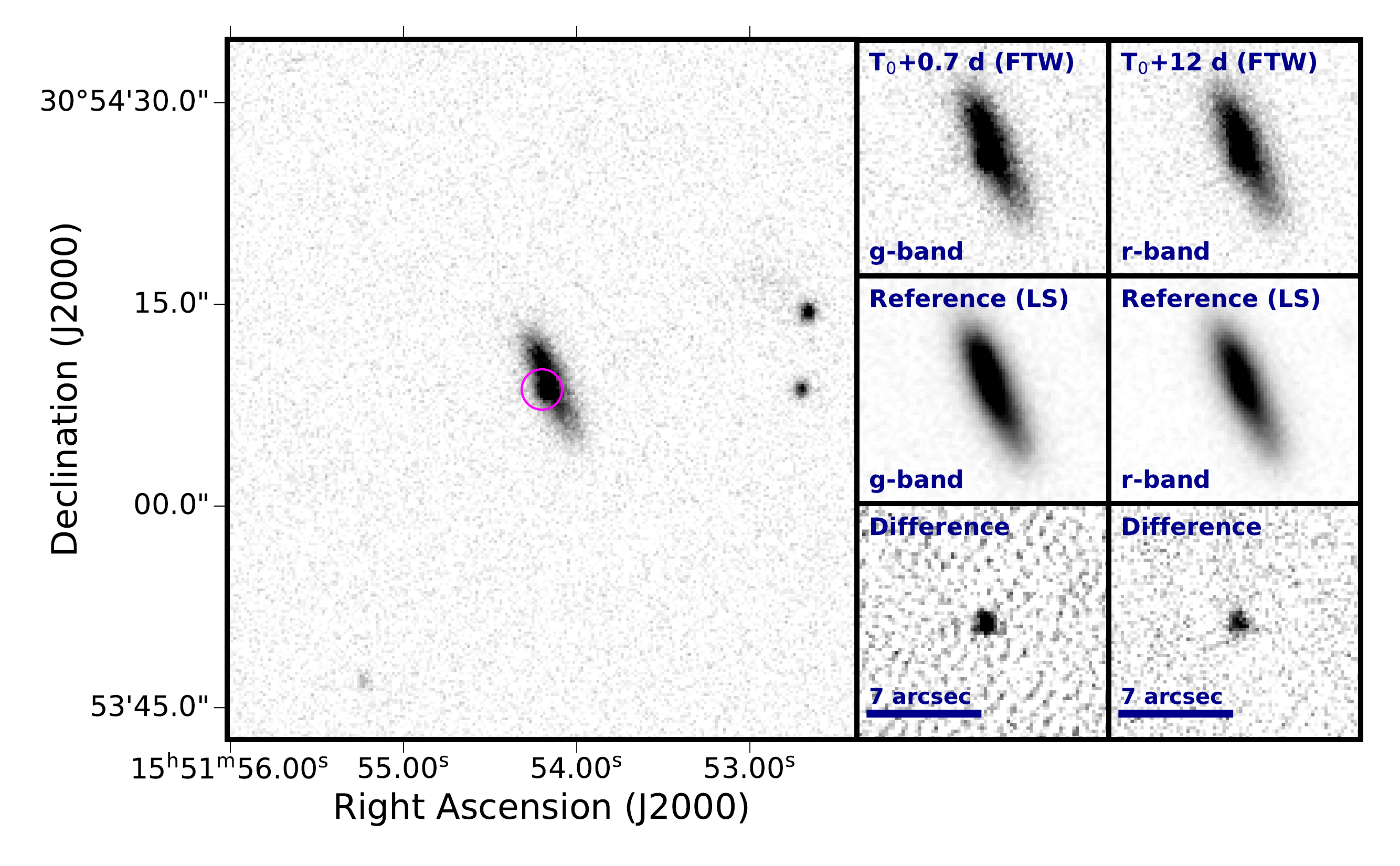}
    \caption{Images taken with the FTW 3KK instrument at $\text{T}_0 + 0.7 \text{d}$ (during the initial decline) and $\text{T}_0 + 12 \text{d}$ (during the rebrightening period). We show images in both the $g$ and $r$-bands. The transient, AT2025ulz, can clearly been seen and its location is circled in pink. It is visible in both filters bulging out of the galaxy and once the image is difference relative to the template image from the LS \citep{2019AJ....157..168D}, a clear PSF becomes visible.}
    \label{fig:FC_2025ulz_diff}
\end{figure*}

We show an example of image subtraction for this target which successfully removes the host galaxy contribution using \texttt{SFFT} \citep{Hu2022}. This is displayed in Figure \ref{fig:FC_2025ulz_diff}.

\section{Photometry}\label{app:phot}

We present the photometry that we acquired with FTW and Gemini-North between 0.70 and 36.61 days after the LVK trigger. 

\startlongtable
\begin{deluxetable*}{lCcccCC}
\tablecaption{Photometric observations of AT 2025ulz obtained with FTW/3KK and Gemini-North/GMOS. $\Delta t$ is relative to the time of the merger. Upper limits are given at the $3\sigma$-level. Magnitudes are not corrected for Galactic extinction $A_\lambda$ \citep{schlafly_measuring_2011}.}
\tablehead{
	\colhead{Start Time (UT)} & \colhead{$\Delta t$ (d)} & \colhead{Telescope} & \colhead{Instrument} & \colhead{Filter} & \colhead{AB magnitude} & \colhead{$A_\lambda$ (mag)}
}
\startdata
2025-08-18T19:35:11 & 0.70           & FTW          & 3KK        & g    & 21.25 \pm 0.10 & 0.09              \\
2025-08-18T19:35:11 & 0.70           & FTW          & 3KK        & i    & 21.44 \pm 0.10 & 0.05              \\
2025-08-18T19:35:25 & 0.70           & FTW          & 3KK        & J    & {>21.0}        & 0.03              \\
2025-08-18T20:15:50 & 0.73           & FTW          & 3KK        & i    & 21.34 \pm 0.10 & 0.05              \\
2025-08-18T20:15:50 & 0.73           & FTW          & 3KK        & g    & 21.41 \pm 0.10 & 0.09              \\
2025-08-18T21:15:53 & 0.76           & FTW          & 3KK        & z    & 21.50 \pm 0.13 & 0.04              \\
2025-08-18T21:15:53 & 0.76           & FTW          & 3KK        & r    & 21.64 \pm 0.10 & 0.06              \\
2025-08-18T21:15:53 & 0.76           & FTW          & 3KK        & z    & 21.79 \pm 0.16 & 0.04              \\
2025-08-18T21:51:31 & 0.79           & FTW          & 3KK        & g    & 21.29 \pm 0.10 & 0.09              \\
2025-08-18T21:51:31 & 0.79           & FTW          & 3KK        & i    & 21.54 \pm 0.11 & 0.05              \\
2025-08-19T19:22:06 & 1.61           & FTW          & 3KK        & z    & 21.91 \pm 0.16 & 0.04              \\
2025-08-19T19:22:06 & 1.61           & FTW          & 3KK        & z    & 21.92 \pm 0.19 & 0.04              \\
2025-08-19T19:22:06 & 1.61           & FTW          & 3KK        & r    & 22.03 \pm 0.10 & 0.06              \\
2025-08-19T19:22:20 & 1.61           & FTW          & 3KK        & J    & {>20.9}        & 0.03              \\
2025-08-19T20:09:46 & 1.64           & FTW          & 3KK        & g    & 22.05 \pm 0.10 & 0.09              \\
2025-08-19T20:09:46 & 1.64           & FTW          & 3KK        & i    & 21.65 \pm 0.10 & 0.05              \\
2025-08-20T05:31:06 & 2.00           & Gemini North & GMOS       & z    & 21.55 \pm 0.09 & 0.04              \\
2025-08-20T05:50:26 & 2.02           & Gemini North & GMOS       & i    & 21.67 \pm 0.07 & 0.05              \\
2025-08-20T05:58:23 & 2.02           & Gemini North & GMOS       & r    & 22.39 \pm 0.12 & 0.06              \\
2025-08-20T06:07:03 & 2.03           & Gemini North & GMOS       & g    & 22.59 \pm 0.10 & 0.09              \\
2025-08-21T06:03:57 & 2.95           & Gemini North & GMOS       & z    & 22.53 \pm 0.15 & 0.04              \\
2025-08-21T08:06:07 & 3.02           & Gemini North & GMOS       & g    & 23.01 \pm 0.10 & 0.09              \\
2025-08-22T05:30:58 & 3.85           & Gemini North & GMOS       & z    & 22.48 \pm 0.09 & 0.04              \\
2025-08-22T06:08:38 & 3.87           & Gemini North & GMOS       & r    & 22.90 \pm 0.10 & 0.06              \\
2025-08-22T06:19:22 & 3.88           & Gemini North & GMOS       & g    & 22.92 \pm 0.10 & 0.09              \\
2025-08-23T05:42:11 & 4.78           & Gemini North & GMOS       & g    & 23.08 \pm 0.10 & 0.09              \\
2025-08-24T19:33:25 & 6.23           & FTW          & 3KK        & g    & 22.87 \pm 0.14 & 0.09              \\
2025-08-24T19:33:25 & 6.23           & FTW          & 3KK        & i    & 21.90 \pm 0.12 & 0.05              \\
2025-08-24T19:33:39 & 6.23           & FTW          & 3KK        & J    & {>21.0}        & 0.03              \\
2025-08-25T05:30:27 & 6.61           & Gemini North & GMOS       & z    & 22.43 \pm 0.09 & 0.04              \\
2025-08-25T05:35:51 & 6.61           & Gemini North & GMOS       & i    & 21.72 \pm 0.10 & 0.05              \\
2025-08-25T05:41:09 & 6.62           & Gemini North & GMOS       & r    & 22.42 \pm 0.10 & 0.06              \\
2025-08-25T19:39:00 & 7.15           & FTW          & 3KK        & g    & 22.67 \pm 0.10 & 0.09              \\
2025-08-25T19:39:00 & 7.15           & FTW          & 3KK        & g    & 22.56 \pm 0.10 & 0.09              \\
2025-08-25T19:39:00 & 7.15           & FTW          & 3KK        & i    & 21.57 \pm 0.10 & 0.05              \\
2025-08-25T19:39:15 & 7.16           & FTW          & 3KK        & J    & {>21.0}        & 0.03              \\
2025-08-25T20:14:51 & 7.18           & FTW          & 3KK        & r    & 22.10 \pm 0.11 & 0.06              \\
2025-08-25T20:14:51 & 7.18           & FTW          & 3KK        & r    & 22.14 \pm 0.10 & 0.06              \\
2025-08-25T20:14:51 & 7.18           & FTW          & 3KK        & z    & 22.31 \pm 0.20 & 0.04              \\
2025-08-25T20:14:51 & 7.18           & FTW          & 3KK        & z    & 21.97 \pm 0.13 & 0.04              \\
2025-08-26T19:22:43 & 8.07           & FTW          & 3KK        & i    & 21.45 \pm 0.17 & 0.05              \\
2025-08-26T19:22:58 & 8.07           & FTW          & 3KK        & J    & {>21.0}        & 0.03              \\
2025-08-27T19:35:10 & 9.00           & FTW          & 3KK        & g    & 22.28 \pm 0.17 & 0.09              \\
2025-08-27T19:35:10 & 9.00           & FTW          & 3KK        & i    & 21.19 \pm 0.10 & 0.05              \\
2025-08-27T19:35:27 & 9.00           & FTW          & 3KK        & J    & {>20.8}        & 0.03              \\
2025-08-28T23:29:39 & 10.92          & Blanco       & DECam      & z    & 21.5 \pm 0.10  & 21.8              \\ 
2025-08-28T23:33:39 & 10.93          & Blanco       & DECam      & r    & 21.5 \pm 0.10  & 22.8              \\ 
2025-08-28T23:36:29 & 10.93          & Blanco       & DECam      & g    & 21.9 \pm 0.10  & 23.1              \\ 
2025-08-29T23:04:30 & 10.97          & FTW          & 3KK        & i    & 21.15 \pm 0.24 & 0.05              \\
2025-08-29T23:04:40 & 10.97          & FTW          & 3KK        & J    & {>19.7}        & 0.03              \\
2025-08-30T19:32:15 & 11.76          & FTW          & 3KK        & i    & 20.96 \pm 0.10 & 0.05              \\
2025-08-30T19:32:15 & 11.76          & FTW          & 3KK        & g    & 21.77 \pm 0.10 & 0.09              \\
2025-08-30T19:32:27 & 11.76          & FTW          & 3KK        & J    & {>21.0}        & 0.03              \\
2025-08-30T20:10:45 & 11.78          & FTW          & 3KK        & r    & 21.30 \pm 0.10 & 0.06              \\
2025-08-30T20:10:45 & 11.78          & FTW          & 3KK        & z    & 21.39 \pm 0.16 & 0.04              \\
2025-08-30T20:10:45 & 11.78          & FTW          & 3KK        & z    & 21.10 \pm 0.12 & 0.04              \\
2025-08-31T19:28:18 & 12.68          & FTW          & 3KK        & r    & 21.33 \pm 0.10 & 0.06              \\
2025-08-31T19:28:18 & 12.68          & FTW          & 3KK        & i    & 20.92 \pm 0.10 & 0.05              \\
2025-08-31T19:28:30 & 12.68          & FTW          & 3KK        & J    & {>21.0}        & 0.03              \\
2025-08-31T20:04:03 & 12.70          & FTW          & 3KK        & z    & 21.47 \pm 0.16 & 0.04              \\
2025-09-03T19:11:44 & 15.43          & FTW          & 3KK        & g    & 21.63 \pm 0.10 & 0.09              \\
2025-09-03T19:11:44 & 15.43          & FTW          & 3KK        & i    & 20.79 \pm 0.10 & 0.05              \\
2025-09-03T19:11:57 & 15.43          & FTW          & 3KK        & J    & {>21.0}        & 0.03              \\
2025-09-03T19:29:51 & 15.44          & FTW          & 3KK        & z    & 21.28 \pm 0.12 & 0.04              \\
2025-09-03T19:29:51 & 15.44          & FTW          & 3KK        & r    & 21.14 \pm 0.10 & 0.06              \\
2025-09-05T19:20:13 & 17.28          & FTW          & 3KK        & r    & 21.07 \pm 0.18 & 0.06              \\
2025-09-05T19:27:28 & 17.29          & FTW          & 3KK        & J    & {>20.2}        & 0.03              \\
2025-09-05T19:38:29 & 17.29          & FTW          & 3KK        & i    & 20.69 \pm 0.33 & 0.05              \\
2025-09-06T18:59:58 & 18.19          & FTW          & 3KK        & r    & 20.94 \pm 0.10 & 0.06              \\
2025-09-06T18:59:58 & 18.19          & FTW          & 3KK        & z    & 21.04 \pm 0.10 & 0.04              \\
2025-09-06T18:59:58 & 18.19          & FTW          & 3KK        & z    & 20.89 \pm 0.08 & 0.04              \\
2025-09-06T19:00:09 & 18.19          & FTW          & 3KK        & J    & 20.83 \pm 0.29 & 0.03              \\
2025-09-06T19:37:40 & 18.21          & FTW          & 3KK        & g    & 21.71 \pm 0.24 & 0.09              \\
2025-09-06T19:37:40 & 18.21          & FTW          & 3KK        & i    & 20.76 \pm 0.10 & 0.05              \\
2025-09-07T18:58:16 & 19.11          & FTW          & 3KK        & g    & 21.57 \pm 0.10 & 0.09              \\
2025-09-07T18:58:16 & 19.11          & FTW          & 3KK        & i    & 20.56 \pm 0.10 & 0.05              \\
2025-09-07T18:58:28 & 19.11          & FTW          & 3KK        & J    & {>21.0}        & 0.03              \\
2025-09-07T19:34:06 & 19.13          & FTW          & 3KK        & z    & 21.03 \pm 0.08 & 0.04              \\
2025-09-07T19:34:07 & 19.13          & FTW          & 3KK        & r    & 21.06 \pm 0.10 & 0.06              \\
2025-09-07T19:34:07 & 19.13          & FTW          & 3KK        & z    & 20.98 \pm 0.10 & 0.04              \\
2025-09-14T18:53:15 & 25.56          & FTW          & 3KK        & z    & 21.41 \pm 0.14 & 0.04              \\
2025-09-14T18:53:15 & 25.56          & FTW          & 3KK        & r    & 21.61 \pm 0.10 & 0.06              \\
2025-09-14T18:53:27 & 25.56          & FTW          & 3KK        & J    & 20.81 \pm 0.34 & 0.03              \\
2025-09-14T19:29:14 & 25.58          & FTW          & 3KK        & i    & 20.99 \pm 0.23 & 0.05              \\
2025-09-18T18:17:11 & 29.22          & FTW          & 3KK        & z    & 21.30 \pm 0.22 & 0.04              \\
2025-09-18T18:17:11 & 29.22          & FTW          & 3KK        & r    & 21.71 \pm 0.10 & 0.06              \\
2025-09-18T18:17:23 & 29.22          & FTW          & 3KK        & J    & {>20.8}        & 0.03              \\
2025-09-18T19:03:40 & 29.25          & FTW          & 3KK        & g    & 23.11 \pm 0.19 & 0.09              \\
2025-09-18T19:03:40 & 29.25          & FTW          & 3KK        & i    & 21.20 \pm 0.10 & 0.05              \\
2025-09-19T18:27:59 & 30.15          & FTW          & 3KK        & r    & 21.98 \pm 0.10 & 0.06              \\
2025-09-19T18:28:11 & 30.15          & FTW          & 3KK        & J    & {>21.0}        & 0.03              \\
2025-09-19T19:03:56 & 30.17          & FTW          & 3KK        & i    & 21.10 \pm 0.10 & 0.05              \\
2025-09-20T05:08:23 & 30.56          & Gemini North & GMOS       & z    & 21.31 \pm 0.12 & 0.04              \\
2025-09-20T05:19:02 & 30.57          & Gemini North & GMOS       & r    & 22.35 \pm 0.12 & 0.06              \\
2025-09-20T05:27:44 & 30.57          & Gemini North & GMOS       & g    & 22.70 \pm 0.11 & 0.09              \\
2025-09-20T18:42:36 & 31.08          & FTW          & 3KK        & z    & 21.48 \pm 0.30 & 0.04              \\
2025-09-20T18:42:36 & 31.08          & FTW          & 3KK        & r    & 22.06 \pm 0.15 & 0.06              \\
2025-09-20T18:42:49 & 31.08          & FTW          & 3KK        & J    & {>20.9}        & 0.03              \\
2025-09-20T19:00:54 & 31.09          & FTW          & 3KK        & i    & 21.10 \pm 0.10 & 0.05              \\
2025-09-20T19:00:54 & 31.09          & FTW          & 3KK        & g    & 23.23 \pm 0.21 & 0.09              \\
2025-09-21T18:22:50 & 31.99          & FTW          & 3KK        & J    & {>20.3}        & 0.03              \\
2025-09-21T18:40:48 & 32.00          & FTW          & 3KK        & i    & 21.23 \pm 0.17 & 0.05              \\
2025-09-22T18:17:18 & 32.91          & FTW          & 3KK        & J    & {>19.9}        & 0.03              \\
2025-09-23T05:14:12 & 33.33          & Gemini North & GMOS       & z    & 21.64 \pm 0.12 & 0.04              \\
2025-09-23T05:25:04 & 33.34          & Gemini North & GMOS       & i    & 21.68 \pm 0.12 & 0.05              \\
2025-09-23T05:33:45 & 33.34          & Gemini North & GMOS       & r    & 22.08 \pm 0.11 & 0.06              \\
2025-09-23T05:42:27 & 33.35          & Gemini North & GMOS       & g    & 22.82 \pm 0.12 & 0.09              \\
2025-09-26T18:13:51 & 36.59          & FTW          & 3KK        & r    & {>22.7}        & 0.06              \\
2025-09-26T18:13:51 & 36.59          & FTW          & 3KK        & z    & {>21.3}        & 0.04              \\
2025-09-26T18:14:02 & 36.59          & FTW          & 3KK        & J    & {>20.6}        & 0.03              \\
2025-09-26T18:32:07 & 36.61          & FTW          & 3KK        & g    & 23.02 \pm 0.26 & 0.09              \\
2025-09-26T18:32:07 & 36.61          & FTW          & 3KK        & i    & 21.52 \pm 0.14 & 0.05              \\
\enddata
\label{tab:photometry}
\end{deluxetable*}

\section{Lightcurve Model Fits}
\label{app:fits}

\begin{figure*}
    \includegraphics[width=1\textwidth]{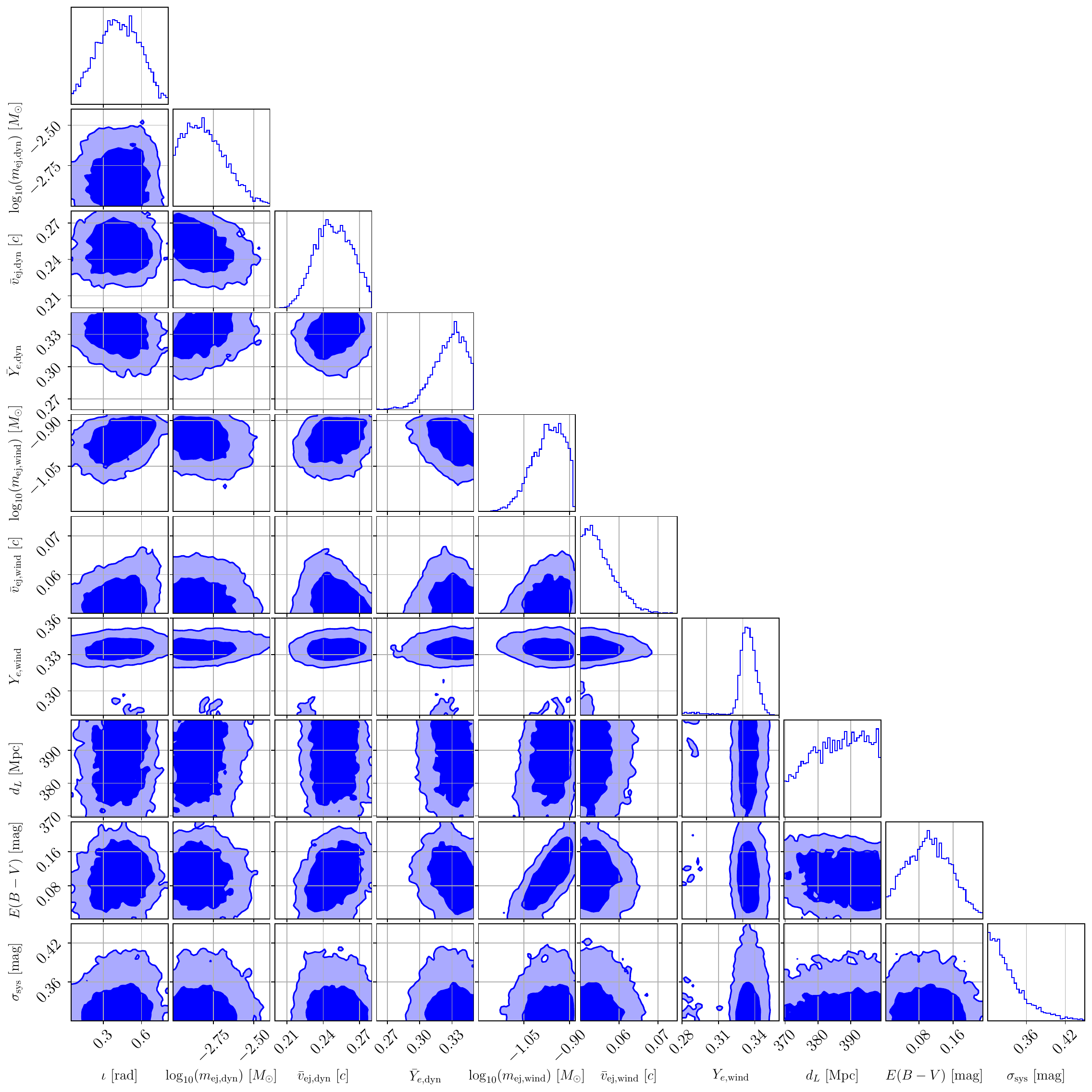}
    \caption{Posterior distribution of the KN analysis after 4 days of observation time.
    The corner plot shows the 68\% and 95\% credible regions.}
    \label{fig:corner_KN}
\end{figure*}

\begin{figure*}
    \includegraphics[width=1\textwidth]{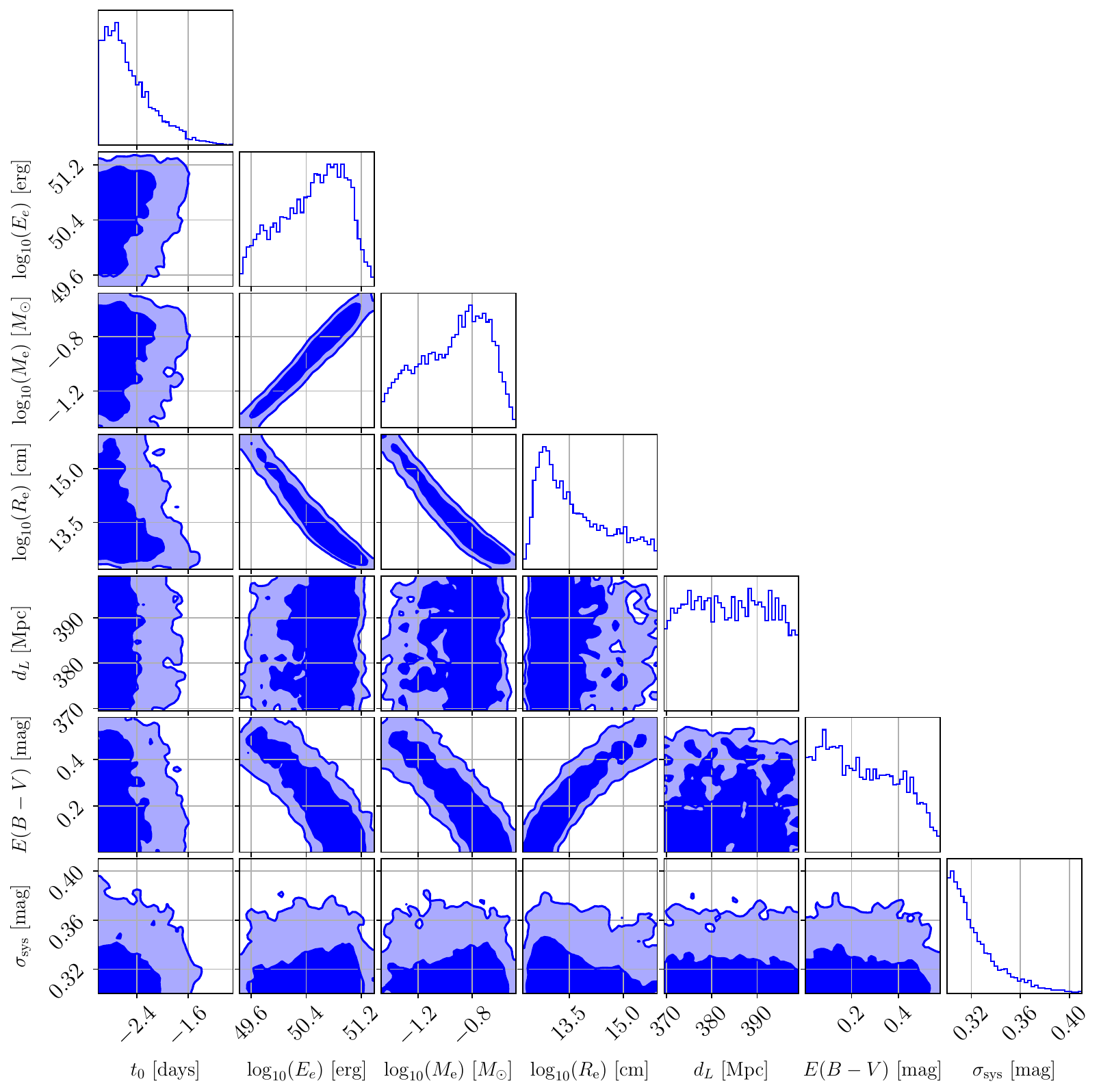}
    \caption{Posterior distribution of the shock cooling analysis after 4 days of observation time.
    The corner plot shows the 68\% and 95\% credible regions.}
    \label{fig:corner_SC}
\end{figure*}

\begin{figure*}
    \includegraphics[width=1\textwidth]{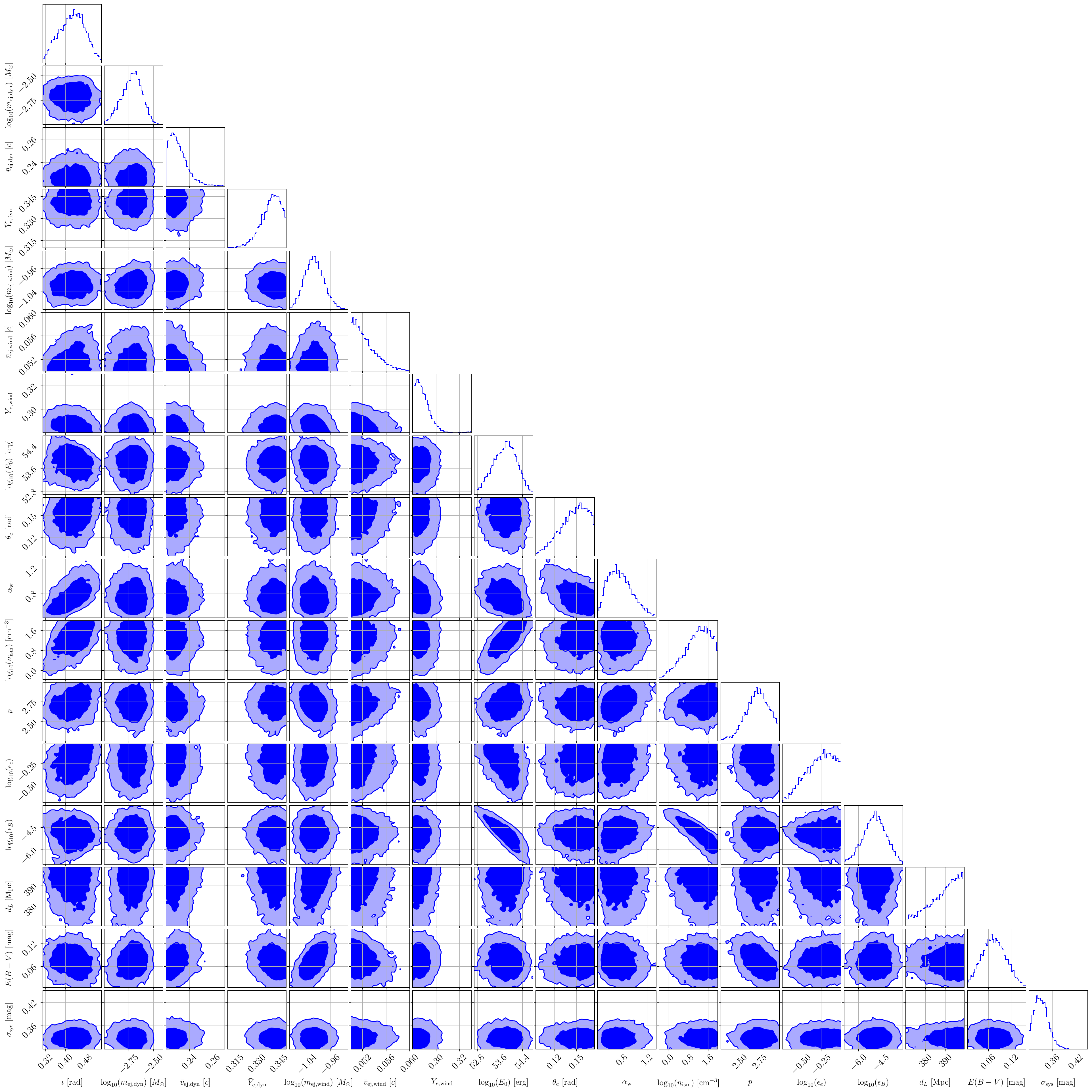}
    \caption{Posterior distribution of the KN plus GRB afterglow analysis for the full data set.
    The corner plot shows the 68\% and 95\% credible regions.}
    \label{fig:corner_KN_GRB}
\end{figure*}

\begin{figure*}
    \includegraphics[width=1\textwidth]{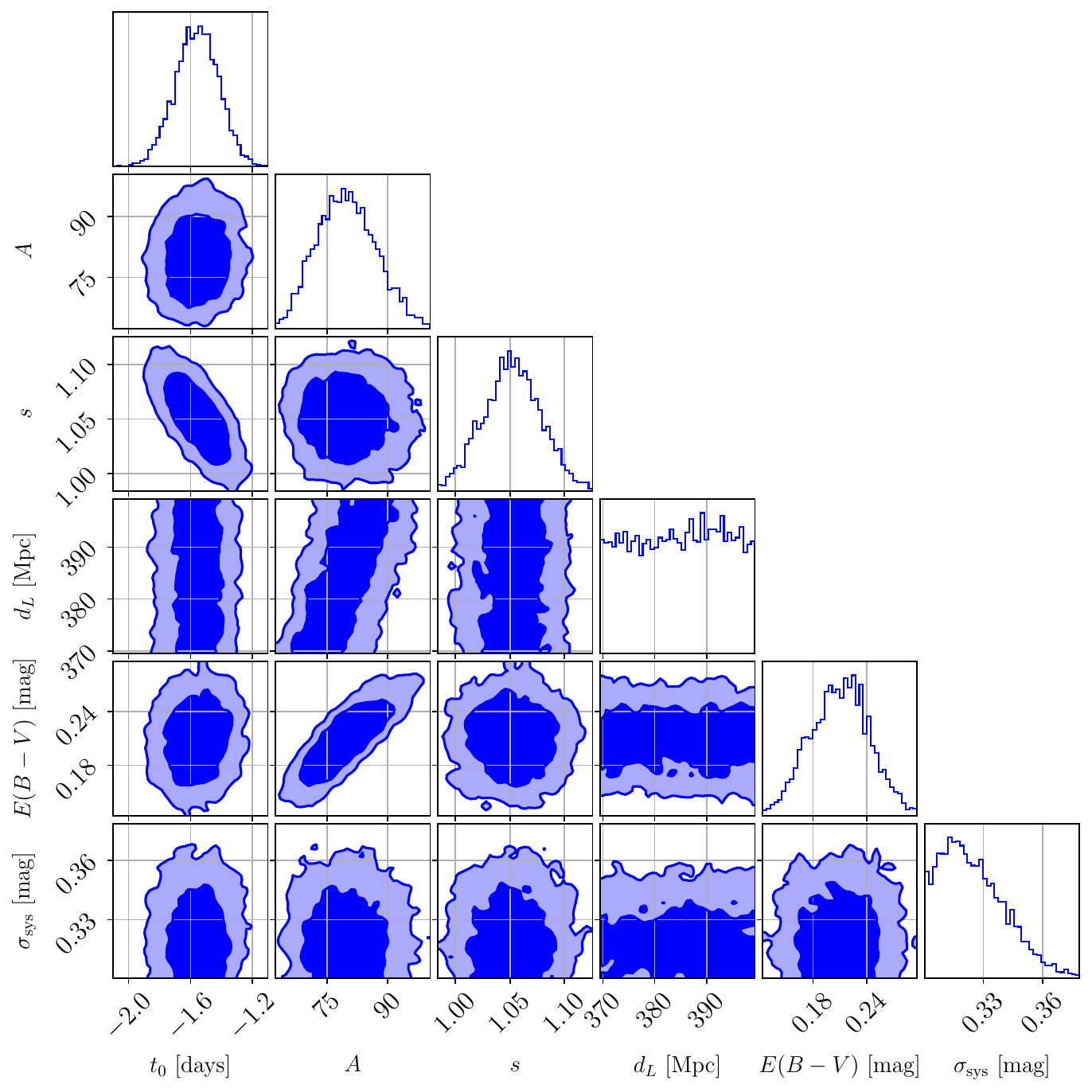}
    \caption{Posterior distribution of the SN 1993J template fit.
    The corner plot shows the 68\% and 95\% credible regions.}
    \label{fig:corner_SN}
\end{figure*}

\begin{table*}[t]
        \centering
    \tabcolsep=0.3cm
    \caption{Parameters and priors used for the Bayesian analyses with the kilonova plus GRB afterglow model and with the SN1993J template. 
    The last column shows the 95\% credible interval for the particular parameter. Uniform priors are marked with $\mathcal{U}$, $\text{Log}$ is a log-uniform prior, and the $\text{Sin}$ prior stands for $\cos(\iota)\sim\mathcal{U}(0,1)$.}
    \begin{tabular}{c c c c c c}
\toprule 
\toprule 
Model & Parameter & Label & Prior & Posterior \\ 
\midrule 
 \multirow{12}{2.5cm}{Kilonova \citep{Koehn:2025zzb} + GRB afterglow \citep{Ryan2019}} & inclination & $\iota$ [rad] & $\text{Sin}(0,\pi/2)$ & $0.43_{-0.10}^{+0.10}$ \\ 
  & dynamical ejecta mass & $\log_{10}(m_{\mathrm{ej,dyn}})$ [$M_\odot$] & $\mathcal{U}(-3, -1.3)$ & $-2.71_{-0.22}^{+0.18}$ \\ 
  & dynamical ejecta velocity & $\bar{v}_{\mathrm{ej,dyn}}$ [$c$] & $\mathcal{U}(0.12, 0.28)$ & $0.23_{-0.01}^{+0.02}$ \\ 
  & dynamical ejecta electron fraction & $\bar{Y}_{e,\mathrm{dyn}}$ & $\mathcal{U}(0.15, 0.35)$ & $0.34_{-0.02}^{+0.01}$ \\ 
  & wind ejecta mass & $\log_{10}(m_{\mathrm{ej,wind}})$ [$M_\odot$] & $\mathcal{U}(-2, -0.9)$ & $-1.02_{-0.06}^{+0.07}$ \\ 
  & wind ejecta velocity & $\bar{v}_{\mathrm{ej,wind}}$ [$c$] & $\mathcal{U}(0.05, 0.15)$ & $0.05_{-0.00}^{+0.01}$ \\ 
  & wind ejecta electron fraction & $Y_{e,\mathrm{wind}}$ & $\mathcal{U}(0.2, 0.4)$ & $0.29_{-0.01}^{+0.04}$ \\ 
  & isotropic jet energy & $\log_{10}(E_0)$ [erg] & $\mathcal{U}(50, 57)$ & $53.80_{-0.90}^{+0.77}$ \\ 
  & jet opening angle & $\theta_{\mathrm{c}}$ [rad] & $\mathcal{U}(0.01, \pi/18)$ & $0.15_{-0.04}^{+0.02}$ \\ 
  & wing factor & $\alpha_{\mathrm{w}}$ & $\mathcal{U}(0.2, 3.5)$ & $0.75_{-0.29}^{+0.47}$ \\ 
  & interstellar medium density & $\log_{10}(n_{\mathrm{ism}})$ [cm$^{-3}$] & $\mathcal{U}(-6, 2)$ & $1.23_{-1.27}^{+0.72}$ \\ 
  & electron power law index & $p$ & $\mathcal{U}(2, 3)$ & $2.72_{-0.35}^{+0.25}$ \\ 
  & electron energy fraction & $\log_{10}(\epsilon_e)$ & $\mathcal{U}(-4, 0)$ & $-0.25_{-0.39}^{+0.24}$ \\ 
  & magnetic energy fraction & $\log_{10}(\epsilon_B)$ & $\mathcal{U}(-8, 0)$ & $-4.91_{-1.68}^{+1.52}$ \\ 
  & luminosity distance & $d_L$ [Mpc] & $\mathcal{U}(369, 399)$ & $391_{-19}^{+8}$ \\ 
  & color excess & $E(B-V)$ [mag] & $\mathcal{U}(0, 0.6)$ & $0.07_{-0.06}^{+0.07}$ \\ 
  & systematic uncertainty & $\sigma_{\rm sys}$ [mag] & $\mathcal{U}(0.3, 2.0)$ & $0.33_{-0.03}^{+0.04}$ \\ 
 & fit residual & $\chi^2$/d.o.f. & - & 0.94 \\ 
\midrule 
 \multirow{9}{2.5cm}{SN 1993J \citep{SN1993J_circular}} & timeshift & $t_0$ [days] & $\mathcal{U}(-3, 0.1)$ & $-1.56_{-0.29}^{+0.27}$ \\ 
  & flux stretch & $A$ & $\text{Log}(0.001, 1000)$ & $79.58_{-13.39}^{+16.01}$ \\ 
  & time stretch & $s$ & $\mathcal{U}(0.05, 4)$ & $1.05_{-0.05}^{+0.05}$ \\ 
  & luminosity distance & $d_L$ [Mpc] & $\mathcal{U}(369, 399)$ & $385_{-15}^{+14}$ \\ 
  & color excess & $E(B-V)$ [mag] & $\mathcal{U}(0, 0.6)$ & $0.21_{-0.07}^{+0.07}$ \\ 
  & systematic uncertainty & $\sigma_{\rm sys}$ [mag] & $\mathcal{U}(0.3, 2.0)$ & $0.32_{-0.02}^{+0.04}$ \\ 
 & fit residual & $\chi^2$/d.o.f. & - & 0.88 \\ 
 \bottomrule
    \end{tabular}
\label{tab:KN_GRB_and_SN}
\end{table*}
In this Appendix, we provide the full posterior distributions from the light curve fits presented in Figure~\ref{fig:lightcurves_nmma}.
As described in \S~\ref{sec:results}, the analyses with the KN or the shock cooling model use data between 0.2 and 4 days of observation time.
Priors are set as listed in Table~\ref{tab:KN_vs_SC}.
Figure~\ref{fig:corner_KN} shows the corner plot of the posterior for the KN analysis.
Likewise, Figure~\ref{fig:corner_SC} shows the shock cooling posterior.

Furthermore, we present posteriors for the joint KN plus GRB afterglow inference in Figure~\ref{fig:corner_KN_GRB} and for the inference with the SN 1993J template in Figure~\ref{fig:corner_SN}.
The priors and confidence intervals are listed in Table~\ref{tab:KN_GRB_and_SN}.
The KN plus GRB afterglow model is able to reproduce the fall and rise of the light curve~\citep[see also][]{Kasliwal2025sn}.
In order to adhere to the X-ray and radio upper limits before 10~days, the model infers a jet inclination angle of $\iota = {25}^{+5}_{-6}\,^\circ$ and a high electron power law index $p = {2.72}^{+0.25}_{-0.33}$ to get a steep spectral slope.
The latter is higher than the canonical value found during particle acceleration simulations of $p\lesssim 2.2$~\citep{Sironi:2015oza}, but lies within the distribution of observed values for GRBs \citep[e.g.,][]{Fong2015}. 
The steep value of $p$ is required to produce a sharp rise of the afterglow lightcurve before peak \citep{Ryan2019}.
While this model provides a reasonable match to the data, it is worth pointing out that further X-ray and radio limits (not utilized here) reported in \citet{OConnor2025ulz} exclude the GRB afterglow fit (i.e., our inferred radio and X-ray lightcurves significantly over predict the flux in those bands when compared to the deep upper limits in \citealt{OConnor2025ulz}).

The inference with the SN 1993J template provides a good fit, except for the early decay of the light curve.
In particular, the time stretching parameter is constrained to $s={1.05}_{-0.05}^{+0.05}$, meaning AT2025ulz evolves on roughly the same temporal scale as SN 1993J.
The start of the explosion is given at $t_0 = {-1.56}_{-0.27}^{+0.27}$, in agreement with \citet{Franz2025}.
The inferences with the SN 1993J template and the KN plus GRB afterglow model achieve a similar $\chi^2$ residual, though it should be noted that the latter analysis includes more data points (e.g., the X-ray and radio limits, which were not included in the SN 1993J analysis).

\section{Spectral Models}\label{app:spec}

We also include modeled spectra at the time of the spectroscopic observations within the first eight days after T0 presented in \citet{Kasliwal2025sn}.

\begin{figure*}
    \centering
    \includegraphics[width=1\linewidth]{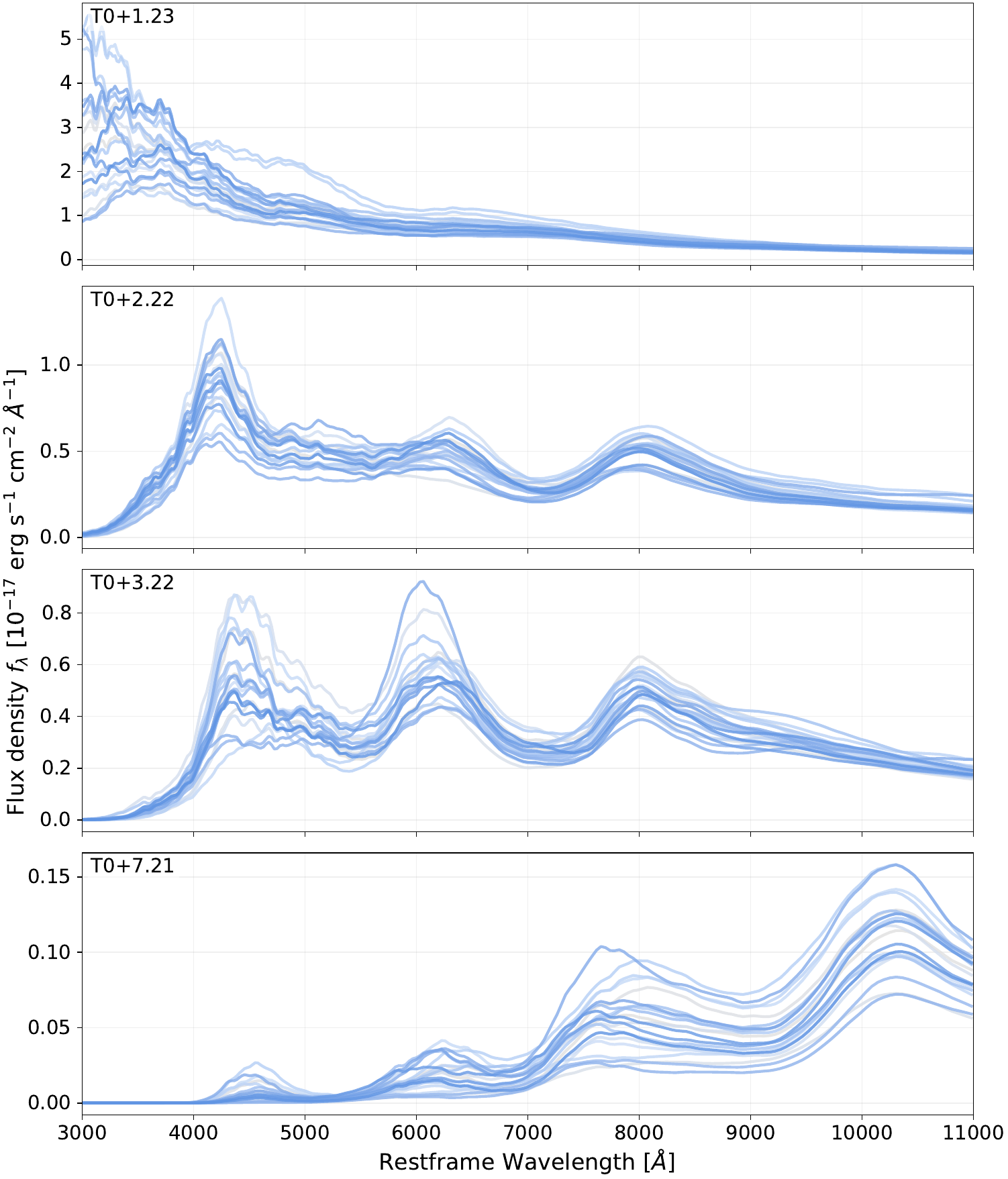}
    \caption{Model spectra of various KN models using the input parameters provided by NMMA (see Table \ref{tab:KN_vs_SC}). Times are chosen to match \citealt{Kasliwal2025sn}.}
    \label{fig:modeled_spectra_kasliwal}
\end{figure*}

\bibliography{bib, sample}{}

\begin{thebibliography}{}
\expandafter\ifx\csname natexlab\endcsname\relax\def\natexlab#1{#1}\fi
\providecommand{\url}[1]{\href{#1}{#1}}
\providecommand{\dodoi}[1]{doi:~\href{http://doi.org/#1}{\nolinkurl{#1}}}
\providecommand{\doeprint}[1]{\href{http://ascl.net/#1}{\nolinkurl{http://ascl.net/#1}}}
\providecommand{\doarXiv}[1]{\href{https://arxiv.org/abs/#1}{\nolinkurl{https://arxiv.org/abs/#1}}}

\bibitem[{B.~P. {Abbott} {et~al.}(2017){Abbott}, {Abbott}, {Abbott}, {Acernese}, {Ackley}, {Adams}, {Adams}, {Addesso}, {Adhikari}, {Adya}, {Affeldt}, {Afrough}, {Agarwal}, {Agathos}, {Agatsuma}, {Aggarwal}, {Aguiar}, {Aiello}, {Ain}, {Ajith}, {Allen}, {Allen}, {Allocca}, {Altin}, {Amato}, {Ananyeva}, {Anderson}, {Anderson}, {Angelova}, {Antier}, {Appert}, {Arai}, {Araya}, {Areeda}, {Arnaud}, {Arun}, {Ascenzi}, {Ashton}, {Ast}, {Aston}, {Astone}, {Atallah}, {Aufmuth}, {Aulbert}, {AultONeal}, {Austin}, {Avila-Alvarez}, {Babak}, {Bacon}, {Bader}, {Bae}, {Bailes}, {Baker}, {Baldaccini}, {Ballardin}, {Ballmer}, {Banagiri}, {Barayoga}, {Barclay}, {Barish}, {Barker}, {Barkett}, {Barone}, {Barr}, {Barsotti}, {Barsuglia}, {Barta}, {Barthelmy}, {Bartlett}, {Bartos}, {Bassiri}, {Basti}, {Batch}, {Bawaj}, {Bayley}, {Bazzan}, {B{\'e}csy}, {Beer}, {Bejger}, {Belahcene}, {Bell}, {Berger}, {Bergmann}, {Bernuzzi}, {Bero}, {Berry}, {Bersanetti}, {Bertolini}, {Betzwieser}, {Bhagwat}, {Bhandare}, {Bilenko}, {Billingsley},
  {Billman}, {Birch}, {Birney}, {Birnholtz}, {Biscans}, {Biscoveanu}, {Bisht}, {Bitossi}, {Biwer}, {Bizouard}, {Blackburn}, {Blackman}, {Blair}, {Blair}, {Blair}, {Bloemen}, {Bock}, {Bode}, {Boer}, {Bogaert}, {Bohe}, {Bondu}, {Bonilla}, {Bonnand}, {Boom}, {Bork}, {Boschi}, {Bose}, {Bossie}, {Bouffanais}, {Bozzi}, {Bradaschia}, {Brady}, {Branchesi}, {Brau}, {Briant}, {Brillet}, {Brinkmann}, {Brisson}, {Brockill}, {Broida}, {Brooks}, {Brown}, {Brown}, {Brunett}, {Buchanan}, {Buikema}, {Bulik}, {Bulten}, {Buonanno}, {Buskulic}, {Buy}, {Byer}, {Cabero}, {Cadonati}, {Cagnoli}, {Cahillane}, {Calder{\'o}n Bustillo}, {Callister}, {Calloni}, {Camp}, {Canepa}, {Canizares}, {Cannon}, {Cao}, {Cao}, {Capano}, {Capocasa}, {Carbognani}, {Caride}, {Carney}, {Carullo}, {Casanueva Diaz}, {Casentini}, {Caudill}, {Cavagli{\`a}}, {Cavalier}, {Cavalieri}, {Cella}, {Cepeda}, {Cerd{\'a}-Dur{\'a}n}, {Cerretani}, {Cesarini}, {Chamberlin}, {Chan}, {Chao}, {Charlton}, {Chase}, {Chassande-Mottin}, {Chatterjee}, {Chatziioannou},
  {Cheeseboro}, {Chen}, {Chen}, {Chen}, {Cheng}, {Chia}, {Chincarini}, {Chiummo}, {Chmiel}, {Cho}, {Cho}, {Chow}, {Christensen}, {Chu}, {Chua}, \& {Chua}}]{abbott_bns_2017}
{Abbott}, B.~P., {Abbott}, R., {Abbott}, T.~D., {et~al.} 2017, \bibinfo{title}{{GW170817: Observation of Gravitational Waves from a Binary Neutron Star Inspiral},} \prl, 119, 161101, \dodoi{10.1103/PhysRevLett.119.161101}

\bibitem[{R. Abbott {et~al.}(2021)Abbott, Abbott, Abraham, Acernese, Ackley, Adams, Adams, Adhikari, Adya, Affeldt, Agarwal, Agathos, Agatsuma, Aggarwal, Aguiar, Aiello, Ain, Ajith, Akutsu, Aleman, Allen, Allocca, Altin, Amato, Anand, Ananyeva, Anderson, Anderson, Ando, Angelova, Ansoldi, Antelis, Antier, Appert, Arai, Arai, Arai, Araki, Araya, Araya, Areeda, Arène, Aritomi, Arnaud, Aronson, Arun, Asada, Asali, Ashton, Aso, Aston, Astone, Aubin, Aufmuth, Aultoneal, Austin, Babak, Badaracco, Bader, Bae, Bae, Baer, Bagnasco, Bai, Baiotti, Baird, Bajpai, Ball, Ballardin, Ballmer, Bals, Balsamo, Baltus, Banagiri, Bankar, Bankar, Barayoga, Barbieri, Barish, Barker, Barneo, Barone, Barr, Barsotti, Barsuglia, Barta, Bartlett, Barton, Bartos, Bassiri, Basti, Bawaj, Bayley, Baylor, Bazzan, Bécsy, Bedakihale, Bejger, Belahcene, Benedetto, Beniwal, Benjamin, Benkel, Bennett, Bentley, Benyaala, Bergamin, Berger, Bernuzzi, Berry, Bersanetti, Bertolini, Betzwieser, Bhandare, Bhandari, Bhattacharjee, Bhaumik, Bidler,
  Bilenko, Billingsley, Birney, Birnholtz, Biscans, Bischi, Biscoveanu, Bisht, Biswas, Bitossi, Bizouard, Blackburn, Blackman, Blair, Blair, Blair, Bobba, Bode, Boer, Bogaert, Boldrini, Bondu, Bonilla, Bonnand, Booker, Boom, Bork, Boschi, Bose, Bose, Bossilkov, Boudart, Bouffanais, Bozzi, Bradaschia, Brady, Bramley, Branch, Branchesi, Brau, Breschi, Briant, Briggs, Brillet, Brinkmann, Brockill, Brooks, Brooks, Brown, Brunett, Bruno, Bruntz, Bryant, Buikema, Bulik, Bulten, Buonanno, Buscicchio, Buskulic, Byer, Cadonati, Caesar, Cagnoli, Cahillane, Cain, Calderón~Bustillo, Callaghan, Callister, Calloni, Camp, Canepa, Cannavacciuolo, Cannon, Cao, Cao, Cao, Capocasa, Capote, Carapella, Carbognani, Carlin, \& Carney}]{abbott_observation_2021}
Abbott, R., Abbott, T.~D., Abraham, S., {et~al.} 2021, \bibinfo{title}{Observation of {Gravitational} {Waves} from {Two} {Neutron} {Star}-{Black} {Hole} {Coalescences},} The Astrophysical Journal, 915, L5, \dodoi{10.3847/2041-8213/ac082e}

\bibitem[{R. Abbott {et~al.}(2023)Abbott, Abbott, Acernese, Ackley, Adams, Adhikari, Adhikari, Adya, Affeldt, Agarwal, Agathos, Agatsuma, Aggarwal, Aguiar, Aiello, Ain, Ajith, Akutsu, de~Alarcón, Akcay, Albanesi, Allocca, Altin, Amato, Anand, Anand, Ananyeva, Anderson, Anderson, Ando, Andrade, Andres, Andrić, Angelova, Ansoldi, Antelis, Antier, Antonini, Appert, Arai, Arai, Arai, Araki, Araya, Araya, Areeda, Arène, Aritomi, Arnaud, Arogeti, Aronson, Arun, Asada, Asali, Ashton, Aso, Assiduo, Aston, Astone, Aubin, Austin, Babak, Badaracco, Bader, Badger, Bae, Bae, Baer, Bagnasco, Bai, Baiotti, Baird, Bajpai, Ball, Ballardin, Ballmer, Balsamo, Baltus, Banagiri, Bankar, Barayoga, Barbieri, Barish, Barker, Barneo, Barone, Barr, Barsotti, Barsuglia, Barta, Bartlett, Barton, Bartos, Bassiri, Basti, Bawaj, Bayley, Baylor, Bazzan, Bécsy, Bedakihale, Bejger, Belahcene, Benedetto, Beniwal, Bennett, Bentley, Benyaala, Bergamin, Berger, Bernuzzi, Berry, Bersanetti, Bertolini, Betzwieser, Beveridge, Bhandare, Bhardwaj,
  Bhattacharjee, Bhaumik, Bilenko, Billingsley, Bini, Birney, Birnholtz, Biscans, Bischi, Biscoveanu, Bisht, Biswas, Bitossi, Bizouard, Blackburn, Blair, Blair, Blair, Bobba, Bode, Boer, Bogaert, Boldrini, Bonavena, Bondu, Bonilla, Bonnand, Booker, Boom, Bork, Boschi, Bose, Bose, Bossilkov, Boudart, Bouffanais, Bozzi, Bradaschia, Brady, Bramley, Branch, Branchesi, Brandt, Brau, Breschi, Briant, Briggs, Brillet, Brinkmann, Brockill, Brooks, Brooks, Brown, Brunett, Bruno, Bruntz, Bryant, Bulik, Bulten, Buonanno, Buscicchio, Buskulic, Buy, Byer, Cadonati, Cagnoli, Cahillane, Bustillo, Callaghan, Callister, Calloni, Cameron, Camp, Canepa, Canevarolo, Cannavacciuolo, Cannon, Cao, Cao, Capocasa, Capote, \& Carapella}]{abbott_population_2023}
Abbott, R., Abbott, T.~D., Acernese, F., {et~al.} 2023, \bibinfo{title}{Population of {Merging} {Compact} {Binaries} {Inferred} {Using} {Gravitational} {Waves} through {GWTC}-3,} Physical Review X, 13, 011048, \dodoi{10.1103/PhysRevX.13.011048}

\bibitem[{N. Aghanim {et~al.}(2020)Aghanim, Akrami, Ashdown, Aumont, Baccigalupi, Ballardini, Banday, Barreiro, Bartolo, Basak, Battye, Benabed, Bernard, Bersanelli, Bielewicz, Bock, Bond, Borrill, Bouchet, Boulanger, Bucher, Burigana, Butler, Calabrese, Cardoso, Carron, Challinor, Chiang, Chluba, Colombo, Combet, Contreras, Crill, Cuttaia, Bernardis, Zotti, Delabrouille, Delouis, Valentino, Diego, Doré, Douspis, Ducout, Dupac, Dusini, Efstathiou, Elsner, Enßlin, Eriksen, Fantaye, Farhang, Fergusson, Fernandez-Cobos, Finelli, Forastieri, Frailis, Fraisse, Franceschi, Frolov, Galeotta, Galli, Ganga, Génova-Santos, Gerbino, Ghosh, González-Nuevo, Górski, Gratton, Gruppuso, Gudmundsson, Hamann, Handley, Hansen, Herranz, Hildebrandt, Hivon, Huang, Jaffe, Jones, Karakci, Keihänen, Keskitalo, Kiiveri, Kim, Kisner, Knox, Krachmalnicoff, Kunz, Kurki-Suonio, Lagache, Lamarre, Lasenby, Lattanzi, Lawrence, Jeune, Lemos, Lesgourgues, Levrier, Lewis, Liguori, Lilje, Lilley, Lindholm, López-Caniego, Lubin, Ma,
  Macías-Pérez, Maggio, Maino, Mandolesi, Mangilli, Marcos-Caballero, Maris, Martin, Martinelli, Martínez-González, Matarrese, Mauri, McEwen, Meinhold, Melchiorri, Mennella, Migliaccio, Millea, Mitra, Miville-Deschênes, Molinari, Montier, Morgante, Moss, Natoli, Nørgaard-Nielsen, Pagano, Paoletti, Partridge, Patanchon, Peiris, Perrotta, Pettorino, Piacentini, Polastri, Polenta, Puget, Rachen, Reinecke, Remazeilles, Renzi, Rocha, Rosset, Roudier, Rubiño-Martín, Ruiz-Granados, Salvati, Sandri, Savelainen, Scott, Shellard, Sirignano, Sirri, Spencer, Sunyaev, Suur-Uski, Tauber, Tavagnacco, Tenti, Toffolatti, Tomasi, Trombetti, Valenziano, Valiviita, Tent, Vibert, Vielva, Villa, Vittorio, Wandelt, Wehus, White, White, Zacchei, \& Zonca}]{aghanim_planck_2020}
Aghanim, N., Akrami, Y., Ashdown, M., {et~al.} 2020, \bibinfo{title}{Planck 2018 results - {VI}. {Cosmological} parameters,} Astronomy \& Astrophysics, 641, A6, \dodoi{10.1051/0004-6361/201833910}

\bibitem[{I. {Agudo} {et~al.}(2023){Agudo}, {Amati}, {An}, {Bauer}, {Benetti}, {Bernardini}, {Beswick}, {Bhirombhakdi}, {de Boer}, {Branchesi}, {Brennan}, {Brocato}, {Caballero-Garc{\'\i}a}, {Cappellaro}, {Castro Rodr{\'\i}guez}, {Castro-Tirado}, {Chambers}, {Chassande-Mottin}, {Chaty}, {Chen}, {Coleiro}, {Covino}, {D'Ammando}, {D'Avanzo}, {D'Elia}, {Fiore}, {Fl{\"o}rs}, {Fraser}, {Frey}, {Frohmaier}, {Fulton}, {Galbany}, {Gall}, {Gao}, {Garc{\'\i}a-Rojas}, {Ghirlanda}, {Giarratana}, {Gillanders}, {Giroletti}, {Gompertz}, {Gromadzki}, {Heintz}, {Hjorth}, {Hu}, {Huber}, {Inkenhaag}, {Izzo}, {Jin}, {Jonker}, {Kann}, {Kool}, {Kotak}, {Leloudas}, {Levan}, {Lin}, {Lyman}, {Magnier}, {Maguire}, {Mandel}, {Marcote}, {Mata S{\'a}nchez}, {Mattila}, {Melandri}, {Micha{\l}owski}, {Moldon}, {Nicholl}, {Nicuesa Guelbenzu}, {Oates}, {Onori}, {Orienti}, {Paladino}, {Paragi}, {Perez-Torres}, {Pian}, {Pignata}, {Piranomonte}, {Quirola-V{\'a}squez}, {Ragosta}, {Rau}, {Ronchini}, {Rossi}, {S{\'a}nchez-Ram{\'\i}rez}, {Salafia},
  {Schulze}, {Smartt}, {Smith}, {Sollerman}, {Srivastav}, {Starling}, {Steeghs}, {Stevance}, {Tanvir}, {Testa}, {Torres}, {Valeev}, {Vergani}, {Vescovi}, {Wainscost}, {Watson}, {Wiersema}, {Wyrzykowski}, {Yang}, {Yang}, \& {Young}}]{Agudo2023}
{Agudo}, I., {Amati}, L., {An}, T., {et~al.} 2023, \bibinfo{title}{{Panning for gold, but finding helium: Discovery of the ultra-stripped supernova SN 2019wxt from gravitational-wave follow-up observations},} \aap, 675, A201, \dodoi{10.1051/0004-6361/202244751}

\bibitem[{H. {Aihara} {et~al.}(2022){Aihara}, {AlSayyad}, {Ando}, {Armstrong}, {Bosch}, {Egami}, {Furusawa}, {Furusawa}, {Harasawa}, {Harikane}, {Hsieh}, {Ikeda}, {Ito}, {Iwata}, {Kodama}, {Koike}, {Kokubo}, {Komiyama}, {Li}, {Liang}, {Lin}, {Lupton}, {Lust}, {MacArthur}, {Mawatari}, {Mineo}, {Miyatake}, {Miyazaki}, {More}, {Morishima}, {Murayama}, {Nakajima}, {Nakata}, {Nishizawa}, {Oguri}, {Okabe}, {Okura}, {Ono}, {Osato}, {Ouchi}, {Pan}, {Plazas Malag{\'o}n}, {Price}, {Reed}, {Rykoff}, {Shibuya}, {Simunovic}, {Strauss}, {Sugimori}, {Suto}, {Suzuki}, {Takada}, {Takagi}, {Takata}, {Takita}, {Tanaka}, {Tang}, {Taranu}, {Terai}, {Toba}, {Turner}, {Uchiyama}, {Vijarnwannaluk}, {Waters}, {Yamada}, {Yamamoto}, \& {Yamashita}}]{2022PASJ...74..247A}
{Aihara}, H., {AlSayyad}, Y., {Ando}, M., {et~al.} 2022, \bibinfo{title}{{Third data release of the Hyper Suprime-Cam Subaru Strategic Program},} \pasj, 74, 247, \dodoi{10.1093/pasj/psab122}

\bibitem[{S. Anand {et~al.}(2023)Anand {et~al.}}]{Anand:2023jbz}
Anand, S., {et~al.} 2023, \bibinfo{title}{{Chemical Distribution of the Dynamical Ejecta in the Neutron Star Merger GW170817},} \doarXiv{2307.11080}

\bibitem[{C. Andrade {et~al.}(2025)Andrade, Alserkal, Salazar~Manzano, Martin, Andreoni, Coughlin, Guessoum, \& Rivera~Sandoval}]{andrade_effect_2025}
Andrade, C., Alserkal, R., Salazar~Manzano, L., {et~al.} 2025, \bibinfo{title}{The {Effect} of {Vera} {C}. {Rubin} {Observatory} {Cadence} {Selections} on {Kilonova} {Detectability},} Publications of the Astronomical Society of the Pacific, 137, 034102, \dodoi{10.1088/1538-3873/adbfbc}

\bibitem[{I. {Andreoni} {et~al.}(2022{\natexlab{a}}){Andreoni}, {Margutti}, {Salafia}, {Parazin}, {Villar}, {Coughlin}, {Yoachim}, {Mortensen}, {Brethauer}, {Smartt}, {Kasliwal}, {Alexander}, {Anand}, {Berger}, {Bernardini}, {Bianco}, {Blanchard}, {Bloom}, {Brocato}, {Bulla}, {Cartier}, {Cenko}, {Chornock}, {Copperwheat}, {Corsi}, {D'Ammando}, {D'Avanzo}, {H{\'e}l{\`e}ne Datrier}, {Foley}, {Ghirlanda}, {Goobar}, {Grindlay}, {Hajela}, {Holz}, {Karambelkar}, {Kool}, {Lamb}, {Laskar}, {Levan}, {Maguire}, {May}, {Melandri}, {Milisavljevic}, {Miller}, {Nicholl}, {Nissanke}, {Palmese}, {Piranomonte}, {Rest}, {Sagu{\'e}s-Carracedo}, {Siellez}, {Singer}, {Smith}, {Steeghs}, \& {Tanvir}}]{Andreoni2022gw}
{Andreoni}, I., {Margutti}, R., {Salafia}, O.~S., {et~al.} 2022{\natexlab{a}}, \bibinfo{title}{{Target-of-opportunity Observations of Gravitational-wave Events with Vera C. Rubin Observatory},} \apjs, 260, 18, \dodoi{10.3847/1538-4365/ac617c}

\bibitem[{I. {Andreoni} {et~al.}(2022{\natexlab{b}}){Andreoni}, {Coughlin}, {Almualla}, {Bellm}, {Bianco}, {Bulla}, {Cucchiara}, {Dietrich}, {Goobar}, {Kool}, {Li}, {Ragosta}, {Sagu{\'e}s-Carracedo}, \& {Singer}}]{Andreoni2022kn}
{Andreoni}, I., {Coughlin}, M.~W., {Almualla}, M., {et~al.} 2022{\natexlab{b}}, \bibinfo{title}{{Optimizing Cadences with Realistic Light-curve Filtering for Serendipitous Kilonova Discovery with Vera Rubin Observatory},} \apjs, 258, 5, \dodoi{10.3847/1538-4365/ac3bae}

\bibitem[{I. {Andreoni} {et~al.}(2024){Andreoni}, {Coughlin}, {Criswell}, {Bulla}, {Toivonen}, {Singer}, {Palmese}, {Burns}, {Gezari}, {Kasliwal}, {Kiendrebeogo}, {Mahabal}, {Moriya}, {Rest}, {Scolnic}, {Simcoe}, {Soon}, {Stein}, \& {Travouillon}}]{AndreoniRoman}
{Andreoni}, I., {Coughlin}, M.~W., {Criswell}, A.~W., {et~al.} 2024, \bibinfo{title}{{Enabling kilonova science with Nancy Grace Roman Space Telescope},} Astroparticle Physics, 155, 102904, \dodoi{10.1016/j.astropartphys.2023.102904}

\bibitem[{J.~J. {Andrews} \& I. {Mandel}(2019){Andrews} \& {Mandel}}]{Andrews2019a}
{Andrews}, J.~J., \& {Mandel}, I. 2019, \bibinfo{title}{{Double Neutron Star Populations and Formation Channels},} \apjl, 880, L8, \dodoi{10.3847/2041-8213/ab2ed1}

\bibitem[{I. {Arcavi} {et~al.}(2017){Arcavi}, {Hosseinzadeh}, {Howell}, {McCully}, {Poznanski}, {Kasen}, {Barnes}, {Zaltzman}, {Vasylyev}, {Maoz}, \& {Valenti}}]{Arcavi2017}
{Arcavi}, I., {Hosseinzadeh}, G., {Howell}, D.~A., {et~al.} 2017, \bibinfo{title}{{Optical emission from a kilonova following a gravitational-wave-detected neutron-star merger},} \nat, 551, 64, \dodoi{10.1038/nature24291}

\bibitem[{G. Ashton {et~al.}(2021)Ashton, Ackley, Hernandez, \& Piotrzkowski}]{ashton_current_2021}
Ashton, G., Ackley, K., Hernandez, I.~M., \& Piotrzkowski, B. 2021, \bibinfo{title}{Current observations are insufficient to confidently associate the binary black hole merger {GW190521} with {AGN} {J124942}.3 + 344929,} Classical and Quantum Gravity, 38, 235004, \dodoi{10.1088/1361-6382/ac33bb}

\bibitem[{ {Astropy Collaboration} {et~al.}(2018){Astropy Collaboration}, {Price-Whelan}, {Sip{\H{o}}cz}, {G{\"u}nther}, {Lim}, {Crawford}, {Conseil}, {Shupe}, {Craig}, {Dencheva}, {Ginsburg}, {VanderPlas}, {Bradley}, {P{\'e}rez-Su{\'a}rez}, {de Val-Borro}, {Aldcroft}, {Cruz}, {Robitaille}, {Tollerud}, {Ardelean}, {Babej}, {Bach}, {Bachetti}, {Bakanov}, {Bamford}, {Barentsen}, {Barmby}, {Baumbach}, {Berry}, {Biscani}, {Boquien}, {Bostroem}, {Bouma}, {Brammer}, {Bray}, {Breytenbach}, {Buddelmeijer}, {Burke}, {Calderone}, {Cano Rodr{\'\i}guez}, {Cara}, {Cardoso}, {Cheedella}, {Copin}, {Corrales}, {Crichton}, {D'Avella}, {Deil}, {Depagne}, {Dietrich}, {Donath}, {Droettboom}, {Earl}, {Erben}, {Fabbro}, {Ferreira}, {Finethy}, {Fox}, {Garrison}, {Gibbons}, {Goldstein}, {Gommers}, {Greco}, {Greenfield}, {Groener}, {Grollier}, {Hagen}, {Hirst}, {Homeier}, {Horton}, {Hosseinzadeh}, {Hu}, {Hunkeler}, {Ivezi{\'c}}, {Jain}, {Jenness}, {Kanarek}, {Kendrew}, {Kern}, {Kerzendorf}, {Khvalko}, {King}, {Kirkby}, {Kulkarni},
  {Kumar}, {Lee}, {Lenz}, {Littlefair}, {Ma}, {Macleod}, {Mastropietro}, {McCully}, {Montagnac}, {Morris}, {Mueller}, {Mumford}, {Muna}, {Murphy}, {Nelson}, {Nguyen}, {Ninan}, {N{\"o}the}, {Ogaz}, {Oh}, {Parejko}, {Parley}, {Pascual}, {Patil}, {Patil}, {Plunkett}, {Prochaska}, {Rastogi}, {Reddy Janga}, {Sabater}, {Sakurikar}, {Seifert}, {Sherbert}, {Sherwood-Taylor}, {Shih}, {Sick}, {Silbiger}, {Singanamalla}, {Singer}, {Sladen}, {Sooley}, {Sornarajah}, {Streicher}, {Teuben}, {Thomas}, {Tremblay}, {Turner}, {Terr{\'o}n}, {van Kerkwijk}, {de la Vega}, {Watkins}, {Weaver}, {Whitmore}, {Woillez}, {Zabalza}, \& {Astropy Contributors}}]{2018AJ....156..123A}
{Astropy Collaboration}, {Price-Whelan}, A.~M., {Sip{\H{o}}cz}, B.~M., {et~al.} 2018, \bibinfo{title}{{The Astropy Project: Building an Open-science Project and Status of the v2.0 Core Package},} \aj, 156, 123, \dodoi{10.3847/1538-3881/aabc4f}

\bibitem[{ {Astropy Collaboration} {et~al.}(2022){Astropy Collaboration}, {Price-Whelan}, {Lim}, {Earl}, {Starkman}, {Bradley}, {Shupe}, {Patil}, {Corrales}, {Brasseur}, {N{\"o}the}, {Donath}, {Tollerud}, {Morris}, {Ginsburg}, {Vaher}, {Weaver}, {Tocknell}, {Jamieson}, {van Kerkwijk}, {Robitaille}, {Merry}, {Bachetti}, {G{\"u}nther}, {Aldcroft}, {Alvarado-Montes}, {Archibald}, {B{\'o}di}, {Bapat}, {Barentsen}, {Baz{\'a}n}, {Biswas}, {Boquien}, {Burke}, {Cara}, {Cara}, {Conroy}, {Conseil}, {Craig}, {Cross}, {Cruz}, {D'Eugenio}, {Dencheva}, {Devillepoix}, {Dietrich}, {Eigenbrot}, {Erben}, {Ferreira}, {Foreman-Mackey}, {Fox}, {Freij}, {Garg}, {Geda}, {Glattly}, {Gondhalekar}, {Gordon}, {Grant}, {Greenfield}, {Groener}, {Guest}, {Gurovich}, {Handberg}, {Hart}, {Hatfield-Dodds}, {Homeier}, {Hosseinzadeh}, {Jenness}, {Jones}, {Joseph}, {Kalmbach}, {Karamehmetoglu}, {Ka{\l}uszy{\'n}ski}, {Kelley}, {Kern}, {Kerzendorf}, {Koch}, {Kulumani}, {Lee}, {Ly}, {Ma}, {MacBride}, {Maljaars}, {Muna}, {Murphy}, {Norman},
  {O'Steen}, {Oman}, {Pacifici}, {Pascual}, {Pascual-Granado}, {Patil}, {Perren}, {Pickering}, {Rastogi}, {Roulston}, {Ryan}, {Rykoff}, {Sabater}, {Sakurikar}, {Salgado}, {Sanghi}, {Saunders}, {Savchenko}, {Schwardt}, {Seifert-Eckert}, {Shih}, {Jain}, {Shukla}, {Sick}, {Simpson}, {Singanamalla}, {Singer}, {Singhal}, {Sinha}, {Sip{\H{o}}cz}, {Spitler}, {Stansby}, {Streicher}, {{\v{S}}umak}, {Swinbank}, {Taranu}, {Tewary}, {Tremblay}, {de Val-Borro}, {Van Kooten}, {Vasovi{\'c}}, {Verma}, {de Miranda Cardoso}, {Williams}, {Wilson}, {Winkel}, {Wood-Vasey}, {Xue}, {Yoachim}, {Zhang}, {Zonca}, \& {Astropy Project Contributors}}]{2022ApJ...935..167A}
{Astropy Collaboration}, {Price-Whelan}, A.~M., {Lim}, P.~L., {et~al.} 2022, \bibinfo{title}{{The Astropy Project: Sustaining and Growing a Community-oriented Open-source Project and the Latest Major Release (v5.0) of the Core Package},} \apj, 935, 167, \dodoi{10.3847/1538-4357/ac7c74}

\bibitem[{H. {Baba} {et~al.}(2002){Baba}, {Yasuda}, {Ichikawa}, {Yagi}, {Iwamoto}, {Takata}, {Horaguchi}, {Taga}, {Watanabe}, {Ozawa}, \& {Hamabe}}]{2002ASPC..281..298B}
{Baba}, H., {Yasuda}, N., {Ichikawa}, S.-I., {et~al.} 2002, in Astronomical Society of the Pacific Conference Series, Vol. 281, Astronomical Data Analysis Software and Systems XI, ed. D.~A. {Bohlender}, D.~{Durand}, \& T.~H. {Handley}, 298

\bibitem[{S. {Banerjee} {et~al.}(2025){Banerjee}, {Botticella}, {Brennan}, {Cappellaro}, {Chen}, {D'Avanzo}, {D'Elia}, {de Pasquale}, {Eyles-Ferris}, {Fraser}, {Gillanders}, {Gompertz}, {Habeeb}, {Izzo}, {Jonker}, {Levan}, {Bj{\o}rn Malesani}, {Martin-Carrillo}, {Nicholl}, {Oates}, {Piranomonte}, {Piro}, {Rossi}, {Sharan Salafia}, {Sarin}, {Schulze}, {Singh}, {Smartt}, {Sneppen}, {Sollerman}, {Steeghs}, {Tanvir}, {Thakur}, \& {Engrave Collaboration}}]{2025GCN.41532....1B}
{Banerjee}, S., {Botticella}, M.-T., {Brennan}, S.~J., {et~al.} 2025, \bibinfo{title}{{LIGO/Virgo/KAGRA S250818k: ENGRAVE observations of SN 2025ulz as a type II supernova},} GRB Coordinates Network, 41532, 1

\bibitem[{S. Banerjee {et~al.}(2025)Banerjee, Botticella, Brennan, Cappellaro, Chen, D'Avanzo, D'Elia, Pasquale, Eyles-Ferris, Fraser, Gillanders, Gompertz, Habeeb, Izzo, Jonker, Levan, Malesani, Martin-Carrillo, Nicholl, Oates, Piranomonte, Piro, Rossi, Salafia, Sarin, Schulze, Singh, Smartt, Sneppen, Sollerman, Steeghs, Tanvir, \& Thakur}]{banerjee_engrave_2025}
Banerjee, S., Botticella, M., Brennan, S.~J., {et~al.} 2025, \bibinfo{title}{{ENGRAVE} {Transient} {Classification} {Report} for 2025-08-25,} Transient Name Server Classification Report, 2025-3373, 1.
\newblock \url{https://ui.adsabs.harvard.edu/abs/2025TNSCR3373....1B}

\bibitem[{T. Barna {et~al.}(2025)Barna, Fremling, Ahumada, Andreoni, Banerjee, Bloom, Bulla, Chen, Coughlin, Dietrich, Hall, Junell, Rusholme, Sollerman, \& Sravan}]{barna_iib_2025}
Barna, T., Fremling, C., Ahumada, T., {et~al.} 2025, \bibinfo{title}{{IIb} or not {IIb}: {A} {Catalog} of {ZTF} {Kilonova} {Imposters},} Publications of the Astronomical Society of the Pacific, 137, 084105, \dodoi{10.1088/1538-3873/adf578}

\bibitem[{R.~L. {Becerra} {et~al.}(2025{\natexlab{a}}){Becerra}, {Troja}, \& {Dichiara}}]{2025GCN.41528....1B}
{Becerra}, R.~L., {Troja}, E., \& {Dichiara}, S. 2025{\natexlab{a}}, \bibinfo{title}{{LIGO/Virgo/KAGRA S250818k: Swift Observations of AT 2025ulz - Second Epoch},} GRB Coordinates Network, 41528, 1

\bibitem[{R.~L. {Becerra} {et~al.}(2025{\natexlab{b}}){Becerra}, {Yang}, {Watson}, {Troja}, \& {Lee}}]{2025GCN.41544....1B}
{Becerra}, R.~L., {Yang}, Y., {Watson}, A.~M., {Troja}, E., \& {Lee}, W.~H. 2025{\natexlab{b}}, \bibinfo{title}{{LIGO/Virgo/KAGRA S250818k: GTC/OSIRIS Confirmation of Rebrightening of AT2025ulz},} GRB Coordinates Network, 41544, 1

\bibitem[{E.~C. {Bellm} {et~al.}(2019){Bellm}, {Kulkarni}, {Graham}, {Dekany}, {Smith}, {Riddle}, {Masci}, {Helou}, {Prince}, {Adams}, {Barbarino}, {Barlow}, {Bauer}, {Beck}, {Belicki}, {Biswas}, {Blagorodnova}, {Bodewits}, {Bolin}, {Brinnel}, {Brooke}, {Bue}, {Bulla}, {Burruss}, {Cenko}, {Chang}, {Connolly}, {Coughlin}, {Cromer}, {Cunningham}, {De}, {Delacroix}, {Desai}, {Duev}, {Eadie}, {Farnham}, {Feeney}, {Feindt}, {Flynn}, {Franckowiak}, {Frederick}, {Fremling}, {Gal-Yam}, {Gezari}, {Giomi}, {Goldstein}, {Golkhou}, {Goobar}, {Groom}, {Hacopians}, {Hale}, {Henning}, {Ho}, {Hover}, {Howell}, {Hung}, {Huppenkothen}, {Imel}, {Ip}, {Ivezi{\'c}}, {Jackson}, {Jones}, {Juric}, {Kasliwal}, {Kaspi}, {Kaye}, {Kelley}, {Kowalski}, {Kramer}, {Kupfer}, {Landry}, {Laher}, {Lee}, {Lin}, {Lin}, {Lunnan}, {Giomi}, {Mahabal}, {Mao}, {Miller}, {Monkewitz}, {Murphy}, {Ngeow}, {Nordin}, {Nugent}, {Ofek}, {Patterson}, {Penprase}, {Porter}, {Rauch}, {Rebbapragada}, {Reiley}, {Rigault}, {Rodriguez}, {van Roestel}, {Rusholme},
  {van Santen}, {Schulze}, {Shupe}, {Singer}, {Soumagnac}, {Stein}, {Surace}, {Sollerman}, {Szkody}, {Taddia}, {Terek}, {Van Sistine}, {van Velzen}, {Vestrand}, {Walters}, {Ward}, {Ye}, {Yu}, {Yan}, \& {Zolkower}}]{Bellm2019}
{Bellm}, E.~C., {Kulkarni}, S.~R., {Graham}, M.~J., {et~al.} 2019, \bibinfo{title}{{The Zwicky Transient Facility: System Overview, Performance, and First Results},} \pasp, 131, 018002, \dodoi{10.1088/1538-3873/aaecbe}

\bibitem[{E. {Berger} {et~al.}(2013){Berger}, {Fong}, \& {Chornock}}]{Berger2013kilonova}
{Berger}, E., {Fong}, W., \& {Chornock}, R. 2013, \bibinfo{title}{{An r-process Kilonova Associated with the Short-hard GRB 130603B},} \apjl, 774, L23, \dodoi{10.1088/2041-8205/774/2/L23}

\bibitem[{E. {Bertin}(2006{\natexlab{a}}){Bertin}}]{2006ASPC..351..112B}
{Bertin}, E. 2006{\natexlab{a}}, in Astronomical Society of the Pacific Conference Series, Vol. 351, Astronomical Data Analysis Software and Systems XV, ed. C.~{Gabriel}, C.~{Arviset}, D.~{Ponz}, \& S.~{Enrique}, 112

\bibitem[{E. {Bertin}(2006{\natexlab{b}}){Bertin}}]{Bertin2006}
{Bertin}, E. 2006{\natexlab{b}}, in Astronomical Society of the Pacific Conference Series, Vol. 351, Astronomical Data Analysis Software and Systems XV, ed. C.~{Gabriel}, C.~{Arviset}, D.~{Ponz}, \& S.~{Enrique}, 112

\bibitem[{E. {Bertin}(2010){Bertin}}]{Bertin2010}
{Bertin}, E. 2010, \bibinfo{title}{{SWarp: Resampling and Co-adding FITS Images Together},} \doeprint{1010.068}

\bibitem[{E. {Bertin} \& S. {Arnouts}(1996{\natexlab{a}}){Bertin} \& {Arnouts}}]{1996A&AS..117..393B}
{Bertin}, E., \& {Arnouts}, S. 1996{\natexlab{a}}, \bibinfo{title}{{SExtractor: Software for source extraction.},} \aaps, 117, 393, \dodoi{10.1051/aas:1996164}

\bibitem[{E. {Bertin} \& S. {Arnouts}(1996{\natexlab{b}}){Bertin} \& {Arnouts}}]{Bertin1996}
{Bertin}, E., \& {Arnouts}, S. 1996{\natexlab{b}}, \bibinfo{title}{{SExtractor: Software for source extraction.},} \aaps, 117, 393, \dodoi{10.1051/aas:1996164}

\bibitem[{E. {Bertin} {et~al.}(2002){Bertin}, {Mellier}, {Radovich}, {Missonnier}, {Didelon}, \& {Morin}}]{2002ASPC..281..228B}
{Bertin}, E., {Mellier}, Y., {Radovich}, M., {et~al.} 2002, in Astronomical Society of the Pacific Conference Series, Vol. 281, Astronomical Data Analysis Software and Systems XI, ed. D.~A. {Bohlender}, D.~{Durand}, \& T.~H. {Handley}, 228

\bibitem[{M. {Bhardwaj} {et~al.}(2024){Bhardwaj}, {Palmese}, {Maga{\~n}a Hernandez}, {D'Emilio}, \& {Morisaki}}]{2024ApJ...977..122B}
{Bhardwaj}, M., {Palmese}, A., {Maga{\~n}a Hernandez}, I., {D'Emilio}, V., \& {Morisaki}, S. 2024, \bibinfo{title}{{Challenges for Fast Radio Bursts as Multimessenger Sources from Binary Neutron Star Mergers},} \apj, 977, 122, \dodoi{10.3847/1538-4357/ad9023}

\bibitem[{J. {Bosch} {et~al.}(2018){Bosch}, {Armstrong}, {Bickerton}, {Furusawa}, {Ikeda}, {Koike}, {Lupton}, {Mineo}, {Price}, {Takata}, {Tanaka}, {Yasuda}, {AlSayyad}, {Becker}, {Coulton}, {Coupon}, {Garmilla}, {Huang}, {Krughoff}, {Lang}, {Leauthaud}, {Lim}, {Lust}, {MacArthur}, {Mandelbaum}, {Miyatake}, {Miyazaki}, {Murata}, {More}, {Okura}, {Owen}, {Swinbank}, {Strauss}, {Yamada}, \& {Yamanoi}}]{2018PASJ...70S...5B}
{Bosch}, J., {Armstrong}, R., {Bickerton}, S., {et~al.} 2018, \bibinfo{title}{{The Hyper Suprime-Cam software pipeline},} \pasj, 70, S5, \dodoi{10.1093/pasj/psx080}

\bibitem[{M. Breschi {et~al.}(2024)Breschi, Gamba, Carullo, Godzieba, Bernuzzi, Perego, \& Radice}]{Breschi:2024qlc}
Breschi, M., Gamba, R., Carullo, G., {et~al.} 2024, \bibinfo{title}{{Bayesian inference of multimessenger astrophysical data: Joint and coherent inference of gravitational waves and kilonovae},} Astron. Astrophys., 689, A51, \dodoi{10.1051/0004-6361/202449173}

\bibitem[{M. Breschi {et~al.}(2021)Breschi, Perego, Bernuzzi, Del~Pozzo, Nedora, Radice, \& Vescovi}]{Breschi:2021tbm}
Breschi, M., Perego, A., Bernuzzi, S., {et~al.} 2021, \bibinfo{title}{{AT2017gfo: Bayesian inference and model selection of multicomponent kilonovae and constraints on the neutron star equation of state},} Mon. Not. Roy. Astron. Soc., 505, 1661, \dodoi{10.1093/mnras/stab1287}

\bibitem[{G. {Bruni} {et~al.}(2025){Bruni}, {Chandra}, {Christy}, {Kale}, {Laskar}, {Mohnani}, {Das}, {Resmi}, {Ricci}, \& {Troja}}]{2025GCN.41577....1B}
{Bruni}, G., {Chandra}, P., {Christy}, C., S.~C., {et~al.} 2025, \bibinfo{title}{{LIGO/Virgo/KAGRA S250818k: uGMRT 1.3 GHz observations of AT2025ulz},} GRB Coordinates Network, 41577, 1

\bibitem[{J. Buchner {et~al.}(2014)Buchner, Georgakakis, Nandra, Hsu, Rangel, Brightman, Merloni, Salvato, Donley, \& Kocevski}]{Buchner:2014nha}
Buchner, J., Georgakakis, A., Nandra, K., {et~al.} 2014, \bibinfo{title}{{X-ray spectral modelling of the AGN obscuring region in the CDFS: Bayesian model selection and catalogue},} Astron. Astrophys., 564, A125, \dodoi{10.1051/0004-6361/201322971}

\bibitem[{M. {Bulla}(2019){Bulla}}]{Bulla2019}
{Bulla}, M. 2019, \bibinfo{title}{{POSSIS: predicting spectra, light curves, and polarization for multidimensional models of supernovae and kilonovae},} \mnras, 489, 5037, \dodoi{10.1093/mnras/stz2495}

\bibitem[{M. {Bulla}(2023){Bulla}}]{Bulla2023}
{Bulla}, M. 2023, \bibinfo{title}{{The critical role of nuclear heating rates, thermalization efficiencies, and opacities for kilonova modelling and parameter inference},} \mnras, 520, 2558, \dodoi{10.1093/mnras/stad232}

\bibitem[{M. Busmann {et~al.}(2025)Busmann, Hall, Gruen, O'Connor, Palmese, \& Kasliwal}]{busmann_ligovirgokagra_2025}
Busmann, M., Hall, X.~J., Gruen, D., {et~al.} 2025, \bibinfo{title}{{LIGO}/{Virgo}/{KAGRA} {S250818k}: {FTW} {Observations} {Show} {Continued} {Reddening} of {AT} 2025ulz,} GRB Coordinates Network, 41535, 1.
\newblock \url{https://ui.adsabs.harvard.edu/abs/2025GCN.41535....1B}

\bibitem[{M. {Busmann} {et~al.}(2025){Busmann}, {O'Connor}, {Sommer}, {Gruen}, {Beniamini}, {Gill}, {Moss}, {Palmese}, {Riffeser}, {Yang}, {Troja}, {Dichiara}, {Ricci}, {Klingler}, {G{\"o}ssl}, {Hu}, {Rau}, {Ries}, {Ryan}, {Schmidt}, {Yadav}, \& {Zeimann}}]{2025arXiv250314588B}
{Busmann}, M., {O'Connor}, B., {Sommer}, J., {et~al.} 2025, \bibinfo{title}{{The curious case of EP241021a: Unraveling the mystery of its exceptional rebrightening},} arXiv e-prints, arXiv:2503.14588, \dodoi{10.48550/arXiv.2503.14588}

\bibitem[{T. {Cabrera} {et~al.}(2024){Cabrera}, {Palmese}, {Hu}, {O'Connor}, {Ford}, {McKernan}, {Andreoni}, {Ahumada}, {Amsellem}, {Busmann}, {Clark}, {Coughlin}, {Dadiani}, {Diaz}, {Graham}, {Gruen}, {Kunnumkai}, {Postiglione}, {Riffeser}, {Sommer}, \& {Valdes}}]{Cabrera2024}
{Cabrera}, T., {Palmese}, A., {Hu}, L., {et~al.} 2024, \bibinfo{title}{{Searching for electromagnetic emission in an AGN from the gravitational wave binary black hole merger candidate S230922g},} \prd, 110, 123029, \dodoi{10.1103/PhysRevD.110.123029}

\bibitem[{T.~A. Callister(2024)Callister}]{callister2024observedgravitationalwavepopulations}
Callister, T.~A. 2024, \bibinfo{title}{Observed Gravitational-Wave Populations,} \doarXiv{2410.19145}

\bibitem[{K.~C. {Chambers} {et~al.}(2016){Chambers}, {Magnier}, {Metcalfe}, {Flewelling}, {Huber}, {Waters}, {Denneau}, {Draper}, {Farrow}, {Finkbeiner}, {Holmberg}, {Koppenhoefer}, {Price}, {Rest}, {Saglia}, {Schlafly}, {Smartt}, {Sweeney}, {Wainscoat}, {Burgett}, {Chastel}, {Grav}, {Heasley}, {Hodapp}, {Jedicke}, {Kaiser}, {Kudritzki}, {Luppino}, {Lupton}, {Monet}, {Morgan}, {Onaka}, {Shiao}, {Stubbs}, {Tonry}, {White}, {Ba{\~n}ados}, {Bell}, {Bender}, {Bernard}, {Boegner}, {Boffi}, {Botticella}, {Calamida}, {Casertano}, {Chen}, {Chen}, {Cole}, {Deacon}, {Frenk}, {Fitzsimmons}, {Gezari}, {Gibbs}, {Goessl}, {Goggia}, {Gourgue}, {Goldman}, {Grant}, {Grebel}, {Hambly}, {Hasinger}, {Heavens}, {Heckman}, {Henderson}, {Henning}, {Holman}, {Hopp}, {Ip}, {Isani}, {Jackson}, {Keyes}, {Koekemoer}, {Kotak}, {Le}, {Liska}, {Long}, {Lucey}, {Liu}, {Martin}, {Masci}, {McLean}, {Mindel}, {Misra}, {Morganson}, {Murphy}, {Obaika}, {Narayan}, {Nieto-Santisteban}, {Norberg}, {Peacock}, {Pier}, {Postman}, {Primak}, {Rae},
  {Rai}, {Riess}, {Riffeser}, {Rix}, {R{\"o}ser}, {Russel}, {Rutz}, {Schilbach}, {Schultz}, {Scolnic}, {Strolger}, {Szalay}, {Seitz}, {Small}, {Smith}, {Soderblom}, {Taylor}, {Thomson}, {Taylor}, {Thakar}, {Thiel}, {Thilker}, {Unger}, {Urata}, {Valenti}, {Wagner}, {Walder}, {Walter}, {Watters}, {Werner}, {Wood-Vasey}, \& {Wyse}}]{Chambers2016}
{Chambers}, K.~C., {Magnier}, E.~A., {Metcalfe}, N., {et~al.} 2016, \bibinfo{title}{{The Pan-STARRS1 Surveys},} arXiv e-prints, arXiv:1612.05560.
\newblock \doarXiv{1612.05560}

\bibitem[{E.~A. {Chase} {et~al.}(2022){Chase}, {O'Connor}, {Fryer}, {Troja}, {Korobkin}, {Wollaeger}, {Ristic}, {Fontes}, {Hungerford}, \& {Herring}}]{Chase2022}
{Chase}, E.~A., {O'Connor}, B., {Fryer}, C.~L., {et~al.} 2022, \bibinfo{title}{{Kilonova Detectability with Wide-field Instruments},} \apj, 927, 163, \dodoi{10.3847/1538-4357/ac3d25}

\bibitem[{S.~S. Chaudhary {et~al.}(2024)Chaudhary, Toivonen, Waratkar, Mo, Chatterjee, Antier, Brockill, Coughlin, Essick, Ghosh, Morisaki, Baral, Baylor, Adhikari, Brady, Cabourn~Davies, Dal~Canton, Cavaglia, Creighton, Choudhary, Chu, Clearwater, Davis, Dent, Drago, Ewing, Godwin, Guo, Hanna, Huxford, Harry, Katsavounidis, Kovalam, Li, Magee, Marx, Meacher, Messick, Morice-Atkinson, Pace, De~Pietri, Piotrzkowski, Roy, Sachdev, Singer, Singh, Szczepanczyk, Tang, Trevor, Tsukada, Villa-Ortega, Wen, \& Wysocki}]{chaudhary_low-latency_2024}
Chaudhary, S.~S., Toivonen, A., Waratkar, G., {et~al.} 2024, \bibinfo{title}{Low-latency gravitational wave alert products and their performance at the time of the fourth {LIGO}-{Virgo}-{KAGRA} observing run,} Proceedings of the National Academy of Science, 121, e2316474121, \dodoi{10.1073/pnas.2316474121}

\bibitem[{H.-Y. {Chen} {et~al.}(2018){Chen}, {Fishbach}, \& {Holz}}]{Chen2018}
{Chen}, H.-Y., {Fishbach}, M., \& {Holz}, D.~E. 2018, \bibinfo{title}{{A two per cent Hubble constant measurement from standard sirens within five years},} \nat, 562, 545, \dodoi{10.1038/s41586-018-0606-0}

\bibitem[{W.-X. Chen \& A.~M. Beloborodov(2007)Chen \& Beloborodov}]{chen_neutrino-cooled_2007}
Chen, W.-X., \& Beloborodov, A.~M. 2007, \bibinfo{title}{Neutrino-cooled {Accretion} {Disks} around {Spinning} {Black} {Holes},} The Astrophysical Journal, 657, 383, \dodoi{10.1086/508923}

\bibitem[{Y.-X. {Chen} \& B.~D. {Metzger}(2025){Chen} \& {Metzger}}]{ChenMetzger2025}
{Chen}, Y.-X., \& {Metzger}, B.~D. 2025, \bibinfo{title}{{Gravitational Instability and Fragmentation in Collapsar Disks Supports the Formation of Sub-Solar Neutron Stars},} arXiv e-prints, arXiv:2508.17183, \dodoi{10.48550/arXiv.2508.17183}

\bibitem[{T.~S. {Chonis} {et~al.}(2014){Chonis}, {Hill}, {Lee}, {Tuttle}, \& {Vattiat}}]{Chonis2014}
{Chonis}, T.~S., {Hill}, G.~J., {Lee}, H., {Tuttle}, S.~E., \& {Vattiat}, B.~L. 2014, in Society of Photo-Optical Instrumentation Engineers (SPIE) Conference Series, Vol. 9147, Ground-based and Airborne Instrumentation for Astronomy V, ed. S.~K. {Ramsay}, I.~S. {McLean}, \& H.~{Takami}, 91470A, \dodoi{10.1117/12.2056005}

\bibitem[{T.~S. {Chonis} {et~al.}(2016){Chonis}, {Hill}, {Lee}, {Tuttle}, {Vattiat}, {Drory}, {Indahl}, {Peterson}, \& {Ramsey}}]{Chonis2016}
{Chonis}, T.~S., {Hill}, G.~J., {Lee}, H., {et~al.} 2016, in Society of Photo-Optical Instrumentation Engineers (SPIE) Conference Series, Vol. 9908, Ground-based and Airborne Instrumentation for Astronomy VI, ed. C.~J. {Evans}, L.~{Simard}, \& H.~{Takami}, 99084C, \dodoi{10.1117/12.2232209}

\bibitem[{D.~A. {Coulter} {et~al.}(2017){Coulter}, {Foley}, {Kilpatrick}, {Drout}, {Piro}, {Shappee}, {Siebert}, {Simon}, {Ulloa}, {Kasen}, {Madore}, {Murguia-Berthier}, {Pan}, {Prochaska}, {Ramirez-Ruiz}, {Rest}, \& {Rojas-Bravo}}]{Coulter2017}
{Coulter}, D.~A., {Foley}, R.~J., {Kilpatrick}, C.~D., {et~al.} 2017, \bibinfo{title}{{Swope Supernova Survey 2017a (SSS17a), the optical counterpart to a gravitational wave source},} Science, 358, 1556, \dodoi{10.1126/science.aap9811}

\bibitem[{P.~S. {Cowperthwaite} {et~al.}(2017){Cowperthwaite}, {Berger}, {Villar}, {Metzger}, {Nicholl}, {Chornock}, {Blanchard}, {Fong}, {Margutti}, {Soares-Santos}, {Alexander}, {Allam}, {Annis}, {Brout}, {Brown}, {Butler}, {Chen}, {Diehl}, {Doctor}, {Drout}, {Eftekhari}, {Farr}, {Finley}, {Foley}, {Frieman}, {Fryer}, {Garc{\'\i}a-Bellido}, {Gill}, {Guillochon}, {Herner}, {Holz}, {Kasen}, {Kessler}, {Marriner}, {Matheson}, {Neilsen}, {Quataert}, {Palmese}, {Rest}, {Sako}, {Scolnic}, {Smith}, {Tucker}, {Williams}, {Balbinot}, {Carlin}, {Cook}, {Durret}, {Li}, {Lopes}, {Louren{\c{c}}o}, {Marshall}, {Medina}, {Muir}, {Mu{\~n}oz}, {Sauseda}, {Schlegel}, {Secco}, {Vivas}, {Wester}, {Zenteno}, {Zhang}, {Abbott}, {Banerji}, {Bechtol}, {Benoit-L{\'e}vy}, {Bertin}, {Buckley-Geer}, {Burke}, {Capozzi}, {Carnero Rosell}, {Carrasco Kind}, {Castander}, {Crocce}, {Cunha}, {D'Andrea}, {da Costa}, {Davis}, {DePoy}, {Desai}, {Dietrich}, {Drlica-Wagner}, {Eifler}, {Evrard}, {Fernand ez}, {Flaugher}, {Fosalba}, {Gaztanaga},
  {Gerdes}, {Giannantonio}, {Goldstein}, {Gruen}, {Gruendl}, {Gutierrez}, {Honscheid}, {Jain}, {James}, {Jeltema}, {Johnson}, {Johnson}, {Kent}, {Krause}, {Kron}, {Kuehn}, {Nuropatkin}, {Lahav}, {Lima}, {Lin}, {Maia}, {March}, {Martini}, {McMahon}, {Menanteau}, {Miller}, {Miquel}, {Mohr}, {Neilsen}, {Nichol}, {Ogando}, {Plazas}, {Roe}, {Romer}, {Roodman}, {Rykoff}, {Sanchez}, {Scarpine}, {Schindler}, {Schubnell}, {Sevilla-Noarbe}, {Smith}, {Smith}, {Sobreira}, {Suchyta}, {Swanson}, {Tarle}, {Thomas}, {Thomas}, {Troxel}, {Vikram}, {Walker}, {Wechsler}, {Weller}, {Yanny}, \& {Zuntz}}]{Cowperthwaite2017}
{Cowperthwaite}, P.~S., {Berger}, E., {Villar}, V.~A., {et~al.} 2017, \bibinfo{title}{{The Electromagnetic Counterpart of the Binary Neutron Star Merger LIGO/Virgo GW170817. II. UV, Optical, and Near-infrared Light Curves and Comparison to Kilonova Models},} \apjl, 848, L17, \dodoi{10.3847/2041-8213/aa8fc7}

\bibitem[{C. de~Barra(2025)de~Barra}]{de_barra_ligovirgokagra_2025}
de~Barra, C. 2025, \bibinfo{title}{{LIGO}/{Virgo}/{KAGRA} {S250818k}: {Upper} limits from {Fermi}-{GBM} {Observations},} GRB Coordinates Network, 41441, 1.
\newblock \url{https://ui.adsabs.harvard.edu/abs/2025GCN.41441....1D}

\bibitem[{ {DESI Collaboration} {et~al.}(2025){DESI Collaboration}, Abdul-Karim, Aguilar, Ahlen, Alam, Allen, Prieto, Alves, Anand, Andrade, Armengaud, Aviles, Bailey, Baltay, Bansal, Bault, Behera, BenZvi, Bianchi, Blake, Brieden, Brodzeller, Brooks, Buckley-Geer, Burtin, Calderon, Canning, Rosell, Carrilho, Casas, Castander, Cereskaite, Charles, Chaussidon, Chaves-Montero, Chebat, Chen, Claybaugh, Cole, Cooper, Cuceu, Dawson, de~la Macorra, de~Mattia, Deiosso, Della~Costa, Demina, Dey, Dey, Ding, Doel, Edelstein, Eisenstein, Elbers, Fagrelius, Fanning, Fernández-García, Ferraro, Font-Ribera, Forero-Romero, Frenk, Garcia-Quintero, Garrison, Gaztañaga, Gil-Marín, Gontcho, Gonzalez, Gonzalez-Morales, Gordon, Green, Gutierrez, Guy, Hadzhiyska, Hahn, He, Herbold, Herrera-Alcantar, Ho, Honscheid, Howlett, Huterer, Ishak, Juneau, Kamble, Karaçaylı, Kehoe, Kent, Kim, Kirkby, Kisner, Koposov, Kremin, Krolewski, Lahav, Lamman, Landriau, Lang, Lasker, Goff, Guillou, Leauthaud, Levi, Li, Li, Lodha, Lokken,
  Lozano-Rodríguez, Magneville, Manera, Martini, Matthewson, Meisner, Mena-Fernández, Menegas, Mergulhão, Miquel, Moustakas, Muñoz-Gutiérrez, Muñoz-Santos, Myers, Nadathur, Naidoo, Napolitano, Newman, Niz, Noriega, Paillas, Palanque-Delabrouille, Pan, Peacock, Ibanez, Percival, Pérez-Fernández, Pérez-Ràfols, Pieri, Poppett, Prada, Rabinowitz, Raichoor, Ramírez-Pérez, Rashkovetskyi, Ravoux, Rich, Rocher, Rockosi, Rohlf, Román-Herrera, Ross, Rossi, Ruggeri, Ruhlmann-Kleider, Samushia, Sanchez, Sanders, Schlegel, Schubnell, Seo, Shafieloo, Sharples, Silber, Sinigaglia, Sprayberry, Tan, Tarlé, Taylor, Turner, Ureña-López, Vaisakh, Valdes, Valogiannis, Vargas-Magaña, Verde, Walther, Weaver, Weinberg, White, Wolfson, Yèche, Yu, Zaborowski, Zarrouk, Zhai, Zhang, Zhao, Zhao, Zhou, \& Zou}]{desi_collaboration_desi_2025}
{DESI Collaboration}, Abdul-Karim, M., Aguilar, J., {et~al.} 2025, \bibinfo{title}{{DESI} {DR2} {Results} {II}: {Measurements} of {Baryon} {Acoustic} {Oscillations} and {Cosmological} {Constraints},} \url{https://arxiv.org/abs/2503.14738v2}

\bibitem[{A. {Dey} {et~al.}(2019){Dey}, {Schlegel}, {Lang}, {Blum}, {Burleigh}, {Fan}, {Findlay}, {Finkbeiner}, {Herrera}, {Juneau}, {Landriau}, {Levi}, {McGreer}, {Meisner}, {Myers}, {Moustakas}, {Nugent}, {Patej}, {Schlafly}, {Walker}, {Valdes}, {Weaver}, {Y{\`e}che}, {Zou}, {Zhou}, {Abareshi}, {Abbott}, {Abolfathi}, {Aguilera}, {Alam}, {Allen}, {Alvarez}, {Annis}, {Ansarinejad}, {Aubert}, {Beechert}, {Bell}, {BenZvi}, {Beutler}, {Bielby}, {Bolton}, {Brice{\~n}o}, {Buckley-Geer}, {Butler}, {Calamida}, {Carlberg}, {Carter}, {Casas}, {Castander}, {Choi}, {Comparat}, {Cukanovaite}, {Delubac}, {DeVries}, {Dey}, {Dhungana}, {Dickinson}, {Ding}, {Donaldson}, {Duan}, {Duckworth}, {Eftekharzadeh}, {Eisenstein}, {Etourneau}, {Fagrelius}, {Farihi}, {Fitzpatrick}, {Font-Ribera}, {Fulmer}, {G{\"a}nsicke}, {Gaztanaga}, {George}, {Gerdes}, {Gontcho}, {Gorgoni}, {Green}, {Guy}, {Harmer}, {Hernandez}, {Honscheid}, {Huang}, {James}, {Jannuzi}, {Jiang}, {Joyce}, {Karcher}, {Karkar}, {Kehoe}, {Kneib}, {Kueter-Young}, {Lan},
  {Lauer}, {Le Guillou}, {Le Van Suu}, {Lee}, {Lesser}, {Perreault Levasseur}, {Li}, {Mann}, {Marshall}, {Mart{\'\i}nez-V{\'a}zquez}, {Martini}, {du Mas des Bourboux}, {McManus}, {Meier}, {M{\'e}nard}, {Metcalfe}, {Mu{\~n}oz-Guti{\'e}rrez}, {Najita}, {Napier}, {Narayan}, {Newman}, {Nie}, {Nord}, {Norman}, {Olsen}, {Paat}, {Palanque-Delabrouille}, {Peng}, {Poppett}, {Poremba}, {Prakash}, {Rabinowitz}, {Raichoor}, {Rezaie}, {Robertson}, {Roe}, {Ross}, {Ross}, {Rudnick}, {Safonova}, {Saha}, {S{\'a}nchez}, {Savary}, {Schweiker}, {Scott}, {Seo}, {Shan}, {Silva}, {Slepian}, {Soto}, {Sprayberry}, {Staten}, {Stillman}, {Stupak}, {Summers}, {Sien Tie}, {Tirado}, {Vargas-Maga{\~n}a}, {Vivas}, {Wechsler}, {Williams}, {Yang}, {Yang}, {Yapici}, {Zaritsky}, {Zenteno}, {Zhang}, {Zhang}, {Zhou}, \& {Zhou}}]{2019AJ....157..168D}
{Dey}, A., {Schlegel}, D.~J., {Lang}, D., {et~al.} 2019, \bibinfo{title}{{Overview of the DESI Legacy Imaging Surveys},} \aj, 157, 168, \dodoi{10.3847/1538-3881/ab089d}

\bibitem[{T. Dietrich {et~al.}(2021)Dietrich, Hinderer, \& Samajdar}]{dietrich_interpreting_2021}
Dietrich, T., Hinderer, T., \& Samajdar, A. 2021, \bibinfo{title}{Interpreting binary neutron star mergers: describing the binary neutron star dynamics, modelling gravitational waveforms, and analyzing detections,} General Relativity and Gravitation, 53, 27, \dodoi{10.1007/s10714-020-02751-6}

\bibitem[{T. Dietrich \& M. Ujevic(2017)Dietrich \& Ujevic}]{Dietrich:2016fpt}
Dietrich, T., \& Ujevic, M. 2017, \bibinfo{title}{{Modeling dynamical ejecta from binary neutron star mergers and implications for electromagnetic counterparts},} Class. Quant. Grav., 34, 105014, \dodoi{10.1088/1361-6382/aa6bb0}

\bibitem[{M.~R. {Drout} {et~al.}(2017){Drout}, {Piro}, {Shappee}, {Kilpatrick}, {Simon}, {Contreras}, {Coulter}, {Foley}, {Siebert}, {Morrell}, {Boutsia}, {Di Mille}, {Holoien}, {Kasen}, {Kollmeier}, {Madore}, {Monson}, {Murguia-Berthier}, {Pan}, {Prochaska}, {Ramirez-Ruiz}, {Rest}, {Adams}, {Alatalo}, {Ba{\~n}ados}, {Baughman}, {Beers}, {Bernstein}, {Bitsakis}, {Campillay}, {Hansen}, {Higgs}, {Ji}, {Maravelias}, {Marshall}, {Moni Bidin}, {Prieto}, {Rasmussen}, {Rojas-Bravo}, {Strom}, {Ulloa}, {Vargas-Gonz{\'a}lez}, {Wan}, \& {Whitten}}]{Drout2017}
{Drout}, M.~R., {Piro}, A.~L., {Shappee}, B.~J., {et~al.} 2017, \bibinfo{title}{{Light curves of the neutron star merger GW170817/SSS17a: Implications for r-process nucleosynthesis},} Science, 358, 1570, \dodoi{10.1126/science.aaq0049}

\bibitem[{S. Dye {et~al.}(2018)Dye, Lawrence, Read, Fan, Kerr, Varricatt, Furnell, Edge, Irwin, Hambly, Lucas, Almaini, Chambers, Green, Hewett, Liu, McGreer, Best, Zhang, Sutorius, Froebrich, Magnier, Hasinger, Lederer, Bold, \& Tedds}]{dye_ukirt_2018}
Dye, S., Lawrence, A., Read, M.~A., {et~al.} 2018, \bibinfo{title}{The {UKIRT} {Hemisphere} {Survey}: definition and {J}-band data release,} Monthly Notices of the Royal Astronomical Society, 473, 5113, \dodoi{10.1093/mnras/stx2622}

\bibitem[{P.~A. {Evans} {et~al.}(2017){Evans}, {Cenko}, {Kennea}, {Emery}, {Kuin}, {Korobkin}, {Wollaeger}, {Fryer}, {Madsen}, {Harrison}, {Xu}, {Nakar}, {Hotokezaka}, {Lien}, {Campana}, {Oates}, {Troja}, {Breeveld}, {Marshall}, {Barthelmy}, {Beardmore}, {Burrows}, {Cusumano}, {D'A{\`\i}}, {D'Avanzo}, {D'Elia}, {de Pasquale}, {Even}, {Fontes}, {Forster}, {Garcia}, {Giommi}, {Grefenstette}, {Gronwall}, {Hartmann}, {Heida}, {Hungerford}, {Kasliwal}, {Krimm}, {Levan}, {Malesani}, {Melandri}, {Miyasaka}, {Nousek}, {O'Brien}, {Osborne}, {Pagani}, {Page}, {Palmer}, {Perri}, {Pike}, {Racusin}, {Rosswog}, {Siegel}, {Sakamoto}, {Sbarufatti}, {Tagliaferri}, {Tanvir}, \& {Tohuvavohu}}]{Evans2017}
{Evans}, P.~A., {Cenko}, S.~B., {Kennea}, J.~A., {et~al.} 2017, \bibinfo{title}{{Swift and NuSTAR observations of GW170817: Detection of a blue kilonova},} Science, 358, 1565, \dodoi{10.1126/science.aap9580}

\bibitem[{E.~L. Fitzpatrick(1999)Fitzpatrick}]{fitzpatrick_correcting_1999}
Fitzpatrick, E.~L. 1999, \bibinfo{title}{Correcting for the {Effects} of {Interstellar} {Extinction},} Publications of the Astronomical Society of the Pacific, 111, 63, \dodoi{10.1086/316293}

\bibitem[{W. {Fong} \& E. {Berger}(2013){Fong} \& {Berger}}]{FongBerger2013}
{Fong}, W., \& {Berger}, E. 2013, \bibinfo{title}{{The Locations of Short Gamma-Ray Bursts as Evidence for Compact Object Binary Progenitors},} \apj, 776, 18, \dodoi{10.1088/0004-637X/776/1/18}

\bibitem[{W. {Fong} {et~al.}(2015){Fong}, {Berger}, {Margutti}, \& {Zauderer}}]{Fong2015}
{Fong}, W., {Berger}, E., {Margutti}, R., \& {Zauderer}, B.~A. 2015, \bibinfo{title}{{A Decade of Short-duration Gamma-Ray Burst Broadband Afterglows: Energetics, Circumburst Densities, and Jet Opening Angles},} \apj, 815, 102, \dodoi{10.1088/0004-637X/815/2/102}

\bibitem[{W. {Fong} {et~al.}(2021){Fong}, {Laskar}, {Rastinejad}, {Escorial}, {Schroeder}, {Barnes}, {Kilpatrick}, {Paterson}, {Berger}, {Metzger}, {Dong}, {Nugent}, {Strausbaugh}, {Blanchard}, {Goyal}, {Cucchiara}, {Terreran}, {Alexander}, {Eftekhari}, {Fryer}, {Margalit}, {Margutti}, \& {Nicholl}}]{Fong2021kn}
{Fong}, W., {Laskar}, T., {Rastinejad}, J., {et~al.} 2021, \bibinfo{title}{{The Broadband Counterpart of the Short GRB 200522A at z = 0.5536: A Luminous Kilonova or a Collimated Outflow with a Reverse Shock?},} \apj, 906, 127, \dodoi{10.3847/1538-4357/abc74a}

\bibitem[{W.-f. {Fong} {et~al.}(2022){Fong}, {Nugent}, {Dong}, {Berger}, {Paterson}, {Chornock}, {Levan}, {Blanchard}, {Alexander}, {Andrews}, {Cobb}, {Cucchiara}, {Fox}, {Fryer}, {Gordon}, {Kilpatrick}, {Lunnan}, {Margutti}, {Miller}, {Milne}, {Nicholl}, {Perley}, {Rastinejad}, {Escorial}, {Schroeder}, {Smith}, {Tanvir}, \& {Terreran}}]{Fong2022}
{Fong}, W.-f., {Nugent}, A.~E., {Dong}, Y., {et~al.} 2022, \bibinfo{title}{{Short GRB Host Galaxies. I. Photometric and Spectroscopic Catalogs, Host Associations, and Galactocentric Offsets},} \apj, 940, 56, \dodoi{10.3847/1538-4357/ac91d0}

\bibitem[{F. Foucart {et~al.}(2024)Foucart, Duez, Kidder, Pfeiffer, \& Scheel}]{Foucart:2024kci}
Foucart, F., Duez, M.~D., Kidder, L.~E., Pfeiffer, H.~P., \& Scheel, M.~A. 2024, \bibinfo{title}{{Dynamical ejecta from binary neutron star mergers: Impact of a small residual eccentricity and of the equation of state implementation},} Phys. Rev. D, 110, 024003, \dodoi{10.1103/PhysRevD.110.024003}

\bibitem[{N. {Franz} {et~al.}(2025){Franz}, {Subrayan}, {Kilpatrick}, {Hosseinzadeh}, {Sand}, {Alexander}, {Fong}, {Christy}, {Pearson}, {Laskar}, {Hsu}, {Rastinejad}, {Lundquist}, {Berger}, {Bostroem}, {Bom}, {Darc}, {Gurwell}, {Hostler Schimpf}, {Keating}, {Noel}, {Ransome}, {Rao}, {Santana-Silva}, {Souza Santos}, {Shrestha}, {Anche}, {Andrews}, {Borthakur}, {Butler}, {Coppejans}, {Daly}, {Daniel}, {Duffell}, {Eftekhari}, {Fields}, {Gagliano}, {Golay}, {Grichener}, {Hamden}, {Hiramatsu}, {Kumar}, {Manikantan}, {Margutti}, {Paschalidis}, {Paterson}, {Reichart}, {Renzo}, {Salmas}, {Schroeder}, {Smith}, {Spekkens}, {Strader}, {Trilling}, {Vieira}, {Weiner}, \& {Williams}}]{Franz2025}
{Franz}, N., {Subrayan}, B., {Kilpatrick}, C.~D., {et~al.} 2025, \bibinfo{title}{{Optimizing Kilonova Searches: A Case Study of the Type IIb SN 2025ulz in the Localization Volume of the Low-Significance Gravitational Wave Event S250818k},} arXiv e-prints, arXiv:2510.17104.
\newblock \doarXiv{2510.17104}

\bibitem[{J. {Freeburn} {et~al.}(2025){Freeburn}, {O'Connor}, {Hall}, {Busmann}, {Andreoni}, {Palmese}, {Gruen}, {Hu}, {Cabrera}, {Kunnumkai}, \& {Amsellem}}]{2025GCN.41507....1F}
{Freeburn}, J., {O'Connor}, B., {Hall}, X.~J., {et~al.} 2025, \bibinfo{title}{{LIGO/Virgo/KAGRA S250818k: Rebrightening detected with Gemini/GMOS},} GRB Coordinates Network, 41507, 1

\bibitem[{D. Frostig {et~al.}(2025)Frostig, Karambelkar, Stein, Lourie, Kasliwal, Simcoe, Bulla, Ahumada, Mo, Purdum, Juneau, Malonis, \& Fűrész}]{frostig_winter_2025}
Frostig, D., Karambelkar, V.~R., Stein, R.~D., {et~al.} 2025, \bibinfo{title}{{WINTER} on {S250206dm}: {A} {Near}-infrared {Search} for an {Electromagnetic} {Counterpart} to a {Gravitational}-wave {Event},} Publications of the Astronomical Society of the Pacific, 137, 074203, \dodoi{10.1088/1538-3873/ade478}

\bibitem[{S. Fujibayashi {et~al.}(2020)Fujibayashi, Wanajo, Kiuchi, Kyutoku, Sekiguchi, \& Shibata}]{fujibayashi_postmerger_2020}
Fujibayashi, S., Wanajo, S., Kiuchi, K., {et~al.} 2020, \bibinfo{title}{Postmerger {Mass} {Ejection} of {Low}-mass {Binary} {Neutron} {Stars},} The Astrophysical Journal, 901, 122, \dodoi{10.3847/1538-4357/abafc2}

\bibitem[{ {Gaia Collaboration}(2020){Gaia Collaboration}}]{gaiaEDR3}
{Gaia Collaboration}. 2020, \bibinfo{title}{{VizieR Online Data Catalog: Gaia EDR3 (Gaia Collaboration, 2020)},} VizieR Online Data Catalog, I/350

\bibitem[{ {Gaia Collaboration} {et~al.}(2018){Gaia Collaboration}, {Brown}, {Vallenari}, {Prusti}, {de Bruijne}, {Babusiaux}, {Bailer-Jones}, {Biermann}, {Evans}, {Eyer}, {Jansen}, {Jordi}, {Klioner}, {Lammers}, {Lindegren}, {Luri}, {Mignard}, {Panem}, {Pourbaix}, {Randich}, {Sartoretti}, {Siddiqui}, {Soubiran}, {van Leeuwen}, {Walton}, {Arenou}, {Bastian}, {Cropper}, {Drimmel}, {Katz}, {Lattanzi}, {Bakker}, {Cacciari}, {Casta{\~n}eda}, {Chaoul}, {Cheek}, {De Angeli}, {Fabricius}, {Guerra}, {Holl}, {Masana}, {Messineo}, {Mowlavi}, {Nienartowicz}, {Panuzzo}, {Portell}, {Riello}, {Seabroke}, {Tanga}, {Th{\'e}venin}, {Gracia-Abril}, {Comoretto}, {Garcia-Reinaldos}, {Teyssier}, {Altmann}, {Andrae}, {Audard}, {Bellas-Velidis}, {Benson}, {Berthier}, {Blomme}, {Burgess}, {Busso}, {Carry}, {Cellino}, {Clementini}, {Clotet}, {Creevey}, {Davidson}, {De Ridder}, {Delchambre}, {Dell'Oro}, {Ducourant}, {Fern{\'a}ndez-Hern{\'a}ndez}, {Fouesneau}, {Fr{\'e}mat}, {Galluccio}, {Garc{\'\i}a-Torres},
  {Gonz{\'a}lez-N{\'u}{\~n}ez}, {Gonz{\'a}lez-Vidal}, {Gosset}, {Guy}, {Halbwachs}, {Hambly}, {Harrison}, {Hern{\'a}ndez}, {Hestroffer}, {Hodgkin}, {Hutton}, {Jasniewicz}, {Jean-Antoine-Piccolo}, {Jordan}, {Korn}, {Krone-Martins}, {Lanzafame}, {Lebzelter}, {L{\"o}ffler}, {Manteiga}, {Marrese}, {Mart{\'\i}n-Fleitas}, {Moitinho}, {Mora}, {Muinonen}, {Osinde}, {Pancino}, {Pauwels}, {Petit}, {Recio-Blanco}, {Richards}, {Rimoldini}, {Robin}, {Sarro}, {Siopis}, {Smith}, {Sozzetti}, {S{\"u}veges}, {Torra}, {van Reeven}, {Abbas}, {Abreu Aramburu}, {Accart}, {Aerts}, {Altavilla}, {{\'A}lvarez}, {Alvarez}, {Alves}, {Anderson}, {Andrei}, {Anglada Varela}, {Antiche}, {Antoja}, {Arcay}, {Astraatmadja}, {Bach}, {Baker}, {Balaguer-N{\'u}{\~n}ez}, {Balm}, {Barache}, {Barata}, {Barbato}, {Barblan}, {Barklem}, {Barrado}, {Barros}, {Barstow}, {Bartholom{\'e} Mu{\~n}oz}, {Bassilana}, {Becciani}, {Bellazzini}, {Berihuete}, {Bertone}, {Bianchi}, {Bienaym{\'e}}, {Blanco-Cuaresma}, {Boch}, {Boeche}, {Bombrun}, {Borrachero},
  {Bossini}, {Bouquillon}, {Bourda}, {Bragaglia}, {Bramante}, {Breddels}, {Bressan}, {Brouillet}, {Br{\"u}semeister}, {Brugaletta}, {Bucciarelli}, {Burlacu}, {Busonero}, {Butkevich}, {Buzzi}, {Caffau}, {Cancelliere}, {Cannizzaro}, {Cantat-Gaudin}, {Carballo}, {Carlucci}, {Carrasco}, {Casamiquela}, {Castellani}, {Castro-Ginard}, {Charlot}, {Chemin}, {Chiavassa}, {Cocozza}, {Costigan}, {Cowell}, {Crifo}, {Crosta}, {Crowley}, {Cuypers}, {Dafonte}, {Damerdji}, {Dapergolas}, {David}, {David}, {de Laverny}, \& {De Luise}}]{2018A&A...616A...1G}
{Gaia Collaboration}, {Brown}, A.~G.~A., {Vallenari}, A., {et~al.} 2018, \bibinfo{title}{{Gaia Data Release 2. Summary of the contents and survey properties},} \aap, 616, A1, \dodoi{10.1051/0004-6361/201833051}

\bibitem[{ {Gaia Collaboration} {et~al.}(2021){Gaia Collaboration}, {Brown}, {Vallenari}, {Prusti}, {de Bruijne}, {Babusiaux}, {Biermann}, {Creevey}, {Evans}, {Eyer}, {Hutton}, {Jansen}, {Jordi}, {Klioner}, {Lammers}, {Lindegren}, {Luri}, {Mignard}, {Panem}, {Pourbaix}, {Randich}, {Sartoretti}, {Soubiran}, {Walton}, {Arenou}, {Bailer-Jones}, {Bastian}, {Cropper}, {Drimmel}, {Katz}, {Lattanzi}, {van Leeuwen}, {Bakker}, {Cacciari}, {Casta{\~n}eda}, {De Angeli}, {Ducourant}, {Fabricius}, {Fouesneau}, {Fr{\'e}mat}, {Guerra}, {Guerrier}, {Guiraud}, {Jean-Antoine Piccolo}, {Masana}, {Messineo}, {Mowlavi}, {Nicolas}, {Nienartowicz}, {Pailler}, {Panuzzo}, {Riclet}, {Roux}, {Seabroke}, {Sordo}, {Tanga}, {Th{\'e}venin}, {Gracia-Abril}, {Portell}, {Teyssier}, {Altmann}, {Andrae}, {Bellas-Velidis}, {Benson}, {Berthier}, {Blomme}, {Brugaletta}, {Burgess}, {Busso}, {Carry}, {Cellino}, {Cheek}, {Clementini}, {Damerdji}, {Davidson}, {Delchambre}, {Dell'Oro}, {Fern{\'a}ndez-Hern{\'a}ndez}, {Galluccio}, {Garc{\'\i}a-Lario},
  {Garcia-Reinaldos}, {Gonz{\'a}lez-N{\'u}{\~n}ez}, {Gosset}, {Haigron}, {Halbwachs}, {Hambly}, {Harrison}, {Hatzidimitriou}, {Heiter}, {Hern{\'a}ndez}, {Hestroffer}, {Hodgkin}, {Holl}, {Jan{\ss}en}, {Jevardat de Fombelle}, {Jordan}, {Krone-Martins}, {Lanzafame}, {L{\"o}ffler}, {Lorca}, {Manteiga}, {Marchal}, {Marrese}, {Moitinho}, {Mora}, {Muinonen}, {Osborne}, {Pancino}, {Pauwels}, {Petit}, {Recio-Blanco}, {Richards}, {Riello}, {Rimoldini}, {Robin}, {Roegiers}, {Rybizki}, {Sarro}, {Siopis}, {Smith}, {Sozzetti}, {Ulla}, {Utrilla}, {van Leeuwen}, {van Reeven}, {Abbas}, {Abreu Aramburu}, {Accart}, {Aerts}, {Aguado}, {Ajaj}, {Altavilla}, {{\'A}lvarez}, {{\'A}lvarez Cid-Fuentes}, {Alves}, {Anderson}, {Anglada Varela}, {Antoja}, {Audard}, {Baines}, {Baker}, {Balaguer-N{\'u}{\~n}ez}, {Balbinot}, {Balog}, {Barache}, {Barbato}, {Barros}, {Barstow}, {Bartolom{\'e}}, {Bassilana}, {Bauchet}, {Baudesson-Stella}, {Becciani}, {Bellazzini}, {Bernet}, {Bertone}, {Bianchi}, {Blanco-Cuaresma}, {Boch}, {Bombrun}, {Bossini},
  {Bouquillon}, {Bragaglia}, {Bramante}, {Breedt}, {Bressan}, {Brouillet}, {Bucciarelli}, {Burlacu}, {Busonero}, {Butkevich}, {Buzzi}, {Caffau}, {Cancelliere}, {C{\'a}novas}, {Cantat-Gaudin}, {Carballo}, {Carlucci}, {Carnerero}, {Carrasco}, {Casamiquela}, {Castellani}, {Castro-Ginard}, {Castro Sampol}, {Chaoul}, {Charlot}, {Chemin}, {Chiavassa}, {Cioni}, {Comoretto}, {Cooper}, {Cornez}, {Cowell}, {Crifo}, {Crosta}, {Crowley}, {Dafonte}, {Dapergolas}, {David}, {David}, {de Laverny}, {De Luise}, {De March}, {De Ridder}, {de Souza}, {de Teodoro}, {de Torres}, {del Peloso}, {del Pozo}, {Delbo}, {Delgado}, {Delgado}, {Delisle}, {Di Matteo}, {Diakite}, {Diener}, {Distefano}, {Dolding}, {Eappachen}, {Edvardsson}, {Enke}, {Esquej}, {Fabre}, {Fabrizio}, {Faigler}, {Fedorets}, {Fernique}, {Fienga}, {Figueras}, {Fouron}, {Fragkoudi}, {Fraile}, {Franke}, {Gai}, {Garabato}, {Garcia-Gutierrez}, {Garc{\'\i}a-Torres}, {Garofalo}, {Gavras}, {Gerlach}, {Geyer}, {Giacobbe}, {Gilmore}, {Girona}, {Giuffrida}, {Gomel}, {Gomez},
  {Gonzalez-Santamaria}, {Gonz{\'a}lez-Vidal}, {Granvik}, {Guti{\'e}rrez-S{\'a}nchez}, {Guy}, {Hauser}, {Haywood}, {Helmi}, {Hidalgo}, {Hilger}, {H{\l}adczuk}, {Hobbs}, {Holland}, {Huckle}, {Jasniewicz}, {Jonker}, {Juaristi Campillo}, {Julbe}, {Karbevska}, {Kervella}, {Khanna}, {Kochoska}, {Kontizas}, {Kordopatis}, {Korn}, {Kostrzewa-Rutkowska}, {Kruszy{\'n}ska}, {Lambert}, {Lanza}, {Lasne}, {Le Campion}, {Le Fustec}, {Lebreton}, {Lebzelter}, {Leccia}, {Leclerc}, {Lecoeur-Taibi}, {Liao}, {Licata}, {Lindstr{\o}m}, {Lister}, {Livanou}, {Lobel}, {Madrero Pardo}, {Managau}, {Mann}, {Marchant}, {Marconi}, {Marcos Santos}, {Marinoni}, {Marocco}, {Marshall}, {Martin Polo}, {Mart{\'\i}n-Fleitas}, {Masip}, {Massari}, {Mastrobuono-Battisti}, {Mazeh}, {McMillan}, {Messina}, {Michalik}, {Millar}, {Mints}, {Molina}, {Molinaro}, {Moln{\'a}r}, {Montegriffo}, {Mor}, {Morbidelli}, {Morel}, {Morris}, {Mulone}, {Munoz}, {Muraveva}, {Murphy}, {Musella}, {Noval}, {Ord{\'e}novic}, {Orr{\`u}}, {Osinde}, {Pagani}, {Pagano},
  {Palaversa}, {Palicio}, {Panahi}, {Pawlak}, {Pe{\~n}alosa Esteller}, {Penttil{\"a}}, {Piersimoni}, {Pineau}, {Plachy}, {Plum}, {Poggio}, {Poretti}, {Poujoulet}, {Pr{\v{s}}a}, {Pulone}, {Racero}, {Ragaini}, {Rainer}, {Raiteri}, {Rambaux}, {Ramos}, {Ramos-Lerate}, {Re Fiorentin}, {Regibo}, {Reyl{\'e}}, {Ripepi}, {Riva}, {Rixon}, {Robichon}, {Robin}, {Roelens}, {Rohrbasser}, {Romero-G{\'o}mez}, {Rowell}, {Royer}, {Rybicki}, {Sadowski}, {Sagrist{\`a} Sell{\'e}s}, {Sahlmann}, {Salgado}, {Salguero}, {Samaras}, {Sanchez Gimenez}, {Sanna}, {Santove{\~n}a}, {Sarasso}, {Schultheis}, {Sciacca}, {Segol}, {Segovia}, {S{\'e}gransan}, {Semeux}, {Shahaf}, {Siddiqui}, {Siebert}, {Siltala}, {Slezak}, {Smart}, {Solano}, {Solitro}, {Souami}, {Souchay}, {Spagna}, {Spoto}, {Steele}, {Steidelm{\"u}ller}, {Stephenson}, {S{\"u}veges}, {Szabados}, {Szegedi-Elek}, {Taris}, {Tauran}, {Taylor}, {Teixeira}, {Thuillot}, {Tonello}, {Torra}, {Torra}, {Turon}, {Unger}, {Vaillant}, {van Dillen}, {Vanel}, {Vecchiato}, {Viala}, {Vicente},
  {Voutsinas}, {Weiler}, {Wevers}, {Wyrzykowski}, {Yoldas}, {Yvard}, {Zhao}, {Zorec}, {Zucker}, {Zurbach}, \& {Zwitter}}]{Gaia2021}
{Gaia Collaboration}, {Brown}, A.~G.~A., {Vallenari}, A., {et~al.} 2021, \bibinfo{title}{{Gaia Early Data Release 3. Summary of the contents and survey properties},} \aap, 649, A1, \dodoi{10.1051/0004-6361/202039657}

\bibitem[{J.~H. {Gillanders} {et~al.}(2023){Gillanders}, {Troja}, {Fryer}, {Ristic}, {O'Connor}, {Fontes}, {Yang}, {Domoto}, {Rahmouni}, {Tanaka}, {Fox}, \& {Dichiara}}]{Gillanders2023}
{Gillanders}, J.~H., {Troja}, E., {Fryer}, C.~L., {et~al.} 2023, \bibinfo{title}{{Heavy element nucleosynthesis associated with a gamma-ray burst},} arXiv e-prints, arXiv:2308.00633, \dodoi{10.48550/arXiv.2308.00633}

\bibitem[{J.~H. {Gillanders} {et~al.}(2025{\natexlab{a}}){Gillanders}, {Huber}, {Chambers}, {Smartt}, {Smith}, {Srivastav}, {Stoppa}, {Stevance}, {Tweddle}, {Nicholl}, {Young}, {Aamer}, {Angus}, {Fulton}, {Magill}, {McCollum}, {Moore}, {Sim}, {Weston}, {Sheng}, {Chen}, {Shingles}, {Ramsden}, {Schultz}, {de Boer}, {Fairlamb}, {Lin}, {Lowe}, {Magnier}, {Minguez}, {Paek}, {Smith}, {Wainscoat}, {Rest}, \& {Stubbs}}]{2025GCN.41540....1G}
{Gillanders}, J.~H., {Huber}, M.~E., {Chambers}, K.~C., {et~al.} 2025{\natexlab{a}}, \bibinfo{title}{{LIGO/Virgo/KAGRA S250818k: Pan-STARRS imaging confirms re-brightening of SN2025ulz},} GRB Coordinates Network, 41540, 1

\bibitem[{J.~H. {Gillanders} {et~al.}(2025{\natexlab{b}}){Gillanders}, {Huber}, {Nicholl}, {Smartt}, {Smith}, {Chambers}, {Young}, {Tweddle}, {Srivastav}, {Fulton}, {Stoppa}, {Paek}, {Aamer}, {Alarcon}, {Andersson}, {Aryan}, {Auchettl}, {Chen}, {de Boer}, {Kong}, {Licandro}, {Lowe}, {Magill}, {Magnier}, {Minguez}, {Moore}, {Pignata}, {Rest}, {Serra-Ricart}, {Shappee}, {Smith}, {Tucker}, \& {Wainscoat}}]{Gillanders25ulz}
{Gillanders}, J.~H., {Huber}, M.~E., {Nicholl}, M., {et~al.} 2025{\natexlab{b}}, \bibinfo{title}{{Pan-STARRS follow-up of the gravitational-wave event S250818k and the lightcurve of SN 2025ulz},} arXiv e-prints, arXiv:2510.01142.
\newblock \doarXiv{2510.01142}

\bibitem[{A. {Goldstein} {et~al.}(2017){Goldstein}, {Veres}, {Burns}, {Briggs}, {Hamburg}, {Kocevski}, {Wilson-Hodge}, {Preece}, {Poolakkil}, {Roberts}, {Hui}, {Connaughton}, {Racusin}, {von Kienlin}, {Dal Canton}, {Christensen}, {Littenberg}, {Siellez}, {Blackburn}, {Broida}, {Bissaldi}, {Cleveland}, {Gibby}, {Giles}, {Kippen}, {McBreen}, {McEnery}, {Meegan}, {Paciesas}, \& {Stanbro}}]{Goldstein2017}
{Goldstein}, A., {Veres}, P., {Burns}, E., {et~al.} 2017, \bibinfo{title}{{An Ordinary Short Gamma-Ray Burst with Extraordinary Implications: Fermi-GBM Detection of GRB 170817A},} \apjl, 848, L14, \dodoi{10.3847/2041-8213/aa8f41}

\bibitem[{C.~A. {G{\"o}ssl} \& A. {Riffeser}(2002){G{\"o}ssl} \& {Riffeser}}]{2002A&A...381.1095G}
{G{\"o}ssl}, C.~A., \& {Riffeser}, A. 2002, \bibinfo{title}{{Image reduction pipeline for the detection of variable sources in highly crowded fields},} \aap, 381, 1095, \dodoi{10.1051/0004-6361:20011522}

\bibitem[{M.~J. {Graham} {et~al.}(2020){Graham}, {Ford}, {McKernan}, {Ross}, {Stern}, {Burdge}, {Coughlin}, {Djorgovski}, {Drake}, {Duev}, {Kasliwal}, {Mahabal}, {van Velzen}, {Belecki}, {Bellm}, {Burruss}, {Cenko}, {Cunningham}, {Helou}, {Kulkarni}, {Masci}, {Prince}, {Reiley}, {Rodriguez}, {Rusholme}, {Smith}, \& {Soumagnac}}]{Graham2020}
{Graham}, M.~J., {Ford}, K.~E.~S., {McKernan}, B., {et~al.} 2020, \bibinfo{title}{{Candidate Electromagnetic Counterpart to the Binary Black Hole Merger Gravitational-Wave Event S190521g$^{*}$},} \prl, 124, 251102, \dodoi{10.1103/PhysRevLett.124.251102}

\bibitem[{M.~J. {Graham} {et~al.}(2023){Graham}, {McKernan}, {Ford}, {Stern}, {Djorgovski}, {Coughlin}, {Burdge}, {Bellm}, {Helou}, {Mahabal}, {Masci}, {Purdum}, {Rosnet}, \& {Rusholme}}]{Graham2023}
{Graham}, M.~J., {McKernan}, B., {Ford}, K.~E.~S., {et~al.} 2023, \bibinfo{title}{{A Light in the Dark: Searching for Electromagnetic Counterparts to Black Hole-Black Hole Mergers in LIGO/Virgo O3 with the Zwicky Transient Facility},} \apj, 942, 99, \dodoi{10.3847/1538-4357/aca480}

\bibitem[{X.~J. Hall {et~al.}(2025)Hall, Busmann, Gruen, O'Connor, \& Palmese}]{hall_ligovirgokagra_2025}
Hall, X.~J., Busmann, M., Gruen, D., O'Connor, B., \& Palmese, A. 2025, \bibinfo{title}{{LIGO}/{Virgo}/{KAGRA} {S250818k}: {FTW} {Fast} reddening of {AT} 2025ulz,} GRB Coordinates Network, 41433, 1.
\newblock \url{https://ui.adsabs.harvard.edu/abs/2025GCN.41433....1H}

\bibitem[{X.~J. {Hall} {et~al.}(2025{\natexlab{a}}){Hall}, {Stein}, {O'Connor}, \& {Palmese}}]{2025GCN.41453....1H}
{Hall}, X.~J., {Stein}, R., {O'Connor}, B., \& {Palmese}, A. 2025{\natexlab{a}}, \bibinfo{title}{{LIGO/Virgo/KAGRA S250818k: Swift observations of AT 2025ulz},} GRB Coordinates Network, 41453, 1

\bibitem[{X.~J. {Hall} {et~al.}(2025{\natexlab{b}}){Hall}, {Palmese}, {O'Connor}, {Gruen}, {Busmann}, {xxx}, {xxx}, {xxx}, \& {xxx}}]{Hall2025desi}
{Hall}, X.~J., {Palmese}, A., {O'Connor}, B., {et~al.} 2025{\natexlab{b}}, \bibinfo{title}{in prep. AT2025ulz,} in prep.

\bibitem[{G. Hallinan {et~al.}(2017)Hallinan, Corsi, Mooley, Hotokezaka, Nakar, Kasliwal, Kaplan, Frail, Myers, Murphy, De, Dobie, Allison, Bannister, Bhalerao, Chandra, Clarke, Giacintucci, Ho, Horesh, Kassim, Kulkarni, Lenc, Lockman, Lynch, Nichols, Nissanke, Palliyaguru, Peters, Piran, Rana, Sadler, \& Singer}]{hallinan_radio_2017}
Hallinan, G., Corsi, A., Mooley, K.~P., {et~al.} 2017, \bibinfo{title}{A radio counterpart to a neutron star merger,} Science, 358, 1579, \dodoi{10.1126/science.aap9855}

\bibitem[{K. Hayashi {et~al.}(2025)Hayashi, Kiuchi, Kyutoku, Sekiguchi, \& Shibata}]{Hayashi:2024jwt}
Hayashi, K., Kiuchi, K., Kyutoku, K., Sekiguchi, Y., \& Shibata, M. 2025, \bibinfo{title}{{Jet from Binary Neutron Star Merger with Prompt Black Hole Formation},} Phys. Rev. Lett., 134, 211407, \dodoi{10.1103/PhysRevLett.134.211407}

\bibitem[{G.~J. {Hill} {et~al.}(2021){Hill}, {Lee}, {MacQueen}, {Kelz}, {Drory}, {Vattiat}, {Good}, {Ramsey}, {Kriel}, {Peterson}, {DePoy}, {Gebhardt}, {Marshall}, {Tuttle}, {Bauer}, {Chonis}, {Fabricius}, {Froning}, {H{\"a}user}, {Indahl}, {Jahn}, {Landriau}, {Leck}, {Montesano}, {Prochaska}, {Snigula}, {Zeimann}, {Bryant}, {Damm}, {Fowler}, {Janowiecki}, {Martin}, {Mrozinski}, {Odewahn}, {Rostopchin}, {Shetrone}, {Spencer}, {Mentuch Cooper}, {Armandroff}, {Bender}, {Dalton}, {Hopp}, {Komatsu}, {Nicklas}, {Ramsey}, {Roth}, {Schneider}, {Sneden}, \& {Steinmetz}}]{2021AJ....162..298H}
{Hill}, G.~J., {Lee}, H., {MacQueen}, P.~J., {et~al.} 2021, \bibinfo{title}{{The HETDEX Instrumentation: Hobby-Eberly Telescope Wide-field Upgrade and VIRUS},} \aj, 162, 298, \dodoi{10.3847/1538-3881/ac2c02}

\bibitem[{A.~Y.~Q. {Ho} {et~al.}(2023){Ho}, {Perley}, {Gal-Yam}, {Lunnan}, {Sollerman}, {Schulze}, {Das}, {Dobie}, {Yao}, {Fremling}, {Adams}, {Anand}, {Andreoni}, {Bellm}, {Bruch}, {Burdge}, {Castro-Tirado}, {Dahiwale}, {De}, {Dekany}, {Drake}, {Duev}, {Graham}, {Helou}, {Kaplan}, {Karambelkar}, {Kasliwal}, {Kool}, {Kulkarni}, {Mahabal}, {Medford}, {Miller}, {Nordin}, {Ofek}, {Petitpas}, {Riddle}, {Sharma}, {Smith}, {Stewart}, {Taggart}, {Tartaglia}, {Tzanidakis}, \& {Winters}}]{HoFBOTs}
{Ho}, A. Y.~Q., {Perley}, D.~A., {Gal-Yam}, A., {et~al.} 2023, \bibinfo{title}{{A Search for Extragalactic Fast Blue Optical Transients in ZTF and the Rate of AT2018cow-like Transients},} \apj, 949, 120, \dodoi{10.3847/1538-4357/acc533}

\bibitem[{U. {Hopp} {et~al.}(2014){Hopp}, {Bender}, {Grupp}, {Goessl}, {Lang-Bardl}, {Mitsch}, {Riffeser}, \& {Ageorges}}]{2014SPIE.9145E..2DH}
{Hopp}, U., {Bender}, R., {Grupp}, F., {et~al.} 2014, in Society of Photo-Optical Instrumentation Engineers (SPIE) Conference Series, Vol. 9145, Ground-based and Airborne Telescopes V, ed. L.~M. {Stepp}, R.~{Gilmozzi}, \& H.~J. {Hall}, 91452D, \dodoi{10.1117/12.2054498}

\bibitem[{K. Hotokezaka {et~al.}(2013)Hotokezaka, Kiuchi, Kyutoku, Okawa, Sekiguchi, Shibata, \& Taniguchi}]{Hotokezaka:2012ze}
Hotokezaka, K., Kiuchi, K., Kyutoku, K., {et~al.} 2013, \bibinfo{title}{{Mass ejection from the merger of binary neutron stars},} Phys. Rev. D, 87, 024001, \dodoi{10.1103/PhysRevD.87.024001}

\bibitem[{L. Hu {et~al.}(2019)Hu, Wang, Chen, \& Yang}]{hu_image_2022}
Hu, L., Wang, L., Chen, X., \& Yang, J. 2019, \bibinfo{title}{Image Subtraction in Fourier Space,} 936, 157, \dodoi{10.3847/1538-4357/ac7394}

\bibitem[{L. {Hu} {et~al.}(2022){Hu}, {Wang}, {Chen}, \& {Yang}}]{Hu2022}
{Hu}, L., {Wang}, L., {Chen}, X., \& {Yang}, J. 2022, \bibinfo{title}{{Image Subtraction in Fourier Space},} \apj, 936, 157, \dodoi{10.3847/1538-4357/ac7394}

\bibitem[{L. {Hu} {et~al.}(2025){Hu}, {Cabrera}, {Palmese}, {Freeburn}, {Bulla}, {Andreoni}, {Hall}, {O'Connor}, {Amsellem}, {Bom}, {Busmann}, {Fab{\`a}}, {Gassert}, {Kalabalik}, {Kunnumkai}, {Gruen}, {Santana-Silva}, {Santos}, {Ahumada}, {Carney}, {Coughlin}, {Chen}, {Ford}, {Holz}, {Kasliwal}, {Maga{\~n}a Hernandez}, {Mihalenko}, {Perna}, {Riffeser}, {Ries}, {Schnappinger}, {Schmidt}, {Sommer}, {Teague}, {Vega}, {Volchansky}, {Wang}, \& {Zhang}}]{Hu2025}
{Hu}, L., {Cabrera}, T., {Palmese}, A., {et~al.} 2025, \bibinfo{title}{{Kilonova Constraints for the LIGO/Virgo/KAGRA Neutron Star Merger Candidate S250206dm: GW-MMADS Observations},} \apjl, 990, L46, \dodoi{10.3847/2041-8213/adfd49}

\bibitem[{S. Jhawar {et~al.}(2025)Jhawar, Wouters, Pang, Bulla, Coughlin, \& Dietrich}]{Jhawar:2024ezm}
Jhawar, S., Wouters, T., Pang, P. T.~H., {et~al.} 2025, \bibinfo{title}{{Data-driven approach for modeling the temporal and spectral evolution of kilonova systematic uncertainties},} Phys. Rev. D, 111, 043046, \dodoi{10.1103/PhysRevD.111.043046}

\bibitem[{N. {Kaiser} {et~al.}(2010){Kaiser}, {Burgett}, {Chambers}, {Denneau}, {Heasley}, {Jedicke}, {Magnier}, {Morgan}, {Onaka}, \& {Tonry}}]{2010SPIE.7733E..0EK}
{Kaiser}, N., {Burgett}, W., {Chambers}, K., {et~al.} 2010, in Society of Photo-Optical Instrumentation Engineers (SPIE) Conference Series, Vol. 7733, Ground-based and Airborne Telescopes III, ed. L.~M. {Stepp}, R.~{Gilmozzi}, \& H.~J. {Hall}, 77330E, \dodoi{10.1117/12.859188}

\bibitem[{M. {Kasliwal} {et~al.}(2025){Kasliwal}, {Ahumada}, {Stein}, {Karambelkar}, {Hall}, {Singh}, {xxx}, {xxx}, {xxx}, \& {xxx}}]{Kasliwal2025sn}
{Kasliwal}, M., {Ahumada}, T., {Stein}, R., {et~al.} 2025, \bibinfo{title}{submitted. AT2025ulz,} submitted.

\bibitem[{M.~M. {Kasliwal} {et~al.}(2017){Kasliwal}, {Nakar}, {Singer}, {Kaplan}, {Cook}, {Van Sistine}, {Lau}, {Fremling}, {Gottlieb}, {Jencson}, {Adams}, {Feindt}, {Hotokezaka}, {Ghosh}, {Perley}, {Yu}, {Piran}, {Allison}, {Anupama}, {Balasubramanian}, {Bannister}, {Bally}, {Barnes}, {Barway}, {Bellm}, {Bhalerao}, {Bhattacharya}, {Blagorodnova}, {Bloom}, {Brady}, {Cannella}, {Chatterjee}, {Cenko}, {Cobb}, {Copperwheat}, {Corsi}, {De}, {Dobie}, {Emery}, {Evans}, {Fox}, {Frail}, {Frohmaier}, {Goobar}, {Hallinan}, {Harrison}, {Helou}, {Hinderer}, {Ho}, {Horesh}, {Ip}, {Itoh}, {Kasen}, {Kim}, {Kuin}, {Kupfer}, {Lynch}, {Madsen}, {Mazzali}, {Miller}, {Mooley}, {Murphy}, {Ngeow}, {Nichols}, {Nissanke}, {Nugent}, {Ofek}, {Qi}, {Quimby}, {Rosswog}, {Rusu}, {Sadler}, {Schmidt}, {Sollerman}, {Steele}, {Williamson}, {Xu}, {Yan}, {Yatsu}, {Zhang}, \& {Zhao}}]{Kasliwal2017}
{Kasliwal}, M.~M., {Nakar}, E., {Singer}, L.~P., {et~al.} 2017, \bibinfo{title}{{Illuminating gravitational waves: A concordant picture of photons from a neutron star merger},} Science, 358, 1559, \dodoi{10.1126/science.aap9455}

\bibitem[{M.~M. Kasliwal {et~al.}(2025{\natexlab{a}})Kasliwal, Earley, Smith, Guillot, Travouillon, Fucik, Abe, Greffe, Agabi, Ashley, Triaud, Tinyanont, Antier, Bendjoya, Bhattarai, Bertz, Brugger, Burdanov, Caiazzo, Carry, Casagrande, Cenko, Cooke, De, Dekany, Deloupy, Dornic, Fahey, Figer, Freeman, Frostig, Graham, Günther, Hale, Bland-Hawthorn, Illuminati, Jencson, Karambelkar, Key, Lau, Li, Lubin, Neill, Pahuja, Pian, de~Ugarte~Postigo, Roberts, Rodriguez, Rose, Ruiter, Schmider, Simcoe, Stein, Suarez, Taylor, Weber, Wen, de~Wit, Zarzaca, \& Zimmer}]{kasliwal_cryoscope_2025}
Kasliwal, M.~M., Earley, N., Smith, R., {et~al.} 2025{\natexlab{a}}, \bibinfo{title}{Cryoscope: {A} {Cryogenic} {Infrared} {Survey} {Telescope} in {Antarctica},} Publications of the Astronomical Society of the Pacific, 137, 065001, \dodoi{10.1088/1538-3873/adc629}

\bibitem[{M.~M. Kasliwal {et~al.}(2025{\natexlab{b}})Kasliwal, Karambelkar, Fremling, Ahumada, Hall, Perley, Anand, Liu, Das, Bhalerao, Swain, Saikia, {Ztf Collaboration}, \& {Growth Collaboration}}]{kasliwal_ligovirgokagra_2025}
Kasliwal, M.~M., Karambelkar, V., Fremling, C., {et~al.} 2025{\natexlab{b}}, \bibinfo{title}{{LIGO}/{Virgo}/{KAGRA} {S250818k}: {Continued} {Keck} {I} {LRIS} spectroscopy of {ZTF25abjmnps} ({AT2025ulz}),} GRB Coordinates Network, 41538, 1.
\newblock \url{https://ui.adsabs.harvard.edu/abs/2025GCN.41538....1K}

\bibitem[{B.~C. Kelly {et~al.}(2009)Kelly, Bechtold, \& Siemiginowska}]{kelly_are_2009}
Kelly, B.~C., Bechtold, J., \& Siemiginowska, A. 2009, \bibinfo{title}{{ARE} {THE} {VARIATIONS} {IN} {QUASAR} {OPTICAL} {FLUX} {DRIVEN} {BY} {THERMAL} {FLUCTUATIONS}?} The Astrophysical Journal, 698, 895, \dodoi{10.1088/0004-637X/698/1/895}

\bibitem[{H. Koehn {et~al.}(2025)Koehn, Wouters, Pang, Bulla, Rose, Wichern, \& Dietrich}]{Koehn:2025zzb}
Koehn, H., Wouters, T., Pang, P. T.~H., {et~al.} 2025, \bibinfo{title}{{Efficient Bayesian analysis of kilonovae and GRB afterglows with fiesta},} \doarXiv{2507.13807}

\bibitem[{C.~J. Kr{\"u}ger \& F. Foucart(2020)Kr{\"u}ger \& Foucart}]{Kruger:2020gig}
Kr{\"u}ger, C.~J., \& Foucart, F. 2020, \bibinfo{title}{{Estimates for Disk and Ejecta Masses Produced in Compact Binary Mergers},} Phys. Rev. D, 101, 103002, \dodoi{10.1103/PhysRevD.101.103002}

\bibitem[{I. Kullmann {et~al.}(2022)Kullmann, Goriely, Just, Ardevol-Pulpillo, Bauswein, \& Janka}]{Kullmann:2021gvo}
Kullmann, I., Goriely, S., Just, O., {et~al.} 2022, \bibinfo{title}{{Dynamical ejecta of neutron star mergers with nucleonic weak processes I: nucleosynthesis},} Mon. Not. Roy. Astron. Soc., 510, 2804, \dodoi{10.1093/mnras/stab3393}

\bibitem[{K. {Kunnumkai} {et~al.}(2024{\natexlab{a}}){Kunnumkai}, {Palmese}, {Bulla}, {Dietrich}, {Farah}, \& {Pang}}]{kunnumkai_GW230529}
{Kunnumkai}, K., {Palmese}, A., {Bulla}, M., {et~al.} 2024{\natexlab{a}}, \bibinfo{title}{{Kilonova emission from GW230529 and mass gap neutron star-black hole mergers},} arXiv e-prints, arXiv:2409.10651, \dodoi{10.48550/arXiv.2409.10651}

\bibitem[{K. {Kunnumkai} {et~al.}(2024{\natexlab{b}}){Kunnumkai}, {Palmese}, {Farah}, {Bulla}, {Dietrich}, {Pang}, {Anand}, {Andreoni}, {Cabrera}, \& {Connor}}]{kunnumkai_O5}
{Kunnumkai}, K., {Palmese}, A., {Farah}, A.~M., {et~al.} 2024{\natexlab{b}}, \bibinfo{title}{{Detecting electromagnetic counterparts to LIGO/Virgo/KAGRA gravitational wave events with DECam: Neutron Star Mergers},} arXiv e-prints, arXiv:2411.13673, \dodoi{10.48550/arXiv.2411.13673}

\bibitem[{K. {Labrie} {et~al.}(2019{\natexlab{a}}){Labrie}, {Anderson}, {C{\'a}rdenes}, {Simpson}, \& {Turner}}]{2019ASPC..523..321L}
{Labrie}, K., {Anderson}, K., {C{\'a}rdenes}, R., {Simpson}, C., \& {Turner}, J. E.~H. 2019{\natexlab{a}}, in Astronomical Society of the Pacific Conference Series, Vol. 523, Astronomical Data Analysis Software and Systems XXVII, ed. P.~J. {Teuben}, M.~W. {Pound}, B.~A. {Thomas}, \& E.~M. {Warner}, 321

\bibitem[{K. {Labrie} {et~al.}(2019{\natexlab{b}}){Labrie}, {Anderson}, {C{\'a}rdenes}, {Simpson}, \& {Turner}}]{Labrie2019}
{Labrie}, K., {Anderson}, K., {C{\'a}rdenes}, R., {Simpson}, C., \& {Turner}, J. E.~H. 2019{\natexlab{b}}, in Astronomical Society of the Pacific Conference Series, Vol. 523, Astronomical Data Analysis Software and Systems XXVII, ed. P.~J. {Teuben}, M.~W. {Pound}, B.~A. {Thomas}, \& E.~M. {Warner}, 321

\bibitem[{K. {Labrie} {et~al.}(2023){Labrie}, {Simpson}, {Cardenes}, {Turner}, {Soraisam}, {Quint}, {Oberdorf}, {Placco}, {Berke}, {Smirnova}, {Conseil}, {Vacca}, \& {Thomas-Osip}}]{Labrie2023}
{Labrie}, K., {Simpson}, C., {Cardenes}, R., {et~al.} 2023, \bibinfo{title}{{DRAGONS-A Quick Overview},} Research Notes of the American Astronomical Society, 7, 214, \dodoi{10.3847/2515-5172/ad0044}

\bibitem[{F. Lang-Bardl {et~al.}(2016)Lang-Bardl, Bender, Goessl, Grupp, Hess, Kaminski, Hodapp, Hopp, Jacobson, Kravcar, {et~al.}}]{lang2016wendelstein}
Lang-Bardl, F., Bender, R., Goessl, C., {et~al.} 2016, in Ground-based and Airborne Instrumentation for Astronomy VI, Vol. 9908, SPIE, 1295--1302

\bibitem[{Y. Lerner {et~al.}(2025)Lerner, Stone, \& Ofengeim}]{lerner_fragmentation_2025}
Lerner, Y., Stone, N.~C., \& Ofengeim, D.~D. 2025, \bibinfo{title}{Fragmentation in {Collapsar} {Disks}: {Migration}, {Growth}, and {Emission},} arXiv, \dodoi{10.48550/arXiv.2505.21617}

\bibitem[{A. {Levan} {et~al.}(2023){Levan}, {Gompertz}, {Salafia}, {Bulla}, {Burns}, {Hotokezaka}, {Izzo}, {Lamb}, {Malesani}, {Oates}, {Ravasio}, {Rouco Escorial}, {Schneider}, {Sarin}, {Schulze}, {Tanvir}, {Ackley}, {Anderson}, {Brammer}, {Christensen}, {Dhillon}, {Evans}, {Fausnaugh}, {Fong}, {Fruchter}, {Fryer}, {Fynbo}, {Gaspari}, {Heintz}, {Hjorth}, {Kennea}, {Kennedy}, {Laskar}, {Leloudas}, {Mandel}, {Martin-Carrillo}, {Metzger}, {Nicholl}, {Nugent}, {Palmerio}, {Pugliese}, {Rastinejad}, {Rhodes}, {Rossi}, {Smartt}, {Stevance}, {Tohuvavohu}, {van der Horst}, {Vergani}, {Watson}, {Barclay}, {Bhirombhakdi}, {Breedt}, {Breeveld}, {Brown}, {Campana}, {Chrimes}, {D'Avanzo}, {D'Elia}, {De Pasquale}, {Dyer}, {Galloway}, {Garbutt}, {Green}, {Hartmann}, {Jakobsson}, {Kerry}, {Langeroodi}, {Leung}, {Littlefair}, {Munday}, {O'Brien}, {Parsons}, {Pelisoli}, {Saccardi}, {Sahman}, {Salvaterra}, {Sbarufatti}, {Steeghs}, {Tagliaferri}, {Th{\"o}ne}, {de Ugarte Postigo}, \& {Kann}}]{Levan2023}
{Levan}, A., {Gompertz}, B.~P., {Salafia}, O.~S., {et~al.} 2023, \bibinfo{title}{{JWST detection of heavy neutron capture elements in a compact object merger},} arXiv e-prints, arXiv:2307.02098, \dodoi{10.48550/arXiv.2307.02098}

\bibitem[{R.~Z. {Li} {et~al.}(2025){Li}, {Xu}, {Sun}, {Li}, {Liu}, {Yuan}, {Zhang}, \& {Einstein Probe Team}}]{2025GCN.41460....1L}
{Li}, R.~Z., {Xu}, X.~P., {Sun}, H., {et~al.} 2025, \bibinfo{title}{{LIGO/Virgo/KAGRA S250818k: EP/FXT observation of AT 2025ulz},} GRB Coordinates Network, 41460, 1

\bibitem[{ {Ligo Scientific Collaboration} {et~al.}(2025){Ligo Scientific Collaboration}, {VIRGO Collaboration}, \& {Kagra Collaboration}}]{ligo_scientific_collaboration_ligovirgokagra_2025}
{Ligo Scientific Collaboration}, {VIRGO Collaboration}, \& {Kagra Collaboration}. 2025, \bibinfo{title}{{LIGO}/{Virgo}/{KAGRA} {S250818k}: {Properties} of the low-significance {GW} compact binary merger candidate potentially associated with {AT} 2025ulz,} GRB Coordinates Network, 41437, 1.
\newblock \url{https://ui.adsabs.harvard.edu/abs/2025GCN.41437....1L}

\bibitem[{ {LIGO Scientific Collaboration and Virgo Collaboration} {et~al.}(2016){LIGO Scientific Collaboration and Virgo Collaboration}, Abbott, Abbott, Abbott, Abernathy, Acernese, Ackley, Adams, Adams, Addesso, Adhikari, Adya, Affeldt, Agathos, Agatsuma, Aggarwal, Aguiar, Aiello, Ain, Ajith, Allen, Allocca, Altin, Anderson, Anderson, Arai, Arain, Araya, Arceneaux, Areeda, Arnaud, Arun, Ascenzi, Ashton, Ast, Aston, Astone, Aufmuth, Aulbert, Babak, Bacon, Bader, Baker, Baldaccini, Ballardin, Ballmer, Barayoga, Barclay, Barish, Barker, Barone, Barr, Barsotti, Barsuglia, Barta, Bartlett, Barton, Bartos, Bassiri, Basti, Batch, Baune, Bavigadda, Bazzan, Behnke, Bejger, Belczynski, Bell, Bell, Berger, Bergman, Bergmann, Berry, Bersanetti, Bertolini, Betzwieser, Bhagwat, Bhandare, Bilenko, Billingsley, Birch, Birney, Birnholtz, Biscans, Bisht, Bitossi, Biwer, Bizouard, Blackburn, Blair, Blair, Blair, Bloemen, Bock, Bodiya, Boer, Bogaert, Bogan, Bohe, Bojtos, Bond, Bondu, Bonnand, Boom, Bork, Boschi, Bose,
  Bouffanais, Bozzi, Bradaschia, Brady, Braginsky, Branchesi, Brau, Briant, Brillet, Brinkmann, Brisson, Brockill, Brooks, Brown, Brown, Brown, Buchanan, Buikema, Bulik, Bulten, Buonanno, Buskulic, Buy, Byer, Cabero, Cadonati, Cagnoli, Cahillane, Bustillo, Callister, Calloni, Camp, Cannon, Cao, Capano, Capocasa, Carbognani, Caride, Diaz, Casentini, Caudill, Cavaglià, Cavalier, Cavalieri, Cella, Cepeda, Baiardi, Cerretani, Cesarini, Chakraborty, Chalermsongsak, Chamberlin, Chan, Chao, Charlton, Chassande-Mottin, Chen, Chen, Cheng, Chincarini, Chiummo, Cho, Cho, Chow, Christensen, Chu, Chua, Chung, Ciani, Clara, Clark, Cleva, Coccia, Cohadon, Colla, Collette, Cominsky, Constancio, Conte, Conti, Cook, Corbitt, Cornish, Corsi, Cortese, Costa, Coughlin, Coughlin, Coulon, Countryman, Couvares, Cowan, Coward, Cowart, Coyne, Coyne, Craig, Creighton, Creighton, Cripe, Crowder, Cruise, Cumming, Cunningham, Cuoco, Canton, Danilishin, D’Antonio, Danzmann, Darman, Da~Silva~Costa, Dattilo, Dave, Daveloza, Davier,
  Davies, Daw, Day, De, DeBra, Debreczeni, Degallaix, De~Laurentis, Deléglise, Del~Pozzo, Denker, Dent, Dereli, Dergachev, DeRosa, De~Rosa, DeSalvo, Dhurandhar, Díaz, Di~Fiore, Di~Giovanni, Di~Lieto, Di~Pace, Di~Palma, Di~Virgilio, Dojcinoski, Dolique, Donovan, Dooley, Doravari, Douglas, Downes, Drago, Drever, Driggers, Du, Ducrot, Dwyer, Edo, Edwards, Effler, Eggenstein, Ehrens, Eichholz, Eikenberry, Engels, Essick, Etzel, Evans, Evans, Everett, Factourovich, Fafone, Fair, Fairhurst, Fan, Fang, Farinon, Farr, Farr, Favata, Fays, Fehrmann, Fejer, Feldbaum, Ferrante, Ferreira, Ferrini, Fidecaro, Finn, Fiori, Fiorucci, Fisher, Flaminio, Fletcher, Fong, Fournier, Franco, Frasca, Frasconi, Frede, Frei, Freise, Frey, Frey, Fricke, Fritschel, Frolov, Fulda, Fyffe, Gabbard, Gair, Gammaitoni, Gaonkar, Garufi, Gatto, Gaur, Gehrels, Gemme, Gendre, Genin, Gennai, George, Gergely, Germain, Ghosh, Ghosh, Ghosh, Giaime, Giardina, Giazotto, Gill, Glaefke, Gleason, Goetz, Goetz, Gondan, González, Castro, Gopakumar,
  Gordon, Gorodetsky, Gossan, Gosselin, Gouaty, Graef, Graff, Granata, Grant, Gras, Gray, Greco, Green, Greenhalgh, Groot, Grote, Grunewald, Guidi, Guo, Gupta, Gupta, Gushwa, Gustafson, Gustafson, Hacker, Hall, Hall, Hammond, Haney, Hanke, Hanks, Hanna, Hannam, Hanson, Hardwick, Harms, Harry, Harry, Hart, Hartman, Haster, Haughian, Healy, Heefner, Heidmann, Heintze, Heinzel, Heitmann, Hello, Hemming, Hendry, Heng, Hennig, Heptonstall, Heurs, Hild, Hoak, Hodge, Hofman, Hollitt, Holt, Holz, Hopkins, Hosken, Hough, Houston, Howell, Hu, Huang, Huerta, Huet, Hughey, Husa, Huttner, Huynh-Dinh, Idrisy, Indik, Ingram, Inta, Isa, Isac, Isi, Islas, Isogai, Iyer, Izumi, Jacobson, Jacqmin, Jang, Jani, Jaranowski, Jawahar, Jiménez-Forteza, Johnson, Johnson-McDaniel, Jones, Jones, Jonker, Ju, Haris, Kalaghatgi, Kalogera, Kandhasamy, Kang, Kanner, Karki, Kasprzack, Katsavounidis, Katzman, Kaufer, Kaur, Kawabe, Kawazoe, Kéfélian, Kehl, Keitel, Kelley, Kells, Kennedy, Keppel, Key, Khalaidovski, Khalili, Khan, Khan, Khan,
  Khazanov, Kijbunchoo, Kim, Kim, Kim, Kim, Kim, Kim, King, King, Kinzel, Kissel, Kleybolte, Klimenko, Koehlenbeck, Kokeyama, Koley, Kondrashov, Kontos, Koranda, Korobko, Korth, Kowalska, Kozak, Kringel, Krishnan, Królak, Krueger, Kuehn, Kumar, Kumar, Kuo, Kutynia, Kwee, Lackey, Landry, Lange, Lantz, Lasky, Lazzarini, Lazzaro, Leaci, Leavey, Lebigot, Lee, Lee, Lee, Lee, Lenon, Leonardi, Leong, Leroy, Letendre, Levin, Levine, Li, Libson, Littenberg, Lockerbie, Logue, Lombardi, London, Lord, Lorenzini, Loriette, Lormand, Losurdo, Lough, Lousto, Lovelace, Lück, Lundgren, Luo, Lynch, Ma, MacDonald, Machenschalk, MacInnis, Macleod, Magaña-Sandoval, Magee, Mageswaran, Majorana, Maksimovic, Malvezzi, Man, Mandel, Mandic, Mangano, Mansell, Manske, Mantovani, Marchesoni, Marion, Márka, Márka, Markosyan, Maros, Martelli, Martellini, Martin, Martin, Martynov, Marx, Mason, Masserot, Massinger, Masso-Reid, Matichard, Matone, Mavalvala, Mazumder, Mazzolo, McCarthy, McClelland, McCormick, McGuire, McIntyre, McIver,
  McManus, McWilliams, Meacher, Meadors, Meidam, Melatos, Mendell, Mendoza-Gandara, Mercer, Merilh, Merzougui, Meshkov, Messenger, Messick, Meyers, Mezzani, Miao, Michel, Middleton, Mikhailov, Milano, Miller, Millhouse, Minenkov, Ming, Mirshekari, Mishra, Mitra, Mitrofanov, Mitselmakher, Mittleman, Moggi, Mohan, Mohapatra, Montani, Moore, Moore, Moraru, Moreno, Morriss, Mossavi, Mours, Mow-Lowry, Mueller, Mueller, Muir, Mukherjee, Mukherjee, Mukherjee, Mukund, Mullavey, Munch, Murphy, Murray, Mytidis, Nardecchia, Naticchioni, Nayak, Necula, Nedkova, Nelemans, Neri, Neunzert, Newton, Nguyen, Nielsen, Nissanke, Nitz, Nocera, Nolting, Normandin, Nuttall, Oberling, Ochsner, O’Dell, Oelker, Ogin, Oh, Oh, Ohme, Oliver, Oppermann, Oram, O’Reilly, O’Shaughnessy, Ott, Ottaway, Ottens, Overmier, Owen, Pai, Pai, Palamos, Palashov, Palomba, Pal-Singh, Pan, Pan, Pankow, Pannarale, Pant, Paoletti, Paoli, Papa, Paris, Parker, Pascucci, Pasqualetti, Passaquieti, Passuello, Patricelli, Patrick, Pearlstone, Pedraza,
  Pedurand, Pekowsky, Pele, Penn, Perreca, Pfeiffer, Phelps, Piccinni, Pichot, Pickenpack, Piergiovanni, Pierro, Pillant, Pinard, Pinto, Pitkin, Poeld, Poggiani, Popolizio, Post, Powell, Prasad, Predoi, Premachandra, Prestegard, Price, Prijatelj, Principe, Privitera, Prix, Prodi, Prokhorov, Puncken, Punturo, Puppo, Pürrer, Qi, Qin, Quetschke, Quintero, Quitzow-James, Raab, Rabeling, Radkins, Raffai, Raja, Rakhmanov, Ramet, Rapagnani, Raymond, Razzano, Re, Read, Reed, Regimbau, Rei, Reid, Reitze, Rew, Reyes, Ricci, Riles, Robertson, Robie, Robinet, Rocchi, Rolland, Rollins, Roma, Romano, Romano, Romanov, Romie, Rosińska, Rowan, Rüdiger, Ruggi, Ryan, Sachdev, Sadecki, Sadeghian, Salconi, Saleem, Salemi, Samajdar, Sammut, Sampson, Sanchez, Sandberg, Sandeen, Sanders, Sanders, Sassolas, Sathyaprakash, Saulson, Sauter, Savage, Sawadsky, Schale, Schilling, Schmidt, Schmidt, Schnabel, Schofield, Schönbeck, Schreiber, Schuette, Schutz, Scott, Scott, Sellers, Sengupta, Sentenac, Sequino, Sergeev, Serna, Setyawati,
  Sevigny, Shaddock, Shaffer, Shah, Shahriar, Shaltev, Shao, Shapiro, Shawhan, Sheperd, Shoemaker, Shoemaker, Siellez, Siemens, Sigg, Silva, Simakov, Singer, Singer, Singh, Singh, Singhal, Sintes, Slagmolen, Smith, Smith, Smith, Smith, Son, Sorazu, Sorrentino, Souradeep, Srivastava, Staley, Steinke, Steinlechner, Steinlechner, Steinmeyer, Stephens, Stevenson, Stone, Strain, Straniero, Stratta, Strauss, Strigin, Sturani, Stuver, Summerscales, Sun, Sutton, Swinkels, Szczepańczyk, Tacca, Talukder, Tanner, Tápai, Tarabrin, Taracchini, Taylor, Theeg, Thirugnanasambandam, Thomas, Thomas, Thomas, Thorne, Thorne, Thrane, Tiwari, Tiwari, Tokmakov, Tomlinson, Tonelli, Torres, Torrie, Töyrä, Travasso, Traylor, Trifirò, Tringali, Trozzo, Tse, Turconi, Tuyenbayev, Ugolini, Unnikrishnan, Urban, Usman, Vahlbruch, Vajente, Valdes, Vallisneri, van Bakel, van Beuzekom, van~den Brand, Van Den~Broeck, Vander-Hyde, van~der Schaaf, van Heijningen, van Veggel, Vardaro, Vass, Vasúth, Vaulin, Vecchio, Vedovato, Veitch, Veitch,
  Venkateswara, Verkindt, Vetrano, Viceré, Vinciguerra, Vine, Vinet, Vitale, Vo, Vocca, Vorvick, Voss, Vousden, Vyatchanin, Wade, Wade, Wade, Waldman, Walker, Wallace, Walsh, Wang, Wang, Wang, Wang, Wang, Ward, Ward, Warner, Was, Weaver, Wei, Weinert, Weinstein, Weiss, Welborn, Wen, Weßels, Westphal, Wette, Whelan, Whitcomb, White, Whiting, Wiesner, Wilkinson, Willems, Williams, Williams, Williamson, Willis, Willke, Wimmer, Winkelmann, Winkler, Wipf, Wiseman, Wittel, Woan, Worden, Wright, Wu, Yablon, Yakushin, Yam, Yamamoto, Yancey, Yap, Yu, Yvert, Zadrożny, Zangrando, Zanolin, Zendri, Zevin, Zhang, Zhang, Zhang, Zhang, Zhao, Zhou, Zhou, Zhu, Zucker, Zuraw, \& Zweizig}]{ligo_scientific_collaboration_and_virgo_collaboration_observation_2016}
{LIGO Scientific Collaboration and Virgo Collaboration}, Abbott, B., Abbott, R., {et~al.} 2016, \bibinfo{title}{Observation of {Gravitational} {Waves} from a {Binary} {Black} {Hole} {Merger},} Physical Review Letters, 116, 061102, \dodoi{10.1103/PhysRevLett.116.061102}

\bibitem[{L. {Lindegren} {et~al.}(2021){Lindegren}, {Klioner}, {Hern{\'a}ndez}, {Bombrun}, {Ramos-Lerate}, {Steidelm{\"u}ller}, {Bastian}, {Biermann}, {de Torres}, {Gerlach}, {Geyer}, {Hilger}, {Hobbs}, {Lammers}, {McMillan}, {Stephenson}, {Casta{\~n}eda}, {Davidson}, {Fabricius}, {Gracia-Abril}, {Portell}, {Rowell}, {Teyssier}, {Torra}, {Bartolom{\'e}}, {Clotet}, {Garralda}, {Gonz{\'a}lez-Vidal}, {Torra}, {Abbas}, {Altmann}, {Anglada Varela}, {Balaguer-N{\'u}{\~n}ez}, {Balog}, {Barache}, {Becciani}, {Bernet}, {Bertone}, {Bianchi}, {Bouquillon}, {Brown}, {Bucciarelli}, {Busonero}, {Butkevich}, {Buzzi}, {Cancelliere}, {Carlucci}, {Charlot}, {Cioni}, {Crosta}, {Crowley}, {del Peloso}, {del Pozo}, {Drimmel}, {Esquej}, {Fienga}, {Fraile}, {Gai}, {Garcia-Reinaldos}, {Guerra}, {Hambly}, {Hauser}, {Jan{\ss}en}, {Jordan}, {Kostrzewa-Rutkowska}, {Lattanzi}, {Liao}, {Licata}, {Lister}, {L{\"o}ffler}, {Marchant}, {Masip}, {Mignard}, {Mints}, {Molina}, {Mora}, {Morbidelli}, {Murphy}, {Pagani}, {Panuzzo}, {Pe{\~n}alosa
  Esteller}, {Poggio}, {Re Fiorentin}, {Riva}, {Sagrist{\`a} Sell{\'e}s}, {Sanchez Gimenez}, {Sarasso}, {Sciacca}, {Siddiqui}, {Smart}, {Souami}, {Spagna}, {Steele}, {Taris}, {Utrilla}, {van Reeven}, \& {Vecchiato}}]{2021A&A...649A...2L}
{Lindegren}, L., {Klioner}, S.~A., {Hern{\'a}ndez}, J., {et~al.} 2021, \bibinfo{title}{{Gaia Early Data Release 3. The astrometric solution},} \aap, 649, A2, \dodoi{10.1051/0004-6361/202039709}

\bibitem[{N.~P. Lourie {et~al.}(2020)Lourie, Baker, Burruss, Egan, Fżrész, Frostig, Garcia-Zych, Ganciu, Haworth, Hinrichsen, Kasliwal, Karambelkar, Malonis, Simcoe, \& Zolkower}]{lourie_wide-field_2020}
Lourie, N.~P., Baker, J.~W., Burruss, R.~S., {et~al.} 2020, in The wide-field infrared transient explorer ({WINTER}), Vol. 11447, eprint: arXiv:2102.01109, 114479K, \dodoi{10.1117/12.2561210}

\bibitem[{S. MacBride {et~al.}(2025)MacBride, Anand, Howard, Sullivan, Wang, Jones, Yoachim, Bellm, Armstrong, Wood-Vasey, Drlica-Wagner, Herner, Narayan, Acero~Cuellar, Bianco, Sánchez, Banovetz, Mortensen, Venegas, Kang, Utsumi, \& Ribeiro}]{macbride_ligovirgokagra_2025}
MacBride, S., Anand, S., Howard, E., {et~al.} 2025, \bibinfo{title}{{LIGO}/{Virgo}/{KAGRA} {S250725j}: {Observations} with the {NSF}-{DOE} {Vera} {C}. {Rubin} {Observatory},} GRB Coordinates Network, 41595, 1.
\newblock \url{https://ui.adsabs.harvard.edu/abs/2025GCN.41595....1M}

\bibitem[{E.~A. {Magnier} {et~al.}(2020){Magnier}, {Schlafly}, {Finkbeiner}, {Tonry}, {Goldman}, {R{\"o}ser}, {Schilbach}, {Casertano}, {Chambers}, {Flewelling}, {Huber}, {Price}, {Sweeney}, {Waters}, {Denneau}, {Draper}, {Hodapp}, {Jedicke}, {Kaiser}, {Kudritzki}, {Metcalfe}, {Stubbs}, \& {Wainscoat}}]{2020ApJS..251....6M}
{Magnier}, E.~A., {Schlafly}, E.~F., {Finkbeiner}, D.~P., {et~al.} 2020, \bibinfo{title}{{Pan-STARRS Photometric and Astrometric Calibration},} \apjs, 251, 6, \dodoi{10.3847/1538-4365/abb82a}

\bibitem[{R. {Margutti} {et~al.}(2017){Margutti}, {Berger}, {Fong}, {Guidorzi}, {Alexander}, {Metzger}, {Blanchard}, {Cowperthwaite}, {Chornock}, {Eftekhari}, {Nicholl}, {Villar}, {Williams}, {Annis}, {Brown}, {Chen}, {Doctor}, {Frieman}, {Holz}, {Sako}, \& {Soares-Santos}}]{Margutti2017}
{Margutti}, R., {Berger}, E., {Fong}, W., {et~al.} 2017, \bibinfo{title}{{The Electromagnetic Counterpart of the Binary Neutron Star Merger LIGO/Virgo GW170817. V. Rising X-Ray Emission from an Off-axis Jet},} \apjl, 848, L20, \dodoi{10.3847/2041-8213/aa9057}

\bibitem[{I. Markin {et~al.}(2023)Markin, Neuweiler, Abac, Chaurasia, Ujevic, Bulla, \& Dietrich}]{Markin_2023}
Markin, I., Neuweiler, A., Abac, A., {et~al.} 2023, \bibinfo{title}{General-relativistic hydrodynamics simulation of a neutron star–sub-solar-mass black hole merger,} Physical Review D, 108, \dodoi{10.1103/physrevd.108.064025}

\bibitem[{T. Martineau {et~al.}(2024)Martineau, Foucart, Scheel, Duez, Kidder, \& Pfeiffer}]{martineau_black_2024}
Martineau, T., Foucart, F., Scheel, M., {et~al.} 2024, \bibinfo{title}{Black {Hole}-{Neutron} {Star} {Binaries} near {Neutron} {Star} {Disruption} {Limit} in the {Mass} {Regime} of {Event} {GW230529},} arXiv, \dodoi{10.48550/arXiv.2405.06819}

\bibitem[{B.~D. {Metzger} {et~al.}(2024){Metzger}, {Hui}, \& {Cantiello}}]{Metzger2024}
{Metzger}, B.~D., {Hui}, L., \& {Cantiello}, M. 2024, \bibinfo{title}{{Fragmentation in Gravitationally Unstable Collapsar Disks and Subsolar Neutron Star Mergers},} \apjl, 971, L34, \dodoi{10.3847/2041-8213/ad6990}

\bibitem[{S. {Miyazaki} {et~al.}(2018){Miyazaki}, {Komiyama}, {Kawanomoto}, {Doi}, {Furusawa}, {Hamana}, {Hayashi}, {Ikeda}, {Kamata}, {Karoji}, {Koike}, {Kurakami}, {Miyama}, {Morokuma}, {Nakata}, {Namikawa}, {Nakaya}, {Nariai}, {Obuchi}, {Oishi}, {Okada}, {Okura}, {Tait}, {Takata}, {Tanaka}, {Tanaka}, {Terai}, {Tomono}, {Uraguchi}, {Usuda}, {Utsumi}, {Yamada}, {Yamanoi}, {Aihara}, {Fujimori}, {Mineo}, {Miyatake}, {Oguri}, {Uchida}, {Tanaka}, {Yasuda}, {Takada}, {Murayama}, {Nishizawa}, {Sugiyama}, {Chiba}, {Futamase}, {Wang}, {Chen}, {Ho}, {Liaw}, {Chiu}, {Ho}, {Lai}, {Lee}, {Jeng}, {Iwamura}, {Armstrong}, {Bickerton}, {Bosch}, {Gunn}, {Lupton}, {Loomis}, {Price}, {Smith}, {Strauss}, {Turner}, {Suzuki}, {Miyazaki}, {Muramatsu}, {Yamamoto}, {Endo}, {Ezaki}, {Ito}, {Kawaguchi}, {Sofuku}, {Taniike}, {Akutsu}, {Dojo}, {Kasumi}, {Matsuda}, {Imoto}, {Miwa}, {Suzuki}, {Takeshi}, \& {Yokota}}]{2018PASJ...70S...1M}
{Miyazaki}, S., {Komiyama}, Y., {Kawanomoto}, S., {et~al.} 2018, \bibinfo{title}{{Hyper Suprime-Cam: System design and verification of image quality},} \pasj, 70, S1, \dodoi{10.1093/pasj/psx063}

\bibitem[{A. {Moroianu} {et~al.}(2023){Moroianu}, {Wen}, {James}, {Ai}, {Kovalam}, {Panther}, \& {Zhang}}]{Moroianu}
{Moroianu}, A., {Wen}, L., {James}, C.~W., {et~al.} 2023, \bibinfo{title}{{An assessment of the association between a fast radio burst and binary neutron star merger},} Nature Astronomy, 7, 579, \dodoi{10.1038/s41550-023-01917-x}

\bibitem[{V. {Nedora} {et~al.}(2021){Nedora}, {Bernuzzi}, {Radice}, {Daszuta}, {Endrizzi}, {Perego}, {Prakash}, {Safarzadeh}, {Schianchi}, \& {Logoteta}}]{Nedora2021}
{Nedora}, V., {Bernuzzi}, S., {Radice}, D., {et~al.} 2021, \bibinfo{title}{{Numerical Relativity Simulations of the Neutron Star Merger GW170817: Long-term Remnant Evolutions, Winds, Remnant Disks, and Nucleosynthesis},} \apj, 906, 98, \dodoi{10.3847/1538-4357/abc9be}

\bibitem[{A. Neuweiler {et~al.}(2025)Neuweiler {et~al.}}]{Neuweiler:2025klw}
Neuweiler, A., {et~al.} 2025, \bibinfo{title}{{General-relativistic radiation magnetohydrodynamics simulations of binary neutron star mergers: The influence of spin on the multi-messenger picture},} \doarXiv{2510.14850}

\bibitem[{B. {O'Connor} {et~al.}(2025){O'Connor}, {Ricci}, {Troja}, {Palmese}, {xxx}, {xxx}, {xxx}, \& {xxx}}]{OConnor2025ulz}
{O'Connor}, B., {Ricci}, R., {Troja}, E., {et~al.} 2025, \bibinfo{title}{in prep. AT2025ulz,} in prep.

\bibitem[{B. {O'Connor} {et~al.}(2021){O'Connor}, {Troja}, {Dichiara}, {Chase}, {Ryan}, {Cenko}, {Fryer}, {Ricci}, {Marshall}, {Kouveliotou}, {Wollaeger}, {Fontes}, {Korobkin}, {Gatkine}, {Kutyrev}, {Veilleux}, {Kawai}, \& {Sakamoto}}]{OConnor2021kn}
{O'Connor}, B., {Troja}, E., {Dichiara}, S., {et~al.} 2021, \bibinfo{title}{{A tale of two mergers: constraints on kilonova detection in two short GRBs at z {\ensuremath{\sim}} 0.5},} \mnras, 502, 1279, \dodoi{10.1093/mnras/stab132}

\bibitem[{B. {O'Connor} {et~al.}(2022){O'Connor}, {Troja}, {Dichiara}, {Beniamini}, {Cenko}, {Kouveliotou}, {Gonz{\'a}lez}, {Durbak}, {Gatkine}, {Kutyrev}, {Sakamoto}, {S{\'a}nchez-Ram{\'\i}rez}, \& {Veilleux}}]{OConnor2022}
{O'Connor}, B., {Troja}, E., {Dichiara}, S., {et~al.} 2022, \bibinfo{title}{{A deep survey of short GRB host galaxies over z 0-2: implications for offsets, redshifts, and environments},} \mnras, 515, 4890, \dodoi{10.1093/mnras/stac1982}

\bibitem[{A. Palmese {et~al.}(2023)Palmese, Bom, Mucesh, \& Hartley}]{palmese_standard_2023}
Palmese, A., Bom, C.~R., Mucesh, S., \& Hartley, W.~G. 2023, \bibinfo{title}{A {Standard} {Siren} {Measurement} of the {Hubble} {Constant} {Using} {Gravitational}-wave {Events} from the {First} {Three} {LIGO}/{Virgo} {Observing} {Runs} and the {DESI} {Legacy} {Survey},} The Astrophysical Journal, 943, 56, \dodoi{10.3847/1538-4357/aca6e3}

\bibitem[{A. Palmese {et~al.}(2021)Palmese, Fishbach, Burke, Annis, \& Liu}]{palmese_ligovirgo_2021}
Palmese, A., Fishbach, M., Burke, C.~J., Annis, J., \& Liu, X. 2021, \bibinfo{title}{Do {LIGO}/{Virgo} {Black} {Hole} {Mergers} {Produce} {AGN} {Flares}? {The} {Case} of {GW190521} and {Prospects} for {Reaching} a {Confident} {Association},} The Astrophysical Journal, 914, L34, \dodoi{10.3847/2041-8213/ac0883}

\bibitem[{A. {Palmese} \& S. {Mastrogiovanni}(2025){Palmese} \& {Mastrogiovanni}}]{GWCosmology}
{Palmese}, A., \& {Mastrogiovanni}, S. 2025, \bibinfo{title}{{Gravitational Wave Cosmology},} Encyclopedia of Astrophysics, Elsevier, arXiv:2502.00239, \dodoi{10.48550/arXiv.2502.00239}

\bibitem[{P.~T.~H. Pang {et~al.}(2023)Pang, Dietrich, Coughlin, Bulla, Tews, Almualla, Barna, Kiendrebeogo, Kunert, Mansingh, Reed, Sravan, Toivonen, Antier, VandenBerg, Heinzel, Nedora, Salehi, Sharma, Somasundaram, \& Van Den~Broeck}]{pang_updated_2023}
Pang, P. T.~H., Dietrich, T., Coughlin, M.~W., {et~al.} 2023, \bibinfo{title}{An updated nuclear-physics and multi-messenger astrophysics framework for binary neutron star mergers,} Nature Communications, 14, 8352, \dodoi{10.1038/s41467-023-43932-6}

\bibitem[{Y.~C. {Pei}(1992){Pei}}]{Pei1992}
{Pei}, Y.~C. 1992, \bibinfo{title}{{Interstellar Dust from the Milky Way to the Magellanic Clouds},} \apj, 395, 130, \dodoi{10.1086/171637}

\bibitem[{D.~A. {Perley} {et~al.}(2019){Perley}, {Mazzali}, {Yan}, {Cenko}, {Gezari}, {Taggart}, {Blagorodnova}, {Fremling}, {Mockler}, {Singh}, {Tominaga}, {Tanaka}, {Watson}, {Ahumada}, {Anupama}, {Ashall}, {Becerra}, {Bersier}, {Bhalerao}, {Bloom}, {Butler}, {Copperwheat}, {Coughlin}, {De}, {Drake}, {Duev}, {Frederick}, {Gonz{\'a}lez}, {Goobar}, {Heida}, {Ho}, {Horst}, {Hung}, {Itoh}, {Jencson}, {Kasliwal}, {Kawai}, {Khanam}, {Kulkarni}, {Kumar}, {Kumar}, {Kutyrev}, {Lee}, {Maeda}, {Mahabal}, {Murata}, {Neill}, {Ngeow}, {Penprase}, {Pian}, {Quimby}, {Ramirez-Ruiz}, {Richer}, {Rom{\'a}n-Z{\'u}{\~n}iga}, {Sahu}, {Srivastav}, {Socia}, {Sollerman}, {Tachibana}, {Taddia}, {Tinyanont}, {Troja}, {Ward}, {Wee}, \& {Yu}}]{Perley2019}
{Perley}, D.~A., {Mazzali}, P.~A., {Yan}, L., {et~al.} 2019, \bibinfo{title}{{The fast, luminous ultraviolet transient AT2018cow: extreme supernova, or disruption of a star by an intermediate-mass black hole?},} \mnras, 484, 1031, \dodoi{10.1093/mnras/sty3420}

\bibitem[{E. {Pian} {et~al.}(2017){Pian}, {D'Avanzo}, {Benetti}, {Branchesi}, {Brocato}, {Campana}, {Cappellaro}, {Covino}, {D'Elia}, {Fynbo}, {Getman}, {Ghirlanda}, {Ghisellini}, {Grado}, {Greco}, {Hjorth}, {Kouveliotou}, {Levan}, {Limatola}, {Malesani}, {Mazzali}, {Melandri}, {M{\o}ller}, {Nicastro}, {Palazzi}, {Piranomonte}, {Rossi}, {Salafia}, {Selsing}, {Stratta}, {Tanaka}, {Tanvir}, {Tomasella}, {Watson}, {Yang}, {Amati}, {Antonelli}, {Ascenzi}, {Bernardini}, {Bo{\"e}r}, {Bufano}, {Bulgarelli}, {Capaccioli}, {Casella}, {Castro-Tirado}, {Chassande-Mottin}, {Ciolfi}, {Copperwheat}, {Dadina}, {De Cesare}, {di Paola}, {Fan}, {Gendre}, {Giuffrida}, {Giunta}, {Hunt}, {Israel}, {Jin}, {Kasliwal}, {Klose}, {Lisi}, {Longo}, {Maiorano}, {Mapelli}, {Masetti}, {Nava}, {Patricelli}, {Perley}, {Pescalli}, {Piran}, {Possenti}, {Pulone}, {Razzano}, {Salvaterra}, {Schipani}, {Spera}, {Stamerra}, {Stella}, {Tagliaferri}, {Testa}, {Troja}, {Turatto}, {Vergani}, \& {Vergani}}]{Pian2017}
{Pian}, E., {D'Avanzo}, P., {Benetti}, S., {et~al.} 2017, \bibinfo{title}{{Spectroscopic identification of r-process nucleosynthesis in a double neutron-star merger},} \nat, 551, 67, \dodoi{10.1038/nature24298}

\bibitem[{A.~L. {Piro} \& E. {Pfahl}(2007){Piro} \& {Pfahl}}]{PiroPfahl2007}
{Piro}, A.~L., \& {Pfahl}, E. 2007, \bibinfo{title}{{Fragmentation of Collapsar Disks and the Production of Gravitational Waves},} \apj, 658, 1173, \dodoi{10.1086/511672}

\bibitem[{L. {Piro} {et~al.}(2021){Piro}, {Bruni}, {Troja}, {O'Connor}, {Panessa}, {Ricci}, {Zhang}, {Burgay}, {Dichiara}, {Lee}, {Lotti}, {Niu}, {Pilia}, {Possenti}, {Trudu}, {Xu}, {Zhu}, {Kutyrev}, \& {Veilleux}}]{Piro2021}
{Piro}, L., {Bruni}, G., {Troja}, E., {et~al.} 2021, \bibinfo{title}{{The fast radio burst FRB 20201124A in a star-forming region: Constraints to the progenitor and multiwavelength counterparts},} \aap, 656, L15, \dodoi{10.1051/0004-6361/202141903}

\bibitem[{ {Planck Collaboration} {et~al.}(2020){Planck Collaboration}, {Aghanim}, {Akrami}, {Ashdown}, {Aumont}, {Baccigalupi}, {Ballardini}, {Banday}, {Barreiro}, {Bartolo}, {Basak}, {Battye}, {Benabed}, {Bernard}, {Bersanelli}, {Bielewicz}, {Bock}, {Bond}, {Borrill}, {Bouchet}, {Boulanger}, {Bucher}, {Burigana}, {Butler}, {Calabrese}, {Cardoso}, {Carron}, {Challinor}, {Chiang}, {Chluba}, {Colombo}, {Combet}, {Contreras}, {Crill}, {Cuttaia}, {de Bernardis}, {de Zotti}, {Delabrouille}, {Delouis}, {Di Valentino}, {Diego}, {Dor{\'e}}, {Douspis}, {Ducout}, {Dupac}, {Dusini}, {Efstathiou}, {Elsner}, {En{\ss}lin}, {Eriksen}, {Fantaye}, {Farhang}, {Fergusson}, {Fernandez-Cobos}, {Finelli}, {Forastieri}, {Frailis}, {Fraisse}, {Franceschi}, {Frolov}, {Galeotta}, {Galli}, {Ganga}, {G{\'e}nova-Santos}, {Gerbino}, {Ghosh}, {Gonz{\'a}lez-Nuevo}, {G{\'o}rski}, {Gratton}, {Gruppuso}, {Gudmundsson}, {Hamann}, {Handley}, {Hansen}, {Herranz}, {Hildebrandt}, {Hivon}, {Huang}, {Jaffe}, {Jones}, {Karakci}, {Keih{\"a}nen},
  {Keskitalo}, {Kiiveri}, {Kim}, {Kisner}, {Knox}, {Krachmalnicoff}, {Kunz}, {Kurki-Suonio}, {Lagache}, {Lamarre}, {Lasenby}, {Lattanzi}, {Lawrence}, {Le Jeune}, {Lemos}, {Lesgourgues}, {Levrier}, {Lewis}, {Liguori}, {Lilje}, {Lilley}, {Lindholm}, {L{\'o}pez-Caniego}, {Lubin}, {Ma}, {Mac{\'\i}as-P{\'e}rez}, {Maggio}, {Maino}, {Mandolesi}, {Mangilli}, {Marcos-Caballero}, {Maris}, {Martin}, {Martinelli}, {Mart{\'\i}nez-Gonz{\'a}lez}, {Matarrese}, {Mauri}, {McEwen}, {Meinhold}, {Melchiorri}, {Mennella}, {Migliaccio}, {Millea}, {Mitra}, {Miville-Desch{\^e}nes}, {Molinari}, {Montier}, {Morgante}, {Moss}, {Natoli}, {N{\o}rgaard-Nielsen}, {Pagano}, {Paoletti}, {Partridge}, {Patanchon}, {Peiris}, {Perrotta}, {Pettorino}, {Piacentini}, {Polastri}, {Polenta}, {Puget}, {Rachen}, {Reinecke}, {Remazeilles}, {Renzi}, {Rocha}, {Rosset}, {Roudier}, {Rubi{\~n}o-Mart{\'\i}n}, {Ruiz-Granados}, {Salvati}, {Sandri}, {Savelainen}, {Scott}, {Shellard}, {Sirignano}, {Sirri}, {Spencer}, {Sunyaev}, {Suur-Uski}, {Tauber}, {Tavagnacco},
  {Tenti}, {Toffolatti}, {Tomasi}, {Trombetti}, {Valenziano}, {Valiviita}, {Van Tent}, {Vibert}, {Vielva}, {Villa}, {Vittorio}, {Wandelt}, {Wehus}, {White}, {White}, {Zacchei}, \& {Zonca}}]{Planck2020}
{Planck Collaboration}, {Aghanim}, N., {Akrami}, Y., {et~al.} 2020, \bibinfo{title}{{Planck 2018 results. VI. Cosmological parameters},} \aap, 641, A6, \dodoi{10.1051/0004-6361/201833910}

\bibitem[{D. Radice {et~al.}(2016)Radice, Galeazzi, Lippuner, Roberts, Ott, \& Rezzolla}]{Radice:2016dwd}
Radice, D., Galeazzi, F., Lippuner, J., {et~al.} 2016, \bibinfo{title}{{Dynamical Mass Ejection from Binary Neutron Star Mergers},} Mon. Not. Roy. Astron. Soc., 460, 3255, \dodoi{10.1093/mnras/stw1227}

\bibitem[{L.~W. {Ramsey} {et~al.}(1998){Ramsey}, {Adams}, {Barnes}, {Booth}, {Cornell}, {Fowler}, {Gaffney}, {Glaspey}, {Good}, {Hill}, {Kelton}, {Krabbendam}, {Long}, {MacQueen}, {Ray}, {Ricklefs}, {Sage}, {Sebring}, {Spiesman}, \& {Steiner}}]{1998SPIE.3352...34R}
{Ramsey}, L.~W., {Adams}, M.~T., {Barnes}, T.~G., {et~al.} 1998, in Society of Photo-Optical Instrumentation Engineers (SPIE) Conference Series, Vol. 3352, Advanced Technology Optical/IR Telescopes VI, ed. L.~M. {Stepp}, 34--42, \dodoi{10.1117/12.319287}

\bibitem[{J.~C. {Rastinejad} {et~al.}(2025){Rastinejad}, {Fong}, {Kilpatrick}, {Nicholl}, \& {Metzger}}]{Rastinejad2025kn}
{Rastinejad}, J.~C., {Fong}, W., {Kilpatrick}, C.~D., {Nicholl}, M., \& {Metzger}, B.~D. 2025, \bibinfo{title}{{Uniform Modeling of Observed Kilonovae: Implications for Diversity and the Progenitors of Merger-driven Long Gamma-Ray Bursts},} \apj, 979, 190, \dodoi{10.3847/1538-4357/ad9c77}

\bibitem[{J.~C. {Rastinejad} {et~al.}(2021){Rastinejad}, {Fong}, {Kilpatrick}, {Paterson}, {Tanvir}, {Levan}, {Metzger}, {Berger}, {Chornock}, {Cobb}, {Laskar}, {Milne}, {Nugent}, \& {Smith}}]{Rastinejad2021}
{Rastinejad}, J.~C., {Fong}, W., {Kilpatrick}, C.~D., {et~al.} 2021, \bibinfo{title}{{Probing Kilonova Ejecta Properties Using a Catalog of Short Gamma-Ray Burst Observations},} \apj, 916, 89, \dodoi{10.3847/1538-4357/ac04b4}

\bibitem[{J.~C. {Rastinejad} {et~al.}(2022){Rastinejad}, {Gompertz}, {Levan}, {Fong}, {Nicholl}, {Lamb}, {Malesani}, {Nugent}, {Oates}, {Tanvir}, {de Ugarte Postigo}, {Kilpatrick}, {Moore}, {Metzger}, {Ravasio}, {Rossi}, {Schroeder}, {Jencson}, {Sand}, {Smith}, {Ag{\"u}{\'\i} Fern{\'a}ndez}, {Berger}, {Blanchard}, {Chornock}, {Cobb}, {De Pasquale}, {Fynbo}, {Izzo}, {Kann}, {Laskar}, {Marini}, {Paterson}, {Escorial}, {Sears}, \& {Th{\"o}ne}}]{Rastinejad2022}
{Rastinejad}, J.~C., {Gompertz}, B.~P., {Levan}, A.~J., {et~al.} 2022, \bibinfo{title}{{A kilonova following a long-duration gamma-ray burst at 350 Mpc},} \nat, 612, 223, \dodoi{10.1038/s41586-022-05390-w}

\bibitem[{A.~R. {Ricci} {et~al.}(2025){Ricci}, {Yadav}, \& {Troja}}]{Ricci-1}
{Ricci}, A.~R., {Yadav}, M., \& {Troja}, E. 2025, \bibinfo{title}{{LIGO/Virgo/KAGRA S250818k: 10 GHz VLA observations of AT2025ulz},} GRB Coordinates Network, 41464, 1

\bibitem[{R. {Ricci} {et~al.}(2025){Ricci}, {Yadav}, {Troja}, \& {ERC BHianca Team}}]{Ricci-2}
{Ricci}, R., {Yadav}, M., {Troja}, E., \& {ERC BHianca Team}. 2025, \bibinfo{title}{{LIGO/Virgo/KAGRA S250818k: VLA upper limits on AT2025ulz},} GRB Coordinates Network, 41542, 1

\bibitem[{A.~G. Riess {et~al.}(2022)Riess, Yuan, Macri, Scolnic, Brout, Casertano, Jones, Murakami, Anand, Breuval, Brink, Filippenko, Hoffmann, Jha, D’arcy~Kenworthy, Mackenty, Stahl, \& Zheng}]{Riess_2022}
Riess, A.~G., Yuan, W., Macri, L.~M., {et~al.} 2022, \bibinfo{title}{A Comprehensive Measurement of the Local Value of the Hubble Constant with 1 km s−1 Mpc−1 Uncertainty from the Hubble Space Telescope and the SH0ES Team,} The Astrophysical Journal Letters, 934, L7, \dodoi{10.3847/2041-8213/ac5c5b}

\bibitem[{J. {Ripero} {et~al.}(1993){Ripero}, {Garcia}, {Rodriguez}, {Pujol}, {Filippenko}, {Treffers}, {Paik}, {Davis}, {Schlegel}, {Hartwick}, {Balam}, {Zurek}, {Robb}, {Garnavich}, \& {Hong}}]{SN1993J_circular}
{Ripero}, J., {Garcia}, F., {Rodriguez}, D., {et~al.} 1993, \bibinfo{title}{{Supernova 1993J in NGC 3031},} \iaucirc, 5731, 1

\bibitem[{A. {Rossi} {et~al.}(2020){Rossi}, {Stratta}, {Maiorano}, {Spighi}, {Masetti}, {Palazzi}, {Gardini}, {Melandri}, {Nicastro}, {Pian}, {Branchesi}, {Dadina}, {Testa}, {Brocato}, {Benetti}, {Ciolfi}, {Covino}, {D'Elia}, {Grado}, {Izzo}, {Perego}, {Piranomonte}, {Salvaterra}, {Selsing}, {Tomasella}, {Yang}, {Vergani}, {Amati}, \& {Stephen}}]{Rossi2020}
{Rossi}, A., {Stratta}, G., {Maiorano}, E., {et~al.} 2020, \bibinfo{title}{{A comparison between short GRB afterglows and kilonova AT2017gfo: shedding light on kilonovae properties},} \mnras, 493, 3379, \dodoi{10.1093/mnras/staa479}

\bibitem[{S. Rosswog {et~al.}(2025)Rosswog, Sarin, Nakhar, \& Diener}]{Rosswog:2024vfe}
Rosswog, S., Sarin, N., Nakhar, E., \& Diener, P. 2025, \bibinfo{title}{{Fast dynamic ejecta in neutron star mergers},} Mon. Not. Roy. Astron. Soc., 538, 907, \dodoi{10.1093/mnras/staf324}

\bibitem[{G. {Ryan} {et~al.}(2020){Ryan}, {van Eerten}, {Piro}, \& {Troja}}]{Ryan2019}
{Ryan}, G., {van Eerten}, H., {Piro}, L., \& {Troja}, E. 2020, \bibinfo{title}{{Gamma-Ray Burst Afterglows in the Multimessenger Era: Numerical Models and Closure Relations},} \apj, 896, 166, \dodoi{10.3847/1538-4357/ab93cf}

\bibitem[{G. {Ryan} {et~al.}(2024){Ryan}, {van Eerten}, {Troja}, {Piro}, {O'Connor}, \& {Ricci}}]{Ryan2024}
{Ryan}, G., {van Eerten}, H., {Troja}, E., {et~al.} 2024, \bibinfo{title}{{Modeling of Long-term Afterglow Counterparts to Gravitational Wave Events: The Full View of GRB 170817A},} \apj, 975, 131, \dodoi{10.3847/1538-4357/ad6a14}

\bibitem[{N. Sarin \& S. Rosswog(2024)Sarin \& Rosswog}]{Sarin:2024tja}
Sarin, N., \& Rosswog, S. 2024, \bibinfo{title}{{Cautionary Tales on Heating-rate Prescriptions in Kilonovae},} Astrophys. J. Lett., 973, L24, \dodoi{10.3847/2041-8213/ad739d}

\bibitem[{V. {Savchenko} {et~al.}(2017){Savchenko}, {Ferrigno}, {Kuulkers}, {Bazzano}, {Bozzo}, {Brandt}, {Chenevez}, {Courvoisier}, {Diehl}, {Domingo}, {Hanlon}, {Jourdain}, {von Kienlin}, {Laurent}, {Lebrun}, {Lutovinov}, {Martin-Carrillo}, {Mereghetti}, {Natalucci}, {Rodi}, {Roques}, {Sunyaev}, \& {Ubertini}}]{Savchenko2017}
{Savchenko}, V., {Ferrigno}, C., {Kuulkers}, E., {et~al.} 2017, \bibinfo{title}{{INTEGRAL Detection of the First Prompt Gamma-Ray Signal Coincident with the Gravitational-wave Event GW170817},} \apjl, 848, L15, \dodoi{10.3847/2041-8213/aa8f94}

\bibitem[{E.~F. {Schlafly} \& D.~P. {Finkbeiner}(2011){Schlafly} \& {Finkbeiner}}]{Schlafly2011}
{Schlafly}, E.~F., \& {Finkbeiner}, D.~P. 2011, \bibinfo{title}{{Measuring Reddening with Sloan Digital Sky Survey Stellar Spectra and Recalibrating SFD},} \apj, 737, 103, \dodoi{10.1088/0004-637X/737/2/103}

\bibitem[{E.~F. Schlafly \& D.~P. Finkbeiner(2011)Schlafly \& Finkbeiner}]{schlafly_measuring_2011}
Schlafly, E.~F., \& Finkbeiner, D.~P. 2011, \bibinfo{title}{Measuring {Reddening} with {Sloan} {Digital} {Sky} {Survey} {Stellar} {Spectra} and {Recalibrating} {SFD},} The Astrophysical Journal, 737, 103, \dodoi{10.1088/0004-637X/737/2/103}

\bibitem[{B.~F. Schutz(1986)Schutz}]{schutz1986determining}
Schutz, B.~F. 1986, \bibinfo{title}{Determining the Hubble constant from gravitational wave observations,} Nature, 323, 310

\bibitem[{D. {Scolnic} {et~al.}(2018){Scolnic}, {Kessler}, {Brout}, {Cowperthwaite}, {Soares-Santos}, {Annis}, {Herner}, {Chen}, {Sako}, {Doctor}, {Butler}, {Palmese}, {Diehl}, {Frieman}, {Holz}, {Berger}, {Chornock}, {Villar}, {Nicholl}, {Biswas}, {Hounsell}, {Foley}, {Metzger}, {Rest}, {Garc{\'\i}a-Bellido}, {M{\"o}ller}, {Nugent}, {Abbott}, {Abdalla}, {Allam}, {Bechtol}, {Benoit-L{\'e}vy}, {Bertin}, {Brooks}, {Buckley-Geer}, {Carnero Rosell}, {Carrasco Kind}, {Carretero}, {Castander}, {Cunha}, {D'Andrea}, {da Costa}, {Davis}, {Doel}, {Drlica-Wagner}, {Eifler}, {Flaugher}, {Fosalba}, {Gaztanaga}, {Gerdes}, {Gruen}, {Gruendl}, {Gschwend}, {Gutierrez}, {Hartley}, {Honscheid}, {James}, {Johnson}, {Johnson}, {Krause}, {Kuehn}, {Kuhlmann}, {Lahav}, {Li}, {Lima}, {Maia}, {March}, {Marshall}, {Menanteau}, {Miquel}, {Neilsen}, {Plazas}, {Sanchez}, {Scarpine}, {Schubnell}, {Sevilla-Noarbe}, {Smith}, {Smith}, {Sobreira}, {Suchyta}, {Swanson}, {Tarle}, {Thomas}, {Tucker}, {Walker}, \& {DES
  Collaboration}}]{Scolnic2018}
{Scolnic}, D., {Kessler}, R., {Brout}, D., {et~al.} 2018, \bibinfo{title}{{How Many Kilonovae Can Be Found in Past, Present, and Future Survey Data Sets?},} \apjl, 852, L3, \dodoi{10.3847/2041-8213/aa9d82}

\bibitem[{Y. Sekiguchi {et~al.}(2015)Sekiguchi, Kiuchi, Kyutoku, \& Shibata}]{Sekiguchi:2015dma}
Sekiguchi, Y., Kiuchi, K., Kyutoku, K., \& Shibata, M. 2015, \bibinfo{title}{{Dynamical mass ejection from binary neutron star mergers: Radiation-hydrodynamics study in general relativity},} Phys. Rev. D, 91, 064059, \dodoi{10.1103/PhysRevD.91.064059}

\bibitem[{B. Shenhar {et~al.}(2024)Shenhar, Guttman, \& Waxman}]{Shenhar:2024rfm}
Shenhar, B., Guttman, O., \& Waxman, E. 2024, \bibinfo{title}{{An analytic description of electron thermalization in kilonovae ejecta},} Mon. Not. Roy. Astron. Soc., 531, 592, \dodoi{10.1093/mnras/stae1218}

\bibitem[{M. {Shetrone} {et~al.}(2007){Shetrone}, {Cornell}, {Fowler}, {Gaffney}, {Laws}, {Mader}, {Mason}, {Odewahn}, {Roman}, {Rostopchin}, {Schneider}, {Umbarger}, \& {Westfall}}]{2007PASP..119..556S}
{Shetrone}, M., {Cornell}, M.~E., {Fowler}, J.~R., {et~al.} 2007, \bibinfo{title}{{Ten Year Review of Queue Scheduling of the Hobby-Eberly Telescope},} \pasp, 119, 556, \dodoi{10.1086/519291}

\bibitem[{D.~M. Siegel {et~al.}(2022)Siegel, Agarwal, Barnes, Metzger, Renzo, \& Villar}]{siegel_super-kilonovae_2022}
Siegel, D.~M., Agarwal, A., Barnes, J., {et~al.} 2022, \bibinfo{title}{"{Super}-kilonovae" from {Massive} {Collapsars} as {Signatures} of {Black} {Hole} {Birth} in the {Pair}-instability {Mass} {Gap},} The Astrophysical Journal, 941, 100, \dodoi{10.3847/1538-4357/ac8d04}

\bibitem[{D.~M. Siegel {et~al.}(2019)Siegel, Barnes, \& Metzger}]{siegel_collapsars_2019}
Siegel, D.~M., Barnes, J., \& Metzger, B.~D. 2019, \bibinfo{title}{Collapsars as a major source of r-process elements,} Nature, 569, 241, \dodoi{10.1038/s41586-019-1136-0}

\bibitem[{L. Sironi {et~al.}(2015)Sironi, Keshet, \& Lemoine}]{Sironi:2015oza}
Sironi, L., Keshet, U., \& Lemoine, M. 2015, \bibinfo{title}{{Relativistic Shocks: Particle Acceleration and Magnetization},} Space Sci. Rev., 191, 519, \dodoi{10.1007/s11214-015-0181-8}

\bibitem[{M.~F. {Skrutskie} {et~al.}(2006){Skrutskie}, {Cutri}, {Stiening}, {Weinberg}, {Schneider}, {Carpenter}, {Beichman}, {Capps}, {Chester}, {Elias}, {Huchra}, {Liebert}, {Lonsdale}, {Monet}, {Price}, {Seitzer}, {Jarrett}, {Kirkpatrick}, {Gizis}, {Howard}, {Evans}, {Fowler}, {Fullmer}, {Hurt}, {Light}, {Kopan}, {Marsh}, {McCallon}, {Tam}, {Van Dyk}, \& {Wheelock}}]{Skrutskie2006}
{Skrutskie}, M.~F., {Cutri}, R.~M., {Stiening}, R., {et~al.} 2006, \bibinfo{title}{{The Two Micron All Sky Survey (2MASS)},} \aj, 131, 1163, \dodoi{10.1086/498708}

\bibitem[{M.~F. Skrutskie {et~al.}(2006)Skrutskie, Cutri, Stiening, Weinberg, Schneider, Carpenter, Beichman, Capps, Chester, Elias, Huchra, Liebert, Lonsdale, Monet, Price, Seitzer, Jarrett, Kirkpatrick, Gizis, Howard, Evans, Fowler, Fullmer, Hurt, Light, Kopan, Marsh, McCallon, Tam, Van~Dyk, \& Wheelock}]{skrutskie_two_2006}
Skrutskie, M.~F., Cutri, R.~M., Stiening, R., {et~al.} 2006, \bibinfo{title}{The {Two} {Micron} {All} {Sky} {Survey} ({2MASS}),} The Astronomical Journal, 131, 1163, \dodoi{10.1086/498708}

\bibitem[{S.~J. {Smartt} {et~al.}(2017){Smartt}, {Chen}, {Jerkstrand}, {Coughlin}, {Kankare}, {Sim}, {Fraser}, {Inserra}, {Maguire}, {Chambers}, {Huber}, {Kr{\"u}hler}, {Leloudas}, {Magee}, {Shingles}, {Smith}, {Young}, {Tonry}, {Kotak}, {Gal-Yam}, {Lyman}, {Homan}, {Agliozzo}, {Anderson}, {Angus}, {Ashall}, {Barbarino}, {Bauer}, {Berton}, {Botticella}, {Bulla}, {Bulger}, {Cannizzaro}, {Cano}, {Cartier}, {Cikota}, {Clark}, {De Cia}, {Della Valle}, {Denneau}, {Dennefeld}, {Dessart}, {Dimitriadis}, {Elias-Rosa}, {Firth}, {Flewelling}, {Fl{\"o}rs}, {Franckowiak}, {Frohmaier}, {Galbany}, {Gonz{\'a}lez-Gait{\'a}n}, {Greiner}, {Gromadzki}, {Guelbenzu}, {Guti{\'e}rrez}, {Hamanowicz}, {Hanlon}, {Harmanen}, {Heintz}, {Heinze}, {Hernandez}, {Hodgkin}, {Hook}, {Izzo}, {James}, {Jonker}, {Kerzendorf}, {Klose}, {Kostrzewa-Rutkowska}, {Kowalski}, {Kromer}, {Kuncarayakti}, {Lawrence}, {Lowe}, {Magnier}, {Manulis}, {Martin-Carrillo}, {Mattila}, {McBrien}, {M{\"u}ller}, {Nordin}, {O'Neill}, {Onori}, {Palmerio}, {Pastorello},
  {Patat}, {Pignata}, {Podsiadlowski}, {Pumo}, {Prentice}, {Rau}, {Razza}, {Rest}, {Reynolds}, {Roy}, {Ruiter}, {Rybicki}, {Salmon}, {Schady}, {Schultz}, {Schweyer}, {Seitenzahl}, {Smith}, {Sollerman}, {Stalder}, {Stubbs}, {Sullivan}, {Szegedi}, {Taddia}, {Taubenberger}, {Terreran}, {van Soelen}, {Vos}, {Wainscoat}, {Walton}, {Waters}, {Weiland}, {Willman}, {Wiseman}, {Wright}, {Wyrzykowski}, \& {Yaron}}]{Smartt2017}
{Smartt}, S.~J., {Chen}, T.~W., {Jerkstrand}, A., {et~al.} 2017, \bibinfo{title}{{A kilonova as the electromagnetic counterpart to a gravitational-wave source},} \nat, 551, 75, \dodoi{10.1038/nature24303}

\bibitem[{M. {Soares-Santos} {et~al.}(2017){Soares-Santos}, {Holz}, {Annis}, {Chornock}, {Herner}, {Berger}, {Brout}, {Chen}, {Kessler}, {Sako}, {Allam}, {Tucker}, {Butler}, {Palmese}, {Doctor}, {Diehl}, {Frieman}, {Yanny}, {Lin}, {Scolnic}, {Cowperthwaite}, {Neilsen}, {Marriner}, {Kuropatkin}, {Hartley}, {Paz-Chinch{\'o}n}, {Alexander}, {Balbinot}, {Blanchard}, {Brown}, {Carlin}, {Conselice}, {Cook}, {Drlica-Wagner}, {Drout}, {Durret}, {Eftekhari}, {Farr}, {Finley}, {Foley}, {Fong}, {Fryer}, {Garc{\'\i}a-Bellido}, {Gill}, {Gruendl}, {Hanna}, {Kasen}, {Li}, {Lopes}, {Louren{\c{c}}o}, {Margutti}, {Marshall}, {Matheson}, {Medina}, {Metzger}, {Mu{\~n}oz}, {Muir}, {Nicholl}, {Quataert}, {Rest}, {Sauseda}, {Schlegel}, {Secco}, {Sobreira}, {Stebbins}, {Villar}, {Vivas}, {Walker}, {Wester}, {Williams}, {Zenteno}, {Zhang}, {Abbott}, {Abdalla}, {Banerji}, {Bechtol}, {Benoit-L{\'e}vy}, {Bertin}, {Brooks}, {Buckley-Geer}, {Burke}, {Carnero Rosell}, {Carrasco Kind}, {Carretero}, {Castander}, {Crocce}, {Cunha},
  {D'Andrea}, {da Costa}, {Davis}, {Desai}, {Dietrich}, {Doel}, {Eifler}, {Fernandez}, {Flaugher}, {Fosalba}, {Gaztanaga}, {Gerdes}, {Giannantonio}, {Goldstein}, {Gruen}, {Gschwend}, {Gutierrez}, {Honscheid}, {Jain}, {James}, {Jeltema}, {Johnson}, {Johnson}, {Kent}, {Krause}, {Kron}, {Kuehn}, {Kuhlmann}, {Lahav}, {Lima}, {Maia}, {March}, {McMahon}, {Menanteau}, {Miquel}, {Mohr}, {Nichol}, {Nord}, {Ogando}, {Petravick}, {Plazas}, {Romer}, {Roodman}, {Rykoff}, {Sanchez}, {Scarpine}, {Schubnell}, {Sevilla-Noarbe}, {Smith}, {Smith}, {Suchyta}, {Swanson}, {Tarle}, {Thomas}, {Thomas}, {Troxel}, {Vikram}, {Wechsler}, {Weller}, {Dark Energy Survey}, \& {Dark Energy Camera GW-EM Collaboration}}]{SoaresSantos2017}
{Soares-Santos}, M., {Holz}, D.~E., {Annis}, J., {et~al.} 2017, \bibinfo{title}{{The Electromagnetic Counterpart of the Binary Neutron Star Merger LIGO/Virgo GW170817. I. Discovery of the Optical Counterpart Using the Dark Energy Camera},} \apjl, 848, L16, \dodoi{10.3847/2041-8213/aa9059}

\bibitem[{D. {Spergel} {et~al.}(2015){Spergel}, {Gehrels}, {Baltay}, {Bennett}, {Breckinridge}, {Donahue}, {Dressler}, {Gaudi}, {Greene}, {Guyon}, {Hirata}, {Kalirai}, {Kasdin}, {Macintosh}, {Moos}, {Perlmutter}, {Postman}, {Rauscher}, {Rhodes}, {Wang}, {Weinberg}, {Benford}, {Hudson}, {Jeong}, {Mellier}, {Traub}, {Yamada}, {Capak}, {Colbert}, {Masters}, {Penny}, {Savransky}, {Stern}, {Zimmerman}, {Barry}, {Bartusek}, {Carpenter}, {Cheng}, {Content}, {Dekens}, {Demers}, {Grady}, {Jackson}, {Kuan}, {Kruk}, {Melton}, {Nemati}, {Parvin}, {Poberezhskiy}, {Peddie}, {Ruffa}, {Wallace}, {Whipple}, {Wollack}, \& {Zhao}}]{roman}
{Spergel}, D., {Gehrels}, N., {Baltay}, C., {et~al.} 2015, \bibinfo{title}{{Wide-Field InfrarRed Survey Telescope-Astrophysics Focused Telescope Assets WFIRST-AFTA 2015 Report},} arXiv e-prints, arXiv:1503.03757, \dodoi{10.48550/arXiv.1503.03757}

\bibitem[{R. Stein {et~al.}(2025)Stein, Ahumada, Kasliwal, Du~Laz, Pathak, Swain, Salgundi, Bhalerao, Hall, {Ztf Collaboration}, \& {Growth Collaboration}}]{stein_ligovirgokagra_2025}
Stein, R., Ahumada, T., Kasliwal, M., {et~al.} 2025, \bibinfo{title}{{LIGO}/{Virgo}/{KAGRA} {S250818k}: {Candidates} from the {Zwicky} {Transient} {Facility},} GRB Coordinates Network, 41414, 1.
\newblock \url{https://ui.adsabs.harvard.edu/abs/2025GCN.41414....1S}

\bibitem[{S. {Stevenson} {et~al.}(2025){Stevenson}, {M{\"o}ller}, \& {Powell}}]{Stevenson2025}
{Stevenson}, S., {M{\"o}ller}, A., \& {Powell}, J. 2025, \bibinfo{title}{{Strategy for identifying Vera C. Rubin Observatory kilonova candidates for targeted gravitational-wave searches},} arXiv e-prints, arXiv:2510.12932, \dodoi{10.48550/arXiv.2510.12932}

\bibitem[{T. {Sumi} {et~al.}(2025){Sumi}, {Buckley}, {Kutyrev}, {Tamura}, {Bennett}, {Bond}, {Cataldo}, {Durbak}, {Cenko}, {Fixsen}, {Guiffreda}, {Hamada}, {Hirao}, {Idei}, {Kelly}, {Loose}, {Lotkin}, {Lyness}, {Maher}, {Makida}, {Matsunaga}, {Miyazaki}, {Mosby}, {Moseley}, {Nagai}, {Nagano}, {Nakayama}, {Nishio}, {Nunota}, {Ogawa}, {Oishi}, {Okumoto}, {Rattenbury}, {Satoh}, {Sharp}, {Suzuki}, {Tamaoki}, {Troja}, {White}, \& {Yama}}]{Sumi2025}
{Sumi}, T., {Buckley}, D. A.~H., {Kutyrev}, A.~S., {et~al.} 2025, \bibinfo{title}{{The Prime Focus Infrared Microlensing Experiment (PRIME): First Results},} arXiv e-prints, arXiv:2508.14474, \dodoi{10.48550/arXiv.2508.14474}

\bibitem[{O. {Tange}(2024){Tange}}]{2024zndo..14550073T}
{Tange}, O. 2024, \bibinfo{title}{{GNU Parallel 20241222 ('Bashar') [stable]},} Zenodo, \dodoi{10.5281/zenodo.14550073}

\bibitem[{N.~R. {Tanvir} {et~al.}(2013){Tanvir}, {Levan}, {Fruchter}, {Hjorth}, {Hounsell}, {Wiersema}, \& {Tunnicliffe}}]{Tanvir2013}
{Tanvir}, N.~R., {Levan}, A.~J., {Fruchter}, A.~S., {et~al.} 2013, \bibinfo{title}{{A `kilonova' associated with the short-duration {\ensuremath{\gamma}}-ray burst GRB 130603B},} \nat, 500, 547, \dodoi{10.1038/nature12505}

\bibitem[{N.~R. {Tanvir} {et~al.}(2017){Tanvir}, {Levan}, {Gonz{\'a}lez-Fern{\'a}ndez}, {Korobkin}, {Mandel}, {Rosswog}, {Hjorth}, {D'Avanzo}, {Fruchter}, {Fryer}, {Kangas}, {Milvang-Jensen}, {Rosetti}, {Steeghs}, {Wollaeger}, {Cano}, {Copperwheat}, {Covino}, {D'Elia}, {de Ugarte Postigo}, {Evans}, {Even}, {Fairhurst}, {Figuera Jaimes}, {Fontes}, {Fujii}, {Fynbo}, {Gompertz}, {Greiner}, {Hodosan}, {Irwin}, {Jakobsson}, {J{\o}rgensen}, {Kann}, {Lyman}, {Malesani}, {McMahon}, {Melandri}, {O'Brien}, {Osborne}, {Palazzi}, {Perley}, {Pian}, {Piranomonte}, {Rabus}, {Rol}, {Rowlinson}, {Schulze}, {Sutton}, {Th{\"o}ne}, {Ulaczyk}, {Watson}, {Wiersema}, \& {Wijers}}]{Tanvir2017}
{Tanvir}, N.~R., {Levan}, A.~J., {Gonz{\'a}lez-Fern{\'a}ndez}, C., {et~al.} 2017, \bibinfo{title}{{The Emergence of a Lanthanide-rich Kilonova Following the Merger of Two Neutron Stars},} \apjl, 848, L27, \dodoi{10.3847/2041-8213/aa90b6}

\bibitem[{ {The LIGO Scientific Collaboration} {et~al.}(2025){The LIGO Scientific Collaboration}, {the Virgo Collaboration}, {the KAGRA Collaboration}, Abac, Abouelfettouh, Acernese, Ackley, Adamcewicz, Adhicary, Adhikari, Adhikari, Adhikari, Adkins, Afroz, Agarwal, Agathos, Aghaei~Abchouyeh, Aguiar, Ahmadzadeh, Aiello, Ain, Ajith, Akutsu, Albanesi, Alfaidi, Al-Jodah, Alléné, Allocca, Al-Shammari, Altin, Alvarez-Lopez, Amarasinghe, Amato, Amra, Ananyeva, Anderson, Anderson, Andia, Ando, Andrade, Andrés-Carcasona, Andrić, Anglin, Ansoldi, Antelis, Antier, Aoumi, Appavuravther, Appert, Apple, Arai, Araya, Araya, Arca~Sedda, Areeda, Argianas, Aritomi, Armato, Armstrong, Arnaud, Arogeti, Aronson, Arun, Ashton, Aso, Assiduo, Assis~de Souza~Melo, Aston, Astone, Attadio, Aubin, AultONeal, Avallone, Babak, Badaracco, Badger, Bae, Bagnasco, Bagui, Baiotti, Bajpai, Baka, Baker, Ball, Ballardin, Ballmer, Banagiri, Banerjee, Bankar, Baptiste, Baral, Barayoga, Barish, Barker, Barman, Barneo, Barone, Barr, Barsotti,
  Barsuglia, Barta, Bartoletti, Barton, Bartos, Basak, Basalaev, Bassiri, Basti, Bates, Bawaj, Baxi, Bayley, Baylor, Baynard, Bazzan, Bedakihale, Beirnaert, Bejger, Belardinelli, Bell, Bellie, Bellizzi, Beltran-Martinez, Benoit, Bentara, Bentley, Ben~Yaala, Bera, Bergamin, Berger, Bernuzzi, Beroiz, Berry, Bersanetti, Bertolini, Betzwieser, Beveridge, Bevilacqua, Bevins, Bhandare, Bhatt, Bhattacharjee, Bhaumik, Bhowmick, Biancalana, Bianchi, Bilenko, Billingsley, Binetti, Bini, Binu, Birnholtz, Biscoveanu, Bisht, Bitossi, Bizouard, Blaber, Blackburn, Blagg, Blair, Blair, Bobba, Bode, Boileau, Boldrini, Bolingbroke, Bolliand, Bonavena, Bondarescu, Bondu, Bonilla, Bonilla, Bonino, Bonnand, Booker, Borchers, Borhanian, Boschi, Bose, Bossilkov, Boudon, Bozzi, Bradaschia, Brady, Branch, Branchesi, Braun, Briant, Brillet, Brinkmann, Brockill, Brockmueller, Brooks, Brown, Brown, Brozzetti, Brunett, Bruno, Bruntz, \& Bryant}]{the_ligo_scientific_collaboration_gwtc-40_2025}
{The LIGO Scientific Collaboration}, {the Virgo Collaboration}, {the KAGRA Collaboration}, {et~al.} 2025, \bibinfo{title}{{GWTC}-4.0: {Population} {Properties} of {Merging} {Compact} {Binaries},} arXiv, \dodoi{10.48550/arXiv.2508.18083}

\bibitem[{E. {Troja}(2023){Troja}}]{Troja2023}
{Troja}, E. 2023, \bibinfo{title}{{Eighteen Years of Kilonova Discoveries with Swift},} Universe, 9, 245, \dodoi{10.3390/universe9060245}

\bibitem[{E. {Troja} {et~al.}(2017){Troja}, {Piro}, {van Eerten}, {Wollaeger}, {Im}, {Fox}, {Butler}, {Cenko}, {Sakamoto}, {Fryer}, {Ricci}, {Lien}, {Ryan}, {Korobkin}, {Lee}, {Burgess}, {Lee}, {Watson}, {Choi}, {Covino}, {D'Avanzo}, {Fontes}, {Gonz{\'a}lez}, {Khandrika}, {Kim}, {Kim}, {Lee}, {Lee}, {Kutyrev}, {Lim}, {S{\'a}nchez-Ram{\'\i}rez}, {Veilleux}, {Wieringa}, \& {Yoon}}]{Troja2017}
{Troja}, E., {Piro}, L., {van Eerten}, H., {et~al.} 2017, \bibinfo{title}{{The X-ray counterpart to the gravitational-wave event GW170817},} \nat, 551, 71, \dodoi{10.1038/nature24290}

\bibitem[{E. {Troja} {et~al.}(2018){Troja}, {Ryan}, {Piro}, {van Eerten}, {Cenko}, {Yoon}, {Lee}, {Im}, {Sakamoto}, {Gatkine}, {Kutyrev}, \& {Veilleux}}]{Troja150101B}
{Troja}, E., {Ryan}, G., {Piro}, L., {et~al.} 2018, \bibinfo{title}{{A luminous blue kilonova and an off-axis jet from a compact binary merger at z = 0.1341},} Nature Communications, 9, 4089, \dodoi{10.1038/s41467-018-06558-7}

\bibitem[{E. {Troja} {et~al.}(2019){Troja}, {Castro-Tirado}, {Becerra Gonz{\'a}lez}, {Hu}, {Ryan}, {Cenko}, {Ricci}, {Novara}, {S{\'a}nchez-R{\'a}mirez}, {Acosta-Pulido}, {Ackley}, {Caballero Garc{\'\i}a}, {Eikenberry}, {Guziy}, {Jeong}, {Lien}, {M{\'a}rquez}, {Pand ey}, {Park}, {Sakamoto}, {Tello}, {Sokolov}, {Sokolov}, {Tiengo}, {Valeev}, {Zhang}, \& {Veilleux}}]{Troja2019b}
{Troja}, E., {Castro-Tirado}, A.~J., {Becerra Gonz{\'a}lez}, J., {et~al.} 2019, \bibinfo{title}{{The afterglow and kilonova of the short GRB 160821B},} \mnras, 489, 2104, \dodoi{10.1093/mnras/stz2255}

\bibitem[{E. {Troja} {et~al.}(2022){Troja}, {Fryer}, {O'Connor}, {Ryan}, {Dichiara}, {Kumar}, {Ito}, {Gupta}, {Wollaeger}, {Norris}, {Kawai}, {Butler}, {Aryan}, {Misra}, {Hosokawa}, {Murata}, {Niwano}, {Pandey}, {Kutyrev}, {van Eerten}, {Chase}, {Hu}, {Caballero-Garcia}, \& {Castro-Tirado}}]{Troja2022}
{Troja}, E., {Fryer}, C.~L., {O'Connor}, B., {et~al.} 2022, \bibinfo{title}{{A nearby long gamma-ray burst from a merger of compact objects},} \nat, 612, 228, \dodoi{10.1038/s41586-022-05327-3}

\bibitem[{M.-H. Ulrich {et~al.}(1997)Ulrich, Maraschi, \& Urry}]{ulrich_variability_1997}
Ulrich, M.-H., Maraschi, L., \& Urry, C.~M. 1997, \bibinfo{title}{Variability of {Active} {Galactic} {Nuclei},} Annual Review of Astronomy and Astrophysics, 35, 445, \dodoi{10.1146/annurev.astro.35.1.445}

\bibitem[{S. {Valenti} {et~al.}(2017){Valenti}, {Sand}, {Yang}, {Cappellaro}, {Tartaglia}, {Corsi}, {Jha}, {Reichart}, {Haislip}, \& {Kouprianov}}]{Valenti2017}
{Valenti}, S., {Sand}, D.~J., {Yang}, S., {et~al.} 2017, \bibinfo{title}{{The Discovery of the Electromagnetic Counterpart of GW170817: Kilonova AT 2017gfo/DLT17ck},} \apjl, 848, L24, \dodoi{10.3847/2041-8213/aa8edf}

\bibitem[{ {Vera C. Rubin Observatory Science Pipelines Developers}(2025){Vera C. Rubin Observatory Science Pipelines Developers}}]{10.71929/rubin/2570545}
{Vera C. Rubin Observatory Science Pipelines Developers}. 2025, \bibinfo{title}{{The LSST Science Pipelines Software: Optical Survey Pipeline Reduction and Analysis Environment},}, {Project Science Technical Note} PSTN-019, {Vera C. Rubin Observatory}, \dodoi{10.71929/rubin/2570545}

\bibitem[{T. Vincent {et~al.}(2020)Vincent, Foucart, Duez, Haas, Kidder, Pfeiffer, \& Scheel}]{Vincent:2019kor}
Vincent, T., Foucart, F., Duez, M.~D., {et~al.} 2020, \bibinfo{title}{{Unequal Mass Binary Neutron Star Simulations with Neutrino Transport: Ejecta and Neutrino Emission},} Phys. Rev. D, 101, 044053, \dodoi{10.1103/PhysRevD.101.044053}

\bibitem[{C.~Z. Waters {et~al.}(2020)Waters, Magnier, Price, Chambers, Burgett, Draper, Flewelling, Hodapp, Huber, Jedicke, Kaiser, Kudritzki, Lupton, Metcalfe, Rest, Sweeney, Tonry, Wainscoat, \& Wood-Vasey}]{waters_pan-starrs_2020}
Waters, C.~Z., Magnier, E.~A., Price, P.~A., {et~al.} 2020, \bibinfo{title}{Pan-{STARRS} {Pixel} {Processing}: {Detrending}, {Warping}, {Stacking},} The Astrophysical Journal Supplement Series, 251, 4, \dodoi{10.3847/1538-4365/abb82b}

\bibitem[{Y.-H. {Yang} {et~al.}(2024){Yang}, {Troja}, {O'Connor}, {Fryer}, {Im}, {Durbak}, {Paek}, {Ricci}, {Bom}, {Gillanders}, {Castro-Tirado}, {Peng}, {Dichiara}, {Ryan}, {van Eerten}, {Dai}, {Chang}, {Choi}, {De}, {Hu}, {Kilpatrick}, {Kutyrev}, {Jeong}, {Lee}, {Makler}, {Navarete}, \& {P{\'e}rez-Garc{\'\i}a}}]{Yang2024}
{Yang}, Y.-H., {Troja}, E., {O'Connor}, B., {et~al.} 2024, \bibinfo{title}{{A lanthanide-rich kilonova in the aftermath of a long gamma-ray burst},} \nat, 626, 742, \dodoi{10.1038/s41586-023-06979-5}

\bibitem[{Y.-H. {Yang} {et~al.}(2025){Yang}, {Troja}, {Risti{\'c}}, {Yadav}, {El Kabir}, {S{\'a}nchez-Ram{\'\i}rez}, {Becerra}, {Fryer}, {O'Connor}, {Dichiara}, {Castro-Tirado}, {Angulo-Valdez}, {Becerra Gonz{\'a}lez}, {Font}, {Fox}, {Hu}, {Hu}, {Lee}, {Pereyra}, {Sintes}, {Watson}, \& {Oc{\'e}lotl C}}]{Yang2025ulz}
{Yang}, Y.-H., {Troja}, E., {Risti{\'c}}, M., {et~al.} 2025, \bibinfo{title}{{AT2025ulz and S250818k: zooming in with the Hubble Space Telescope},} arXiv e-prints, arXiv:2510.18854.
\newblock \doarXiv{2510.18854}

\bibitem[{J.-P. Zhu {et~al.}(2022)Zhu, Wang, Sun, Yang, Li, Hu, Qin, \& Wu}]{zhu_long-duration_2022}
Zhu, J.-P., Wang, X.~I., Sun, H., {et~al.} 2022, \bibinfo{title}{Long-duration {Gamma}-{Ray} {Burst} and {Associated} {Kilonova} {Emission} from {Fast}-spinning {Black} {Hole}-{Neutron} {Star} {Mergers},} The Astrophysical Journal, 936, L10, \dodoi{10.3847/2041-8213/ac85ad}

\end{thebibliography}
\bibliographystyle{aasjournal}



\end{document}